\documentclass[a4paper,UKenglish,cleveref, autoref, thm-restate, nosubfigcap]{lipics-v2021}

\pdfoutput=1 
\hideLIPIcs  

\graphicspath{{./assets/}}

\title{Multilevel Skeletonization Using Local Separators} 

\author{J. Andreas Bærentzen}{Department of Applied Mathematics and Computer Science, DTU, Denmark}{janba@dtu.dk}{0000-0003-2583-0660}{Partially supported by the Villum Foundation through Villum Investigator Project InnoTop.}

\author{Rasmus {Emil Christensen}}{Department of Applied Mathematics and Computer Science, DTU, Denmark}{}{}{}

\author{Emil {Toftegaard Gæde}}{Department of Applied Mathematics and Computer Science, DTU, Denmark}{etoga@dtu.dk}{0009-0001-9462-6359}{}

\author{Eva Rotenberg}{Department of Applied Mathematics and Computer Science, DTU, Denmark}{erot@dtu.dk}{0000-0001-5853-7909}{}

\authorrunning{J.A. Bærentzen, R.E. Christensen, E.T. Gæde, E. Rotenberg} 

\Copyright{J.A. Bærentzen, R.E. Christensen, E.T. Gæde, E. Rotenberg} 



\category{} 

\relatedversion{} 

\funding{\textit{Emil Toftegaard Gæde} and \textit{Eva Rotenberg}: supported by Eva Rotenberg's Carlsberg Foundation Young Researcher Fellowship CF21-0302 - ``Graph Algorithms with Geometric Applications''.}

\nolinenumbers 

\usepackage{todonotes}
\usepackage{longtable}
\usepackage{multirow}

\newcommand{\CC}{C\nolinebreak\hspace{-.05em}\raisebox{.4ex}{\tiny\bf +}\nolinebreak\hspace{-.10em}\raisebox{.4ex}{\tiny\bf +}}

\usepackage[ruled,noline,noend]{algorithm2e}
\DontPrintSemicolon
\SetFuncSty{textsc}
\SetKwProg{Fn}{}{:}{}
\SetKwFunction{FOverview}{Multilevel-Skeletonization}
\SetKwFunction{FCoarsen}{Coarsen}
\SetKwFunction{FRSS}{Restricted-Separator-Search}
\SetKwFunction{FProj}{Project-Refine}
\SetKwFunction{FNeighbor}{Neighbourhood}
\SetKwFunction{FPack}{Capacity-Pack}
\SetKwFunction{FExtract}{Extract-Skeleton}
\SetKwFunction{FShrink}{Minimise-Separator}

\EventEditors{Erin Chambers and Joachim Gudmundsson}
\EventNoEds{2}
\EventLongTitle{39th International Symposium on Computational Geometry (SoCG 2023)}
\EventShortTitle{SoCG 2023}
\EventAcronym{SoCG}
\EventYear{2023}
\EventDate{June 12--15, 2023}
\EventLocation{Dallas, Texas}
\EventLogo{socg-logo}
\SeriesVolume{42}
\ArticleNo{XX}

\begin{document}

\maketitle

\begin{abstract}

In this paper we give a new, efficient algorithm for computing curve skeletons, based on local separators. Our efficiency stems from a multilevel approach, where we solve small problems across levels of detail and combine these in order to quickly obtain a skeleton. 
We do this in a highly modular fashion, ensuring complete flexibility in adapting the algorithm for specific types of input or for otherwise targeting specific applications.

Separator based skeletonization was first proposed by Bærentzen and Rotenberg in [ACM Tran. Graphics'21], showing high quality output at the cost of running times which become prohibitive for large inputs. Our new approach retains the high quality output, and applicability to any spatially embedded graph, while being orders of magnitude faster for all practical purposes. 

We test our skeletonization algorithm for efficiency and quality in practice, comparing it to local separator skeletonization on the University of Groningen Skeletonization Benchmark [Telea'16].

\end{abstract}

\section{Introduction}
A curve skeleton is a compact simplified representation of a shape, consisting only of curves. The act of \emph{skeletonization}, in this context, is the computation of such a curve skeleton for a given input. For the remainder of this paper \emph{skeleton} refers exclusively to curve skeletons. Various fields, including feature extraction, visualisation and medical imaging, care not only about shapes and objects, but also about their structures and features. In applications such as shape matching, the skeleton acts as a simplified representation of an object, allowing for reduced computation cost~\cite{shapematching}, whereas in virtual navigation the curve skeleton can act as a collision free navigational structure~\cite{virtbronch,virtnav}.

The broad areas of application, and the different roles that skeletons play, lead to differing interpretations of exactly what the skeleton is. Although no widely agreed upon definition of skeletons exist, work has been done on narrowing down desirable properties of skeletons in the general case~\cite{CurveSkeletons}.

\begin{figure}[htb]
    \centering
    \includegraphics[width=0.24\textwidth]{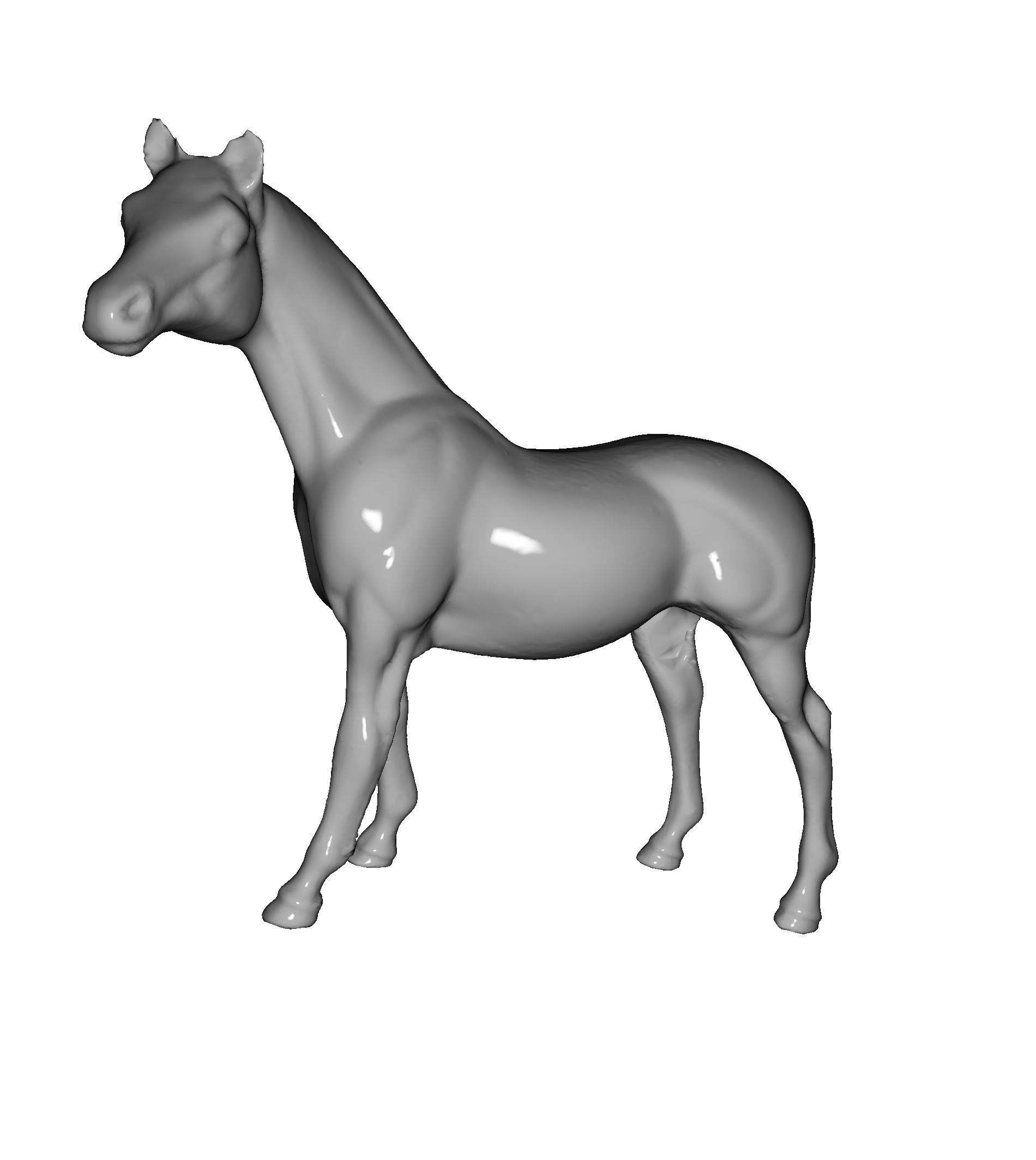}
    \includegraphics[width=0.24\textwidth]{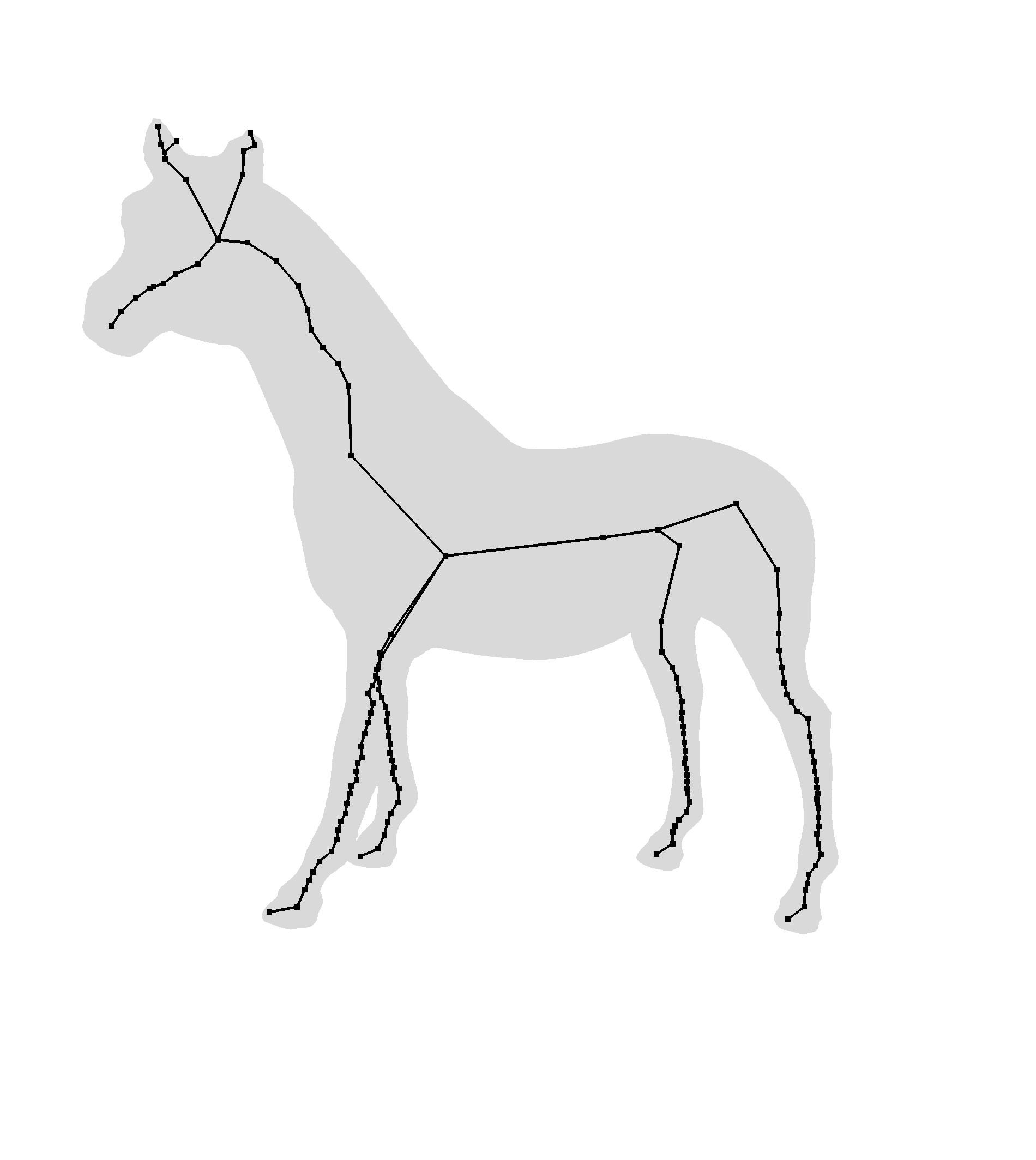}
    \includegraphics[width=0.24\textwidth]{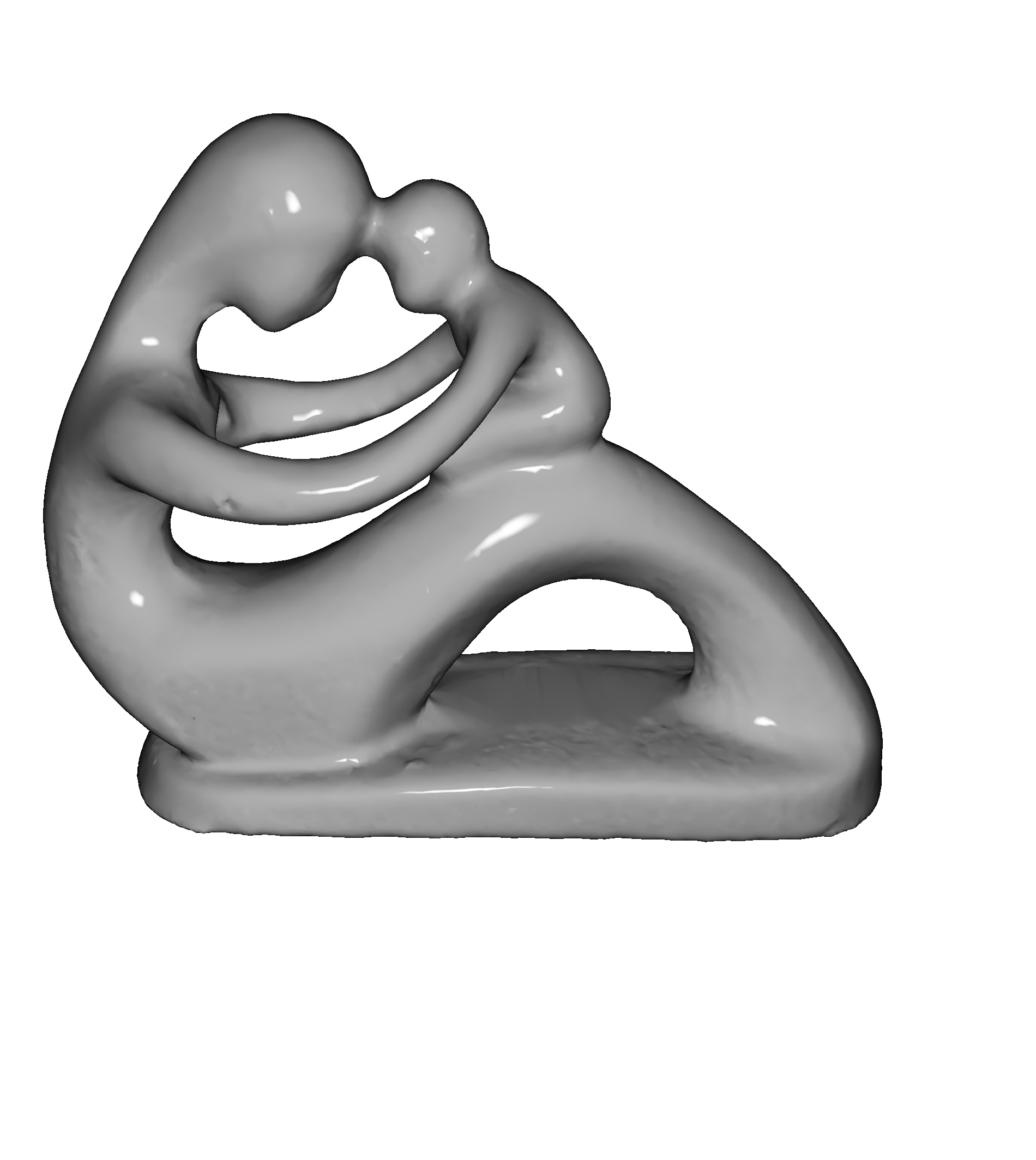}
    \includegraphics[width=0.24\textwidth]{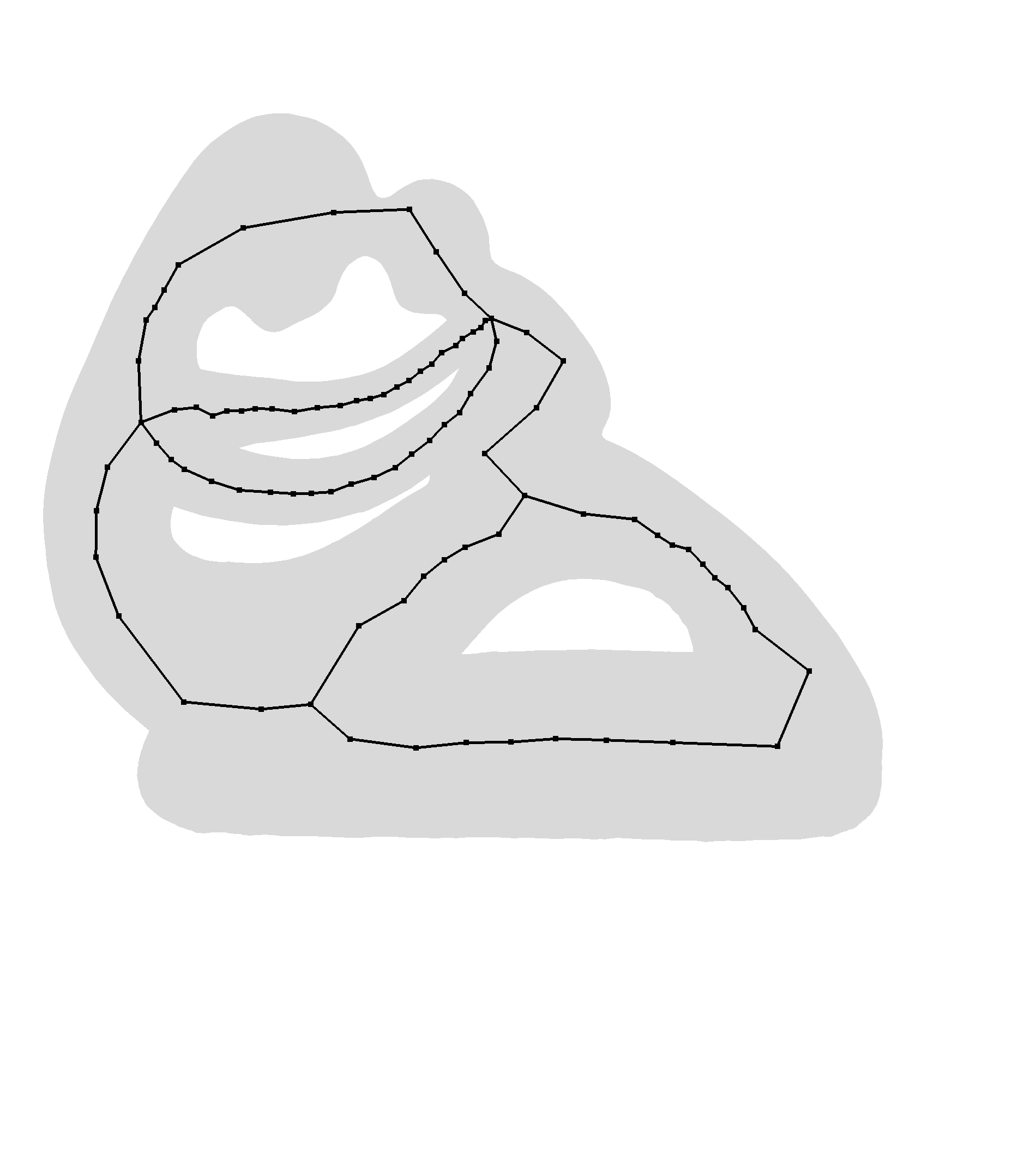}
    \caption{Shaded renders of triangle meshes and skeletons obtained by our algorithm.}
    \label{fig:intro_skel}
\end{figure}

Instead of giving a formal definition, we will base our work on the evocative if imprecise definition of skeletons as simplified curve representations of the underlying structure and topology. In Figure~\ref{fig:intro_skel} we show skeletons of various input, to exemplify our definition.

Many different approaches to skeletonization exist~\cite{skelstar}, such as computing and pruning the medial surface~\cite{medialprune}, computing mean curvature flow~\cite{meancurve} or contracting meshes~\cite{skelcluster}. In a recent paper A. Bærentzen and E. Rotenberg present a new algorithm that bases itself on computing \emph{local separators}~\cite{lsalg}. We refer to this algorithm as the local separator skeletonization algorithm, \emph{LSS}. This approach has the benefit that it requires only that the input be given as a spatially embedded graph, rather than a specific shape representation. This makes the method applicable to a wide variety of inputs, such as meshes, voxel grids or even input that does not necessarily represent a shape. In addition, the skeletons that it generates are of high quality, capturing features that contractive methods tend to miss. 
However, the algorithm is also computationally expensive. 

In this paper we present a multilevel algorithm for computing curve skeletons that we obtain by adapting LSS to a multilevel framework. Below, we start with some preliminaries and then present an overview of our contributions. Next, after a discussion of related work, we describe our approach to graph coarsening, projecting separators onto finer level graphs, and, finally, the multilevel skeletonization algorithm that builds on these components. We provide analyses of the algorithms in the paper and we test our work on a skeletonization benchmark. Our results show that our algorithm is orders of magnitude faster than that proposed by Bærentzen and Rotenberg  while producing skeletons of comparable quality.

\subsection{Preliminaries}
We consider the discrete skeletonization problem, where both the input and output is represented by spatially embedded undirected graphs. Formally, we consider skeletonization of a graph $G=(V,E)$ where each vertex is associated with a geometric position $p_{v\in V}\in\mathbb{R}^3$. Note that we make no other assumptions about the graph, such as whether it is sampled from the surface of a manifold, created from a point cloud, or otherwise.

In graph theory a \emph{vertex separator} is a set of vertices whose removal disconnects the graph. In~\cite{lsalg}, this notion is extended to \emph{local separators}, defined as a subset of vertices, $S\subset V$, that is a vertex separator of the subgraph induced by the closed neighbourhood of $S$. Likewise, the notion of a minimal local separator is defined as a local separator that is a minimal vertex separator of the subgraph induced by the closed neighbourhood. Intuitively, we cannot remove a vertex from a minimal local separator without the remaining set ceasing to be a local separator.
For the rest of this paper, the term \emph{separator} means local separator.

\subsection{Contributions}
\begin{figure}[htb]
    \centering
    \includegraphics[width=0.24\textwidth]{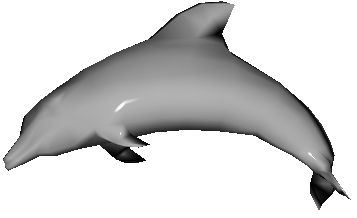}
    \includegraphics[width=0.24\textwidth]{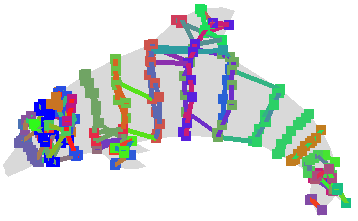}
    \includegraphics[width=0.24\textwidth]{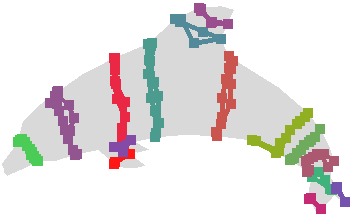}
    \includegraphics[width=0.24\textwidth]{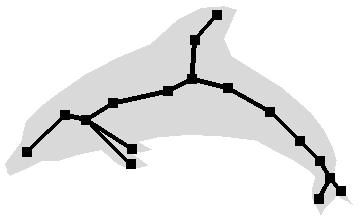}
    \caption{Visualisation of the three phases of the LSS algorithm. From left to right: A shaded render of the input, a number of computed minimal separators, a non-overlapping subset of the separators, and the resulting skeleton after extraction.}
    \label{fig:lssphases}
\end{figure}

The LSS algorithm computes skeletons through a three-phased approach. A large number of minimal local separators is computed, the minimal separators are selected using a greedy packing method, and, lastly, the skeleton is extracted from the packed set of minimal separators. A visualisation of these phases can be seen in Figure~\ref{fig:lssphases}.

As the algorithms for the first two phases play an intrinsic role in our algorithm, we give a brief description of these.

Computing local separators is done through a two-step process. First a region growing approach is used to find a local separator. A vertex is picked, and we iteratively add to the separator an adjacent vertex and check if the neighbourhood is disconnected. We refer to this as \emph{growing} a separator. Once a local separator has been found, it is heuristically minimised by removing vertices that would not destroy the separator. We refer to this as \emph{shrinking} a separator.

Because the running time of LSS is often dominated by the search for local separators, a sampling scheme is used to reduce computation. According to the scheme, vertices are selected for separator computation with probability $2^{-x}$, where $x$ is the number of previously computed separators that contain that vertex.

Unfortunately, sampling only addresses the number of separators that need to be computed and not the time it takes to compute each separator. In this paper we address the latter issue using a multilevel approach. Specifically, we find separators on coarser versions of the graph and project them back up onto the original graph. Importantly, this allows us to set a \emph{patience threshold} for the amount of computation that should be used to find a separator from a given vertex. When the threshold is exceeded we stop the search relying on a separator containing the given vertex to be found on a coarser level.

\subsection{Related Work}

Skeletonization, in terms of computing curve skeletons, is a diverse field not only in terms of interpretations of skeletons, but also in the algorithmic approaches. Several classifications of algorithms exist~\cite{CurveSkeletons,skelstar}, based on underlying traits of the algorithms.

The interpretation that the curve skeleton should lie on the medial surface, gives rise to methods that, in a sense, extract a curve skeleton from the medial surface of the input~\cite{medialprune,teleajalba,teleajalbaGPU,teleareniers,yanchambers}. Since the medial surface is highly sensitive to noise, so are the skeletons generated by these methods. 

A class of algorithms that are resilient to noise are the contractive methods, based on the concept of reducing the volume and surface area of the input until a skeleton is found~\cite{vcontract,mcontract,meancurve}. In their simplest forms these algorithms require that the input be manifold, however it is possible to extend to other types of input~\cite{lapcontract,pcskel}.

A related notion for shape analysis is that of Reeb graphs~\cite{reebg,reebfunc}. These can be used for skeletonization, lending themselves to a topologically driven class of algorithms~\cite{reebpc,reebrobust,reebfast}. The resulting skeletons depend on a parameter, giving some flexibility in targeting specific properties of the output, but also requiring great care in the choice of the parameter.

In addition there are algorithms that fit into classifications not presented here~\cite{skelcluster,skelflow,skelcrust}.

A very successful heuristic approach to the NP-complete problem of graph partitioning is that of multilevel algorithms~\cite{multilevelsurv}. Although the problem considered is different, we employ a similar multilevel scheme for vastly improving practical performance. Such multilevel schemes have been extensively studied~\cite{multi1995,multilevelanal,multipower}.

\section{The Multilevel Framework}

In its most general sense, the multilevel framework is a heuristic approach that aims to solve a problem by obtaining a solution to a smaller problem.

Initially, a series of increasingly simplified approximations of the input is generated. We call this the \emph{coarsening} phase, and the series of simplifications we call \emph{levels}. Since the last level is small, computing a solution is much faster. In graph partitioning literature, this is called the \emph{partitioning} phase; however, we will consider it in terms of solving a \emph{restricted} problem. Then, the solution found on the last level is transformed into a solution on the input through \emph{uncoarsening}. This process is also sometimes called \emph{projection and refinement}, since uncoarsening from one level to the previous is often done by projecting onto the previous level, and then employing some refinement process to improve the solution according to some heuristic.

By design, the multilevel framework is highly flexible. Various coarsening schemes can be used, that may prioritise preserving different properties of the input when simplifying. The restricted problem can be solved by any reasonable approach, and the refinement strategies can be adapted to suit the application.

Our algorithm works by first coarsening the input into several levels of decreasing resolution. The details of this coarsening is described in Section~\ref{sec:coarsening}. Once the hierarchy of graphs has been generated, we do a restricted search for local separators on each level of resolution. The details are covered in Section~\ref{sec:rss}, but the intuition is that searching for large local separators is slow in practice, and by restricting our search we save computation. Since the separators found are small, this does however also mean that we are only able to capture small features of the structure.

\begin{figure}[htb]
    \centering
    \includegraphics[width=0.6\textwidth]{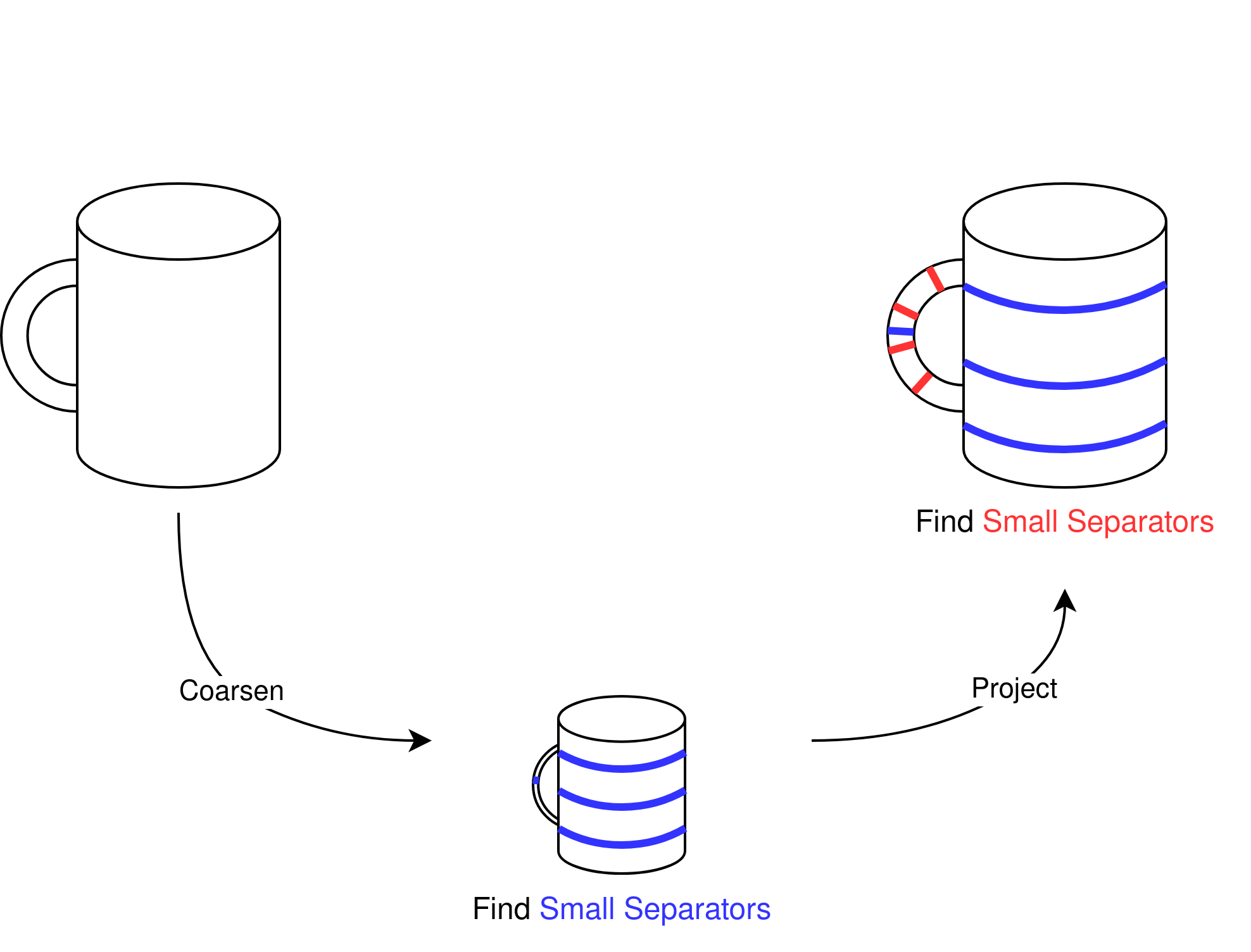}
    \caption{Visualisation of the multilevel skeletonization approach. A solid cylinder with a handle is coarsened until it is of small size. A number of small local separators are found (shown in blue), and then projected back to the original input. Searching for small local separators again yields the separators around the handle (shown in red), but separators are too large at this level to be discovered around the cylinder. We combine the separators to obtain a general solution. }
    \label{fig:multilevelscheme}
\end{figure}

Since small features obtained on low resolution can represent large features on the original input, we obtain separators capturing features of varying sizes by searching for separators across every level.

By projection and refinement, see Section~\ref{sec:uncoarsen}, we then transform the minimal local separators found across the levels into minimal local separators on the input. These can then be packed and extracted by the approach of LSS. The procedure is visualised in Figure~\ref{fig:multilevelscheme}.

\subsection{Coarsening}\label{sec:coarsening}
Given as input a graph, $G=(V,E)$, we construct a sequence of increasingly simplified graphs, $G_0,G_1,\ldots,G_l$ s.t. $G_0\succ G_1\succ\ldots\succ G_l$ where $l=O(\log n)$ and $G_i\succ G_j$ denotes that $G_j$ is a minor of $G_i$, and $G_i=(V_i,E_i)$. Moreover $G_0=G$ and $\forall i\in[0,l),|V_i|\geq2|V_{i+1}|$.

We do this by a matching contraction scheme, in which we repeatedly construct and contract maximal matchings. Various approaches to such coarsening schemes exist in literature~\cite{multilevelanal}, and from these, we choose to consider \emph{light edge matching}.

To construct $G_{i+1}$ from $G_i$, greedily find a maximal matching and contract it. Such a matching can be constructed in $O(|E_i|)$ time by visiting vertices in a random order, matching them to an unmatched neighbour of smallest euclidean distance. We repeat this procedure until the number of vertices has been at least halved.

\begin{figure}[htb]
    \centering
    \includegraphics[width=0.32\textwidth]{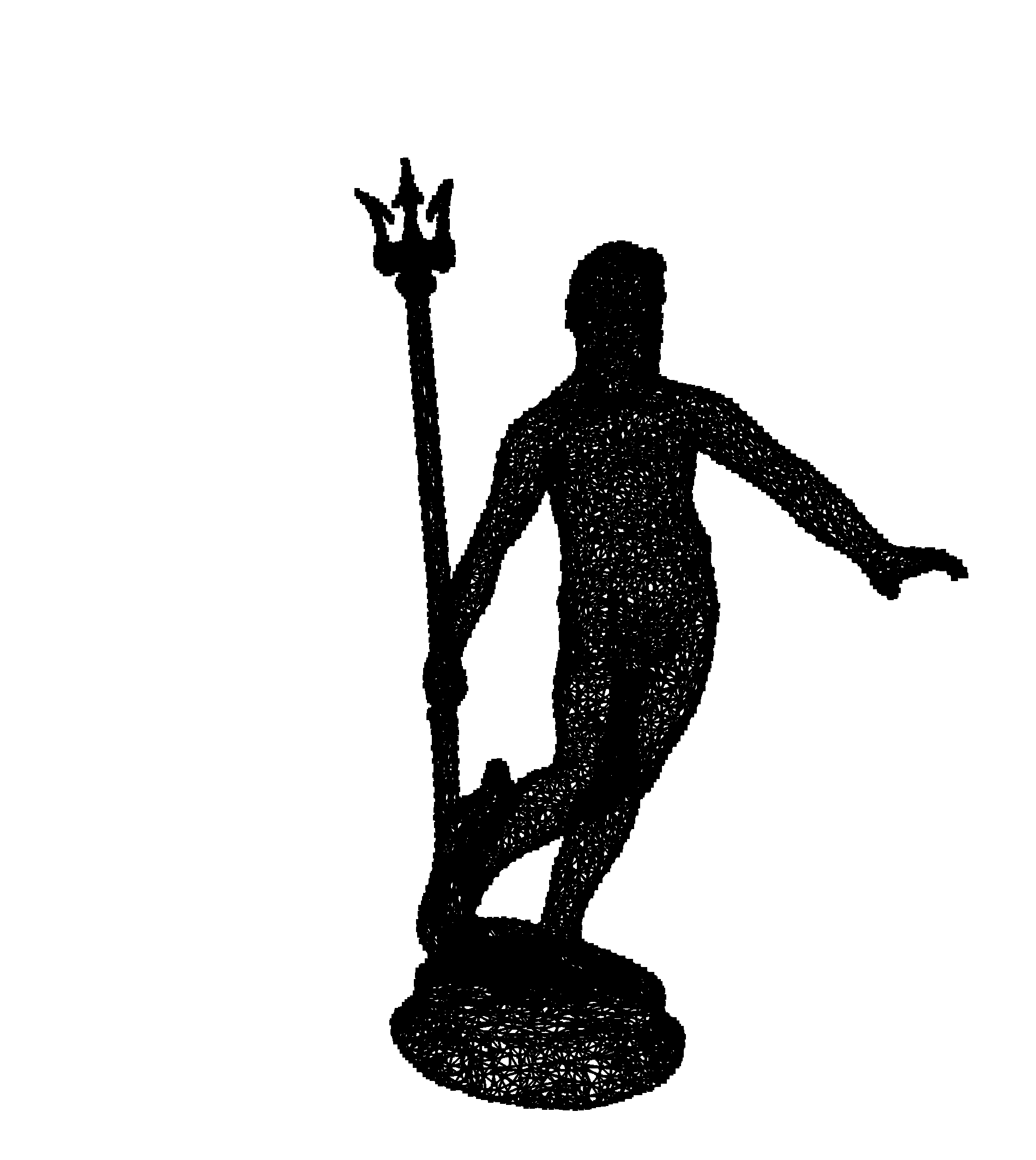}
    \includegraphics[width=0.32\textwidth]{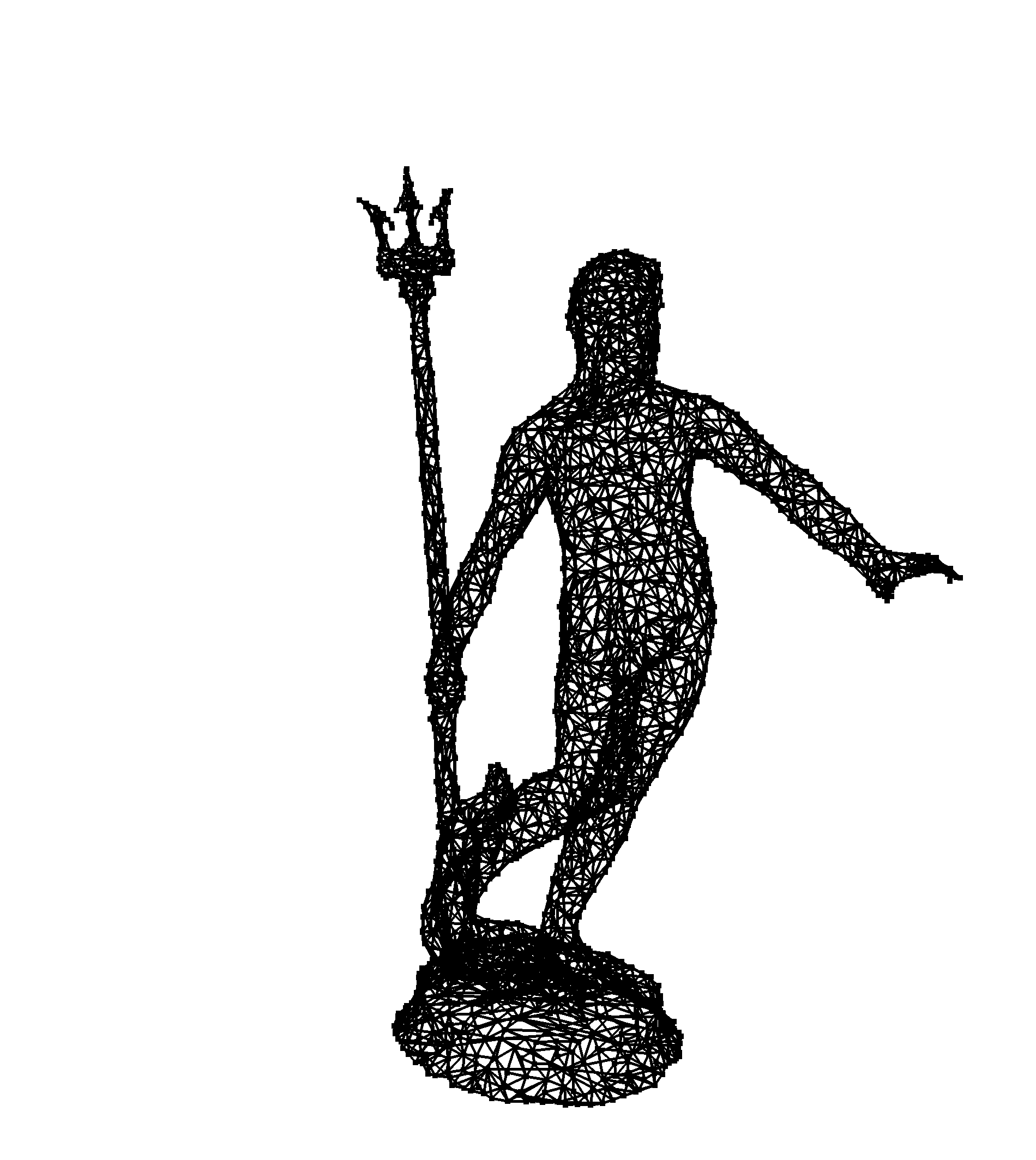}
    \includegraphics[width=0.32\textwidth]{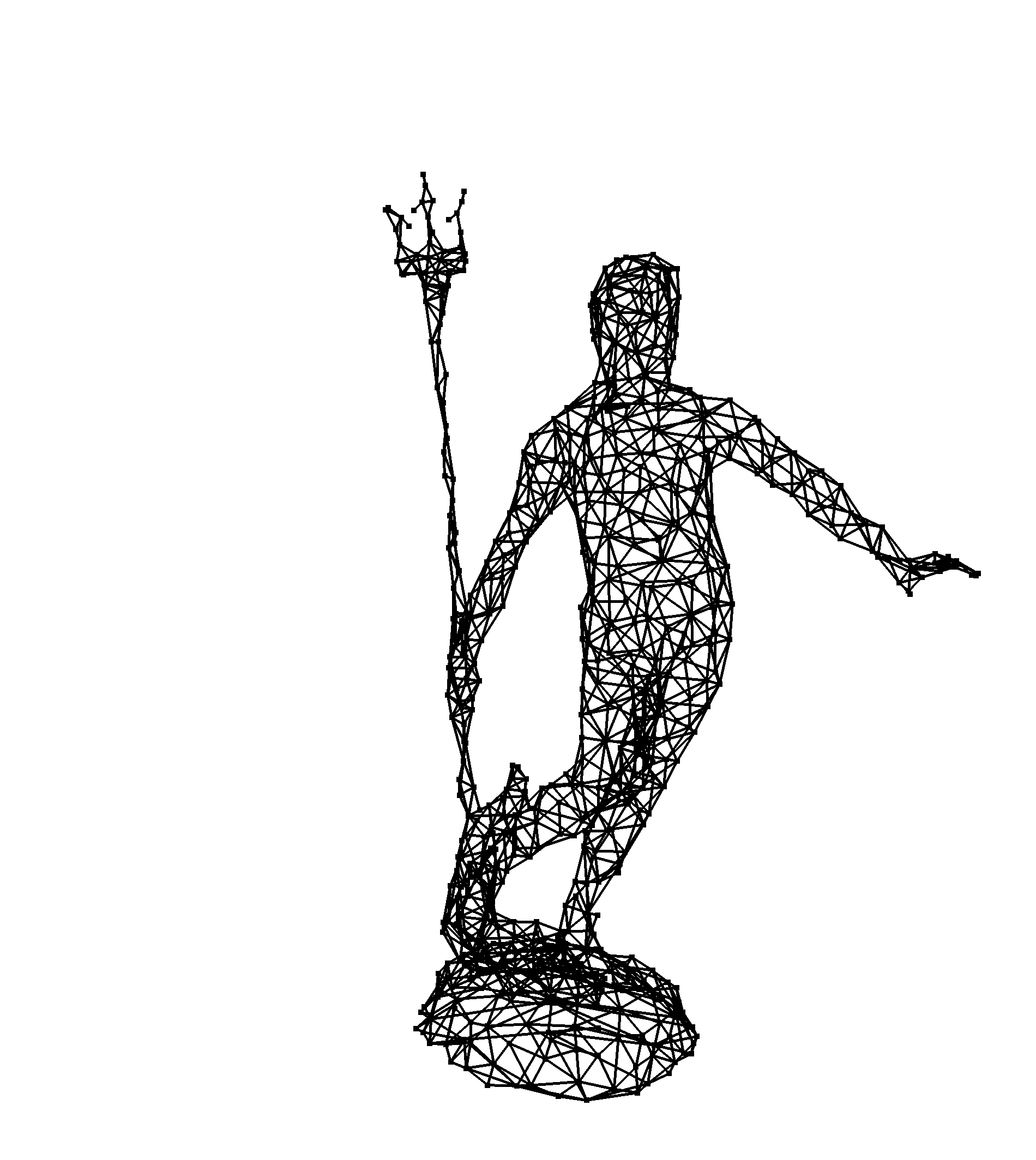}
    \caption{A series of increasingly simplified approximations of \texttt{neptune.ply}, from the Groningen Skeletonization Benchmark, obtained through light edge matching contraction.}
    \label{fig:nep_msg}
\end{figure}

In Figure~\ref{fig:nep_msg} we show some of the graphs obtained during coarsening of a triangle mesh resembling a statue of Neptune.

Note that by contraction we always preserve the number of connected components. This is one of the homotopy-preserving properties of the algorithm.

\paragraph*{Theoretical Analysis}
In the worst case, we may spend $O(|V_i|)$ rounds of contraction in order to reach the desired number of vertices. This is a well known problem of matching contraction schemes on general graphs, but graphs obtained from the world of geometry tend to take a small number of rounds to contract~\cite{multipower}.

A general bound on the time spent on the coarsening phase is then $\sum_{i=0}^l(|V_i||E_i|)=|E|\sum_{i=0}^l|V_i|=O(|V||E|)$. For graphs that are contracted in a constant number of rounds we get $\sum_{i=0}^l|E_i|$ and if we furthermore have $|E_i|=O(|V_i|)$, as is the case for triangle meshes and voxel grids, the bound becomes $O(|V|)$.

\subsection{Restricted Separator Search}\label{sec:rss}
For a given connected set of vertices, $V'$, we refer to the subgraph induced by vertices adjacent to $V'$, that are not in $V'$ themselves, as the \emph{front} of $V'$ and denote it $F(V')$.

A region growing based approach to computing local separators is given in~\cite{lsalg}, where a separator, $\Sigma$, is iteratively grown until $F(\Sigma)$ is disconnected. The approach uses an enclosing ball around the vertices of $\Sigma$ to guide what vertex of $F(\Sigma)$ is added next, and the connectivity of $F(\Sigma)$ is then checked by traversal. As noted by the authors, it is possible to improve performance of the search by using a dynamic connectivity data structure to maintain the front, so that a traversal in every iteration is avoided.

In addition to adapting the algorithm to use a dynamic connectivity data structure, we will also restrict the number of iterations the search performs. Given a vertex, $v$, set $\Sigma_0=\emptyset,F_0=\{v\}$, and then iteratively construct $\Sigma_i=\Sigma_{i-1}\cup\{v_i\in F(\Sigma_{i-1})\}$, where $v_i$ is the closest neighbour of the front to an enclosing sphere around $\Sigma_{i-1}$. Maintain $F(\Sigma_i)$, update the enclosing sphere and repeat until $F(\Sigma_i)$ is disconnected or empty, or $|\Sigma_i|$ exceeds a threshold value. Pseudocode for this restricted separator search is shown in Algorithm~\ref{alg:rss}.

\begin{algorithm}[htb]
\caption{{\normalsize Restricted Separator Search}\\ Given a spatially embedded graph, $G$, a starting vertex, $v_0$, and a thresholding value, $\alpha$, search for a separator of size at most $\alpha$ and return it, or $\emptyset$ if failure. Here $\epsilon$ is a small constant to prevent division by zero.}
\label{alg:rss}
\SetKwFunction{Fconn}{connect}
\SetKwFunction{Fdisc}{remove}
\SetKwFunction{Fnomcom}{number-of-components}
\SetKwFunction{Ffsr}{Front-Size-Ratio}
\Fn{\FRSS{$G,v_0,\alpha$}}{
    $\Sigma = \emptyset$\;
    $F = (\{v_0\},\emptyset)$\;
    $\mathbf{c} = \mathbf{p}_{v_0}$\;
    $i = 0$\;
    $r = 0$\;
    \Repeat{\Fnomcom{$F$}$>1$ \bf{or} $i=\alpha$}{
        $v=\arg\min_{f\in V(F)}\lVert \mathbf{c}-\mathbf{p}_f\rVert$\tcp*[r]{Scan front for closest vertex}
        \If{$\lVert\mathbf{c}-\mathbf{p}_v\rVert>r$}{
            $r=\frac{1}{2}(r+\lVert\mathbf{c}-\mathbf{p}_v\rVert)$\tcp*[r]{Update the sphere}
            $\mathbf{c}=\mathbf{p}_v+\frac{r}{\epsilon+\lVert\mathbf{c}-\mathbf{p}_v\rVert}(\mathbf{c}-\mathbf{p}_v)$\;
        }
        $\Sigma = \Sigma\cup\{v\}$\;
        $\Fdisc{$F,v$}$\;
        \For{$(x,y)\in E(\FNeighbor{G,v}-\Sigma)$}{
            $\Fconn{$F,x,y$}$\tcp*[r]{Maintain the front of $\Sigma$}
        }
        $i=i+1$\;
    }
    \If{$\Fnomcom{$F$}=1$}{\Return{$\emptyset$}}
    \Return{$\Sigma$}
}
\end{algorithm}

\paragraph*{Theoretical Analysis}
We analyse the complexity of this restricted separator search in terms of the graph $G'=\Sigma\cup F(\Sigma)$ with $n'$ vertices and $m'$ edges, using the dynamic connectivity data structure of Holm, de Lichtenberg and Thorup with updates in amortized $O(\log^2n')$ time~\cite{dyncon}.
Since the size of the separator is at most $\alpha$ and we add one vertex each iteration, we use at most $\alpha$ iterations selecting the closest vertex from the front, updating the bounding sphere and maintaining the dynamic connectivity structure. Selecting the closest vertex is done naively with a scan through the front, taking $O(n')$ time and updating the bounding sphere takes $O(1)$ time each iteration. This gives a running time of $O(\alpha n')$. Each edge in the dynamic connectivity structure is inserted and removed at most once, with each operation taking $O(\log^2n')$ amortized time, totalling $O(m'\log^2n')$. The total running time is then $O(\alpha n'+m'\log^2n')$.

In the general case we give no better worst case bound than $O(\alpha|V|+|E|\log^2|V|)$. For graphs of bounded maximum degree we can bound the size of the front. Let $\Delta$ be the maximum degree of $G$, then $n'=O(|\Sigma|\Delta)=O(\alpha\Delta)$ and $m'=O(\alpha\Delta^2)$. This gives a time of $O(\alpha^2\Delta+\alpha\Delta^2\log^2(\alpha\Delta))$. In addition if we choose $\alpha$ to be a small constant, the bound is further improved to $O(\Delta^2\log^2\Delta)$. For graphs where $\Delta=O(1)$ as for voxel grids or knn-graphs, the search then becomes $O(1)$. Note that for this bound to be applicable across the entirety of the algorithm, the degree needs to remain bounded through coarsening.

\subsection{Projection and Refinement}\label{sec:uncoarsen}
For projecting separators to graphs of higher levels of detail, we employ a simple uncoarsening technique. By storing information about what vertices were contracted during coarsening, we can reverse the contractions that gave rise to the vertices of a given separator. Note however that a separator that has been projected in such a way is not guaranteed to be minimal.

\begin{figure}[htb]
    \centering
    \includegraphics[width=\textwidth]{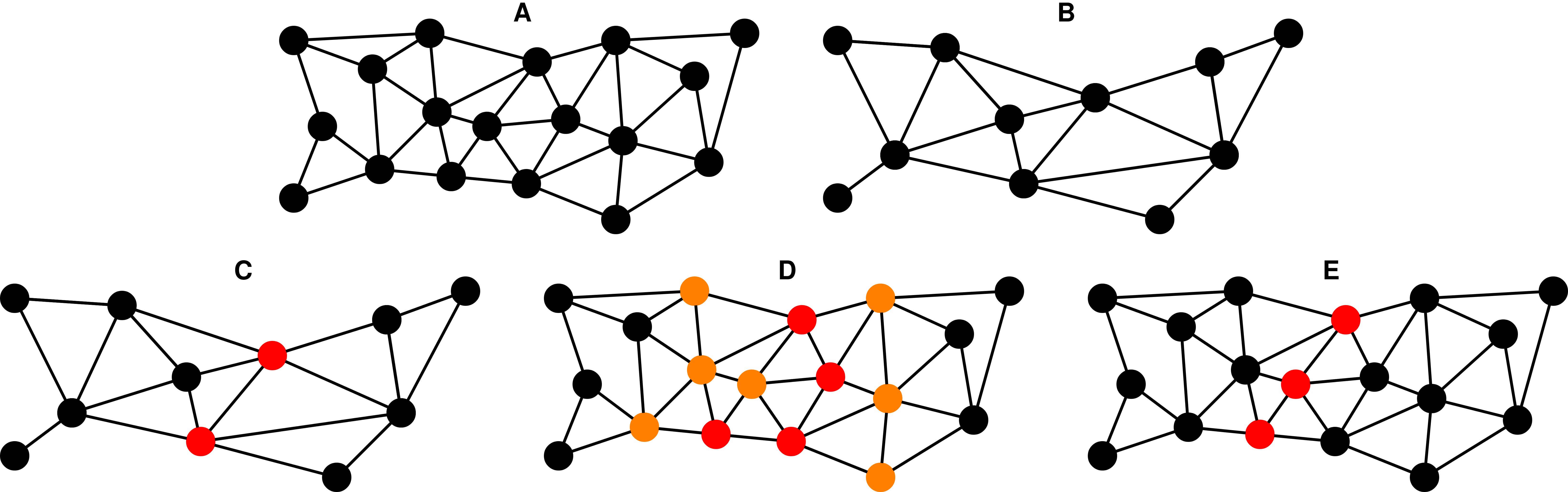}
    \caption{A separator undergoing expansion as part of refinement. (A) shows an input, (B) a coarsened representation, (C) a computed separator denoted by red vertices, (D) the projected separator denoted by red vertices and the added vertices denoted by orange, (E) shows the separator obtained by minimising the thickened separator.}
    \label{fig:thicksep}
\end{figure}

The simplest refinement scheme is thus one that uses the algorithm for minimising separators as in LSS. The minimising algorithm is a heuristic approach that seeks to minimise a separator such that the structure becomes that of a thin band. When used on separators that are obtained through projection, there is not necessarily much room for choice. Therefore we consider a variation of our refinement scheme, we thus choose to "thicken" the separators after projection, by adding the adjacent vertices if it would not destroy the separator. This gives the heuristic minimisation more options for creating separators of shorter length, as visualised in Figure~\ref{fig:thicksep}.

Projecting a separator can be done in linear time proportional to the size of the resulting separator, while the minimisation in worst case takes quadratic time (see Appendix~\ref{sec:analysisofshrinking}).

After processing each separator in this way, we obtain a set of minimal separators for the current level. If we accumulate separators indiscriminately, we will spend time projecting and refining separators that will ultimately be discarded due to overlapping. If we perform set packing on every level, we are going to be too eager in our efforts, discarding things that might not overlap once projected further. Intuitively, we would like to only discard separators if there is a large overlap.

To do this, we associate with each vertex, $v$, of every graph, $G_i$, a capacity, $c_v^i$, equal to the sum of capacities of vertices contracted to obtain it. For vertices of $G_0$ we define the capacities as $1$, formally $\forall v\in V(G_0), c_v^0=1$. In this way, the capacity of a given vertex is the number of vertices of the original input that it represents.

We then modify the greedy set packing algorithm of LSS, so that we include a separator iff it would not cause any vertex to exceed its capacity. Since the capacities of $G_0$ are $1$, this packing is equivalent to the original when applied to the highest level of resolution, and thus we will still have a non-overlapping set of separators at the end.

Note however that this packing allows for duplicates to persist through packing, essentially reducing the capacities of vertices while providing no valuable information. To counteract this, we perform a filtering step using hashing to rid duplicates prior to the packing procedure. Filtering and packing in this way takes time linear in the sum of sizes of separators in the set.

\subsection{The Multilevel Skeletonization Algorithm}

With the details of the phases in place, we can then combine these to construct the multilevel skeletonization algorithm. Given a spatially embedded input graph, $G=(V,E)$, and a threshold value $\alpha$, we construct a curve skeleton by the following:

Generate $G_i$ from $G_{i-1}$ by coarsening, until $|G_l|\leq\alpha$ for some $l$. This generates the sequence of graphs of decreasing resolution $G_0,G_1,\ldots,G_l$ where $l=O(\log|V|)$ and $\sum_{i=0}^l|V_i|=O(|V|)$.

Then, starting at the lowest resolution, $G_l$, find restricted separators. We do this by the restricted separator search, starting at each vertex with probability $2^{-x}$, where $x$ is the number of currently computed separators containing that vertex, using $\alpha$ as the restriction on the size of the search. After computing the separators for a level, we perform capacity packing, and then we project the computed separators to the next level and refine them.
This process is repeated for every level until we arrive at the original graph. At this point, after performing capacity packing, we obtain a non-overlapping set of minimal separators from which we extract the skeleton, using the extraction procedure of LSS~\cite{lsalg}. Pseudocode for this algorithm can be seen in Algorithm~\ref{alg:overview}.

\begin{algorithm}[htb]
\caption{{\normalsize Multilevel Skeletonization}\\ Given a spatially embedded graph, $G$, and a thresholding value, $\alpha$, compute a curve skeleton.}
\label{alg:overview}
\Fn{\FOverview{$G,\alpha$}}{
    $G_0=G$\;
    $l=0$\;
    \Repeat(\tcp*[f]{Coarsening phase}){$|V(G_l)|\leq\alpha$}{
        $l=l+1$\;
        $G_l =\FCoarsen{$G_{l-1}$}$\;
    }
    $S=\emptyset$\tcp*[r]{Maintain set of minimal separators}
    \For(\tcp*[f]{From low to high resolution}){$i=l$ \KwTo $0$}{
        $S'=\emptyset$\;
        \For(\tcp*[f]{Project and refine from previous levels}){$s\in S$}{
            $S'=S'\cup\FProj{$s$}$\;
        }
        $S=S'$\;
        \For(\tcp*[f]{Search on this level}){$v\in V(G_i)$ \bf{with probability} $2^{-x(v)}$}{
            $s=\FRSS{$G_i,v,\alpha$}$\;
            $S=S\cup\{\FShrink{s}\}$\;
        }
        $S=\FPack{$S$}$\;
    }
    \Return{\FExtract{$G,S$}}
}
\end{algorithm}

\paragraph*{Theoretical Analysis}
For completeness' sake we consider then the complexity of the algorithm. Recall that the coarsening phase in the general worst case takes $O(|V||E|)$ time, but for not too irregular input takes $O(|V|)$ time.
We then perform a restricted separator search from each vertex across every level, which is $O(\alpha|V|^2+|V||E|\log^2|V|)$ in the general worst case, but $O(\alpha|V|)$ for graphs that retain constant bounded degree through coarsening. 
We consider then the time a single separator contributes to the total when expanding, filtering and packing. These operations are linear in the size of the separator on each level, which is worst case $O(|V_i|)$ on level $i$. This totals $\sum_{i=0}^lO(|V_i|)=O(|V|)$ for a single separator across all levels. Minimizing a single separator takes worst case $O(|V_i|^2)$ on level $i$, which for a single separator contributes  $\sum_{i=0}^lO(|V_i|^2)=O(|V|^2)$ across all levels.
We also perform packing on each level, linear in the sum of sizes of separators, which in total takes $\sum_{i=0}^lO(|V_i|^2)=O(|V|^2)$ time.
The general worst case bound then becomes $O(|V|^3+|V||E|\log^2|V|)$, which is an improvement over LSS. It is worth mentioning however, that in practice the running time for both LSS and our algorithm is heavily dominated by the search for separators, and that theoretically expensive procedures, such as minimisation, make up only a small fraction of the running time.

\section{Experiments}
\label{sec:experiments}

In this work, our main objective was to make an algorithm that produces the same quality of skeletons as LSS~\cite{lsalg}, only with improved running times, using new algorithmic ideas and algorithm engineering. As we will show in this section, the improvements to practical running times are very satisfactory. 

As for quality, it is our overall assessment that the quality has not been compromised by the speed-up. 

There is, however, no standard for how skeletons should be compared. In~\cite{skelstar} it is remarked that quantifying the quality of skeletons is an open challenge, but we shall instead compare ourselves only to skeletons obtained by LSS, to quantify the deviation obtained by employing the multilevel approach.

To do this, we will measure a number of metrics, namely the number of vertices in the skeleton, the number of leaf nodes, branch nodes, chordless cycles which estimates the genus of the input, and the directed Hausdorff distance in both directions. For our comparisons, it is the relationship between directed Hausdorff distances that matter, rather than the magnitudes. A high distance from an LSS skeleton to our skeleton, with a low distance the other way, could indicate that LSS captures a feature that our skeleton does not. Likewise for the inverse, which might indicate that we are capturing a feature that LSS does not deem to exist. We give our Hausdorff distances divided by the radius of a bounding sphere, to reduce influence from the differing scales of input.

We run our tests on the Groningen Skeletonization Benchmark~\cite{skelbench}, consisting of several triangle meshes of varying structure. The test are executed on HPC Cluster nodes with Xeon Gold 6226R (2.90GHz) CPUs, using 8 cores of a single CPU for each test. For running time measurements, tests are run three times, and the median value is reported.

For comparisons we examine three algorithms, namely the local separator skeletonization algorithm (LSS)~\cite{lsalg}, our multilevel algorithm using light edge matchings, described in Section~\ref{sec:coarsening}, as contraction scheme (LEM) as well as with light edge matchings and thickened separators, described in Section~\ref{sec:uncoarsen}, as the refinement scheme (LEMTS).

We note that the variation of LSS with which we compare our algorithms also includes usage of a dynamic connectivity structure, so that the search procedures are identical up to the threshold parameter.

\paragraph*{Implementation}
Our implementation is written in \CC, built into a fork of the GEL library, and made publicly available~\cite{github}. This is the same library that contains LSS, and as such our algorithms use the same underlying data structures and subroutines. All programs are compiled using \texttt{-O3} optimisation flags.
Details regarding the dynamic connectivity structure are given in Appendix~\ref{sec:dyncon}.
We run the restricted searches in parallel internally on each level, using a simple fork-join pattern, identical to that of LSS. To account for the multilevel structure of our algorithm, we then pack using a single thread, project using at most two threads and repeat the pattern for the next level. The decision to use only two threads for projection stems from the fact that the overhead associated with spinning up threads quickly outweighs the benefits of parallel projection since there are often few separators after packing.

\subsection{Results}
Here we present our results in terms of measurements on the Groningen Skeletonization Benchmark. Initially we argue for our choice of $\alpha$, showing how the threshold impacts both skeleton quality as well as running time. We then present a number of results relating to the skeletons themselves to showcase the quality of output. We then present our measurements of running times, as well as discuss interesting observations.

Additional measurements are presented in Appendix~\ref{sec:time} and Appendix~\ref{sec:skel}.

\paragraph*{The Right Amount of Patience -- Determining a Threshold Value $\alpha$}

\begin{figure}[htb]
    \centering
    \includegraphics[width=0.75\textwidth]{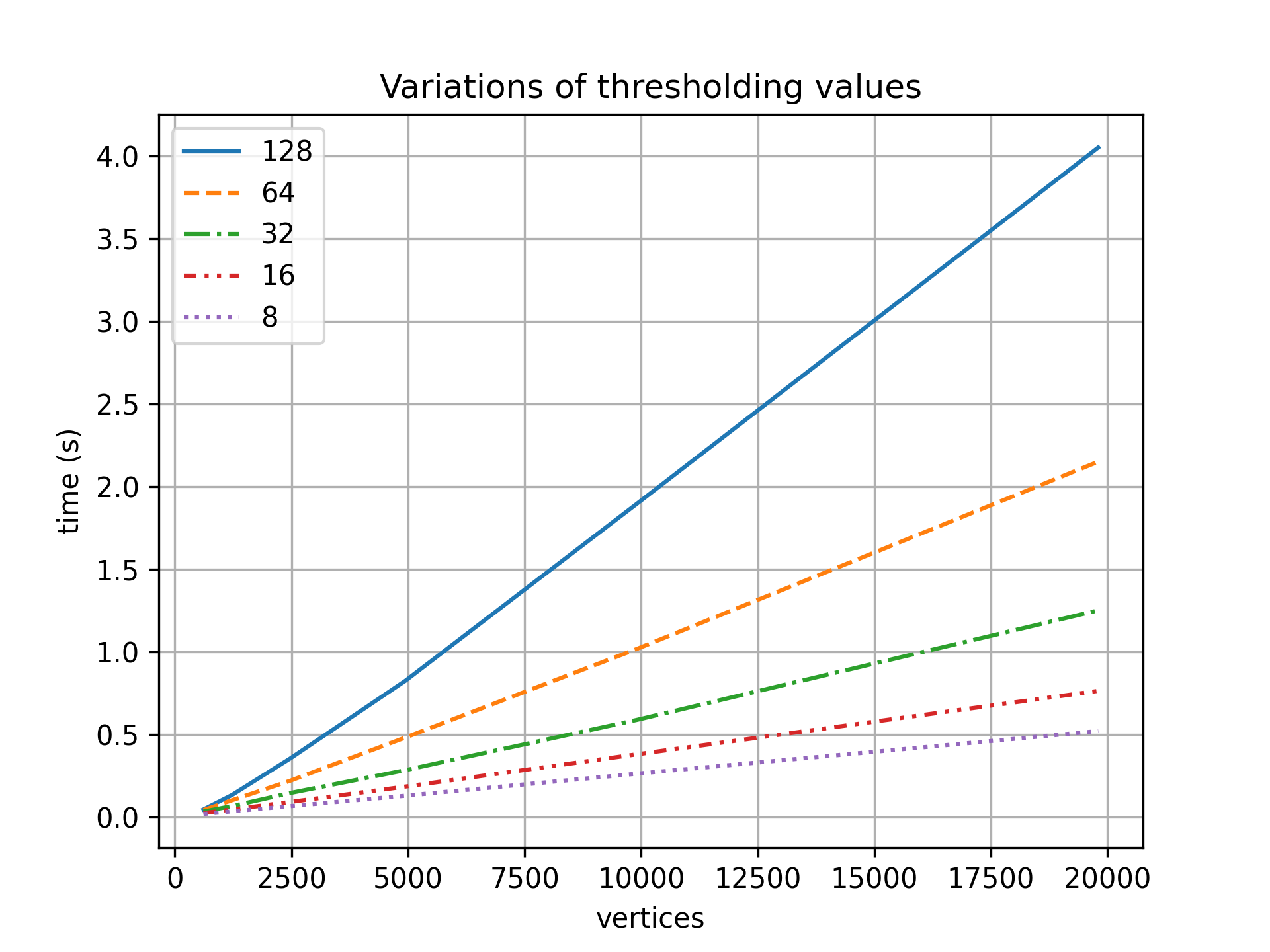}
    \caption{Running time measurements of varying values of $\alpha$ on subdivisions of a mesh.}
    \label{fig:alphatest}
\end{figure}

To show the effects of $\alpha$ on the running time, we consider a small test suite using subdivisions of a mesh to generate various sizes of input. This is done to ensure a similar underlying structure throughout the test. 
We test our multilevel algorithm for $\alpha=8,16,32,64,128$ (see Figure~\ref{fig:alphatest}). Not surprisingly, a lower threshold leads to a lower running time. 

\begin{figure}[htb]
    \centering
    \includegraphics[width=0.19\textwidth]{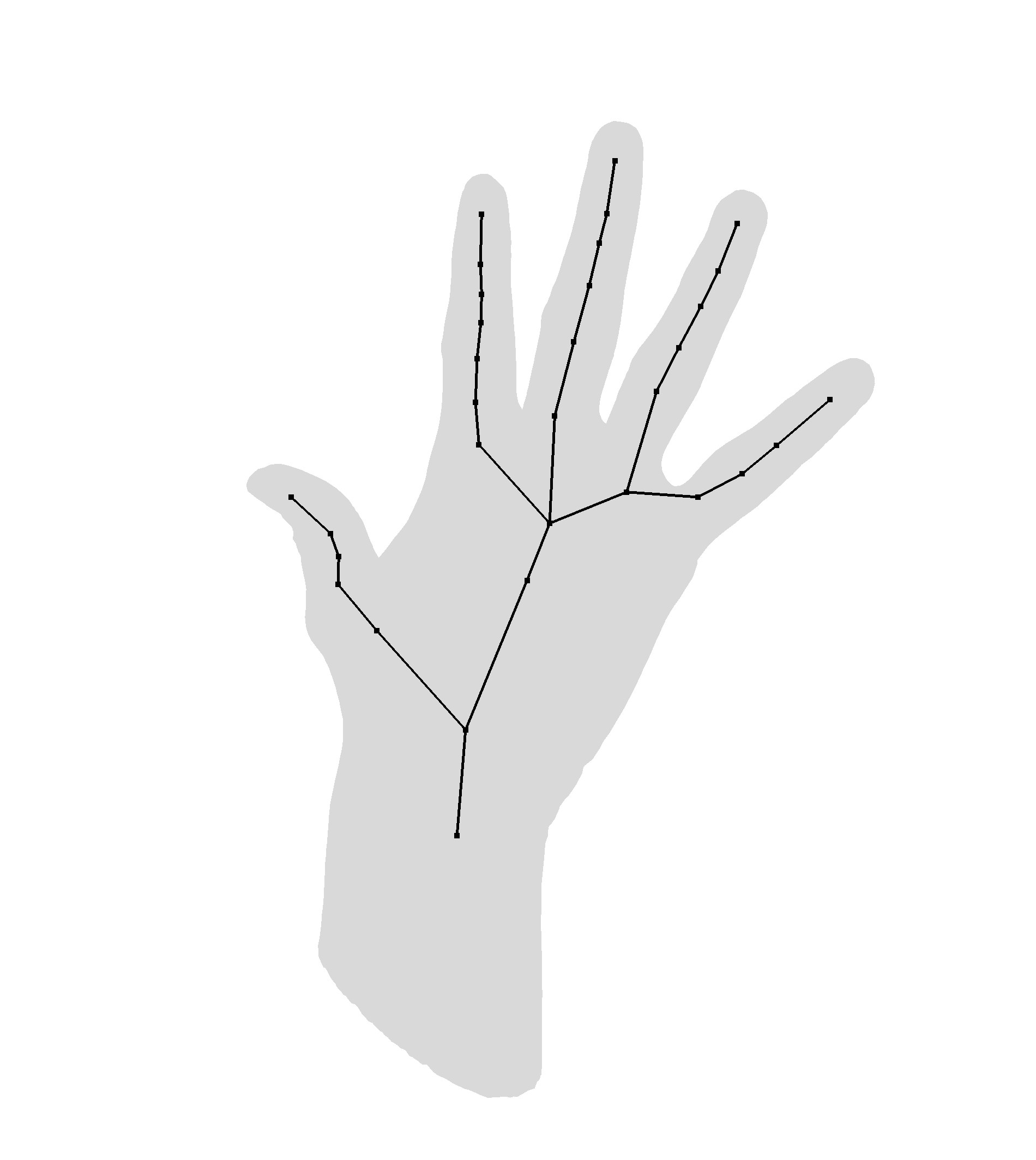}
    \includegraphics[width=0.19\textwidth]{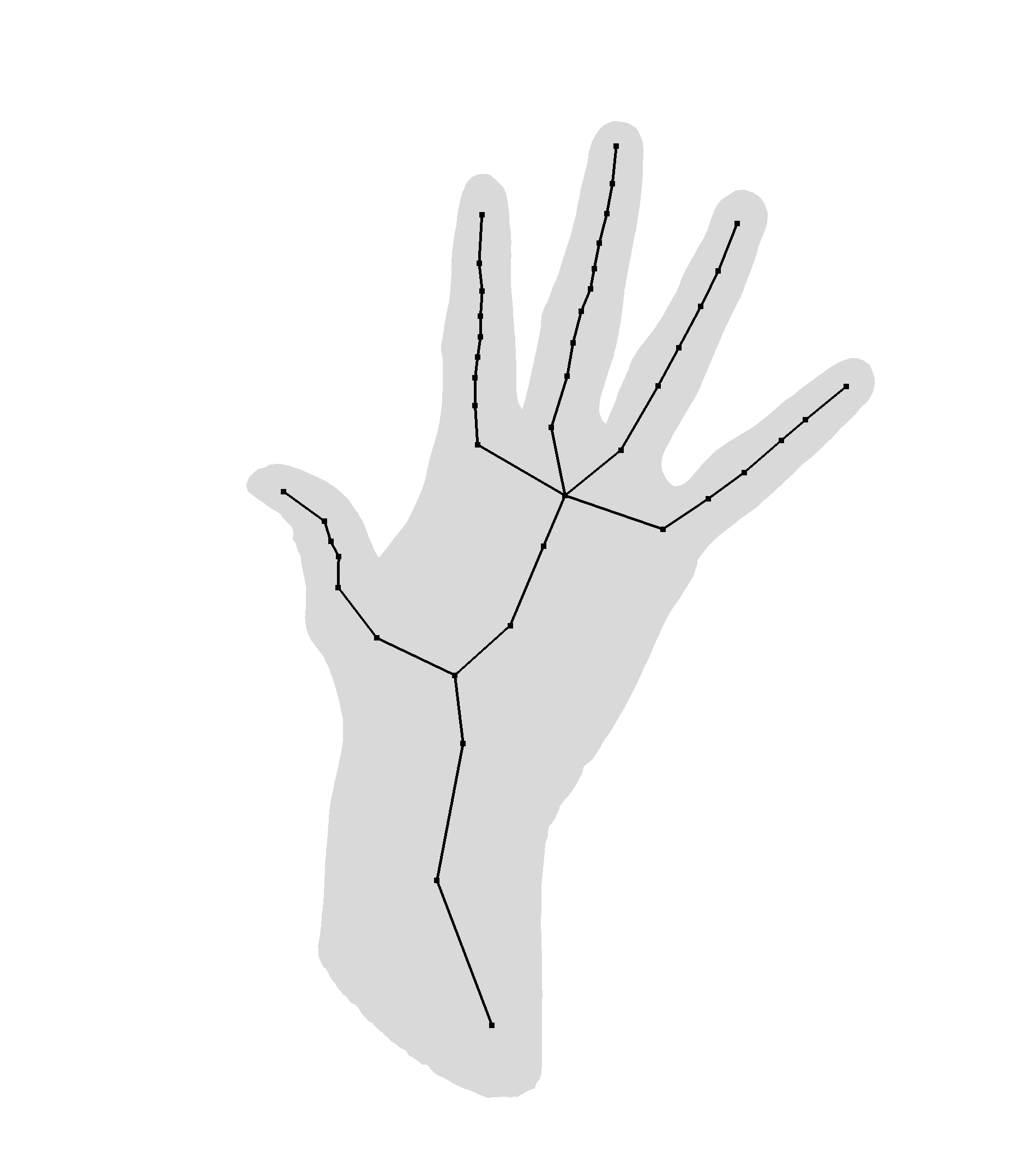}
    \includegraphics[width=0.19\textwidth]{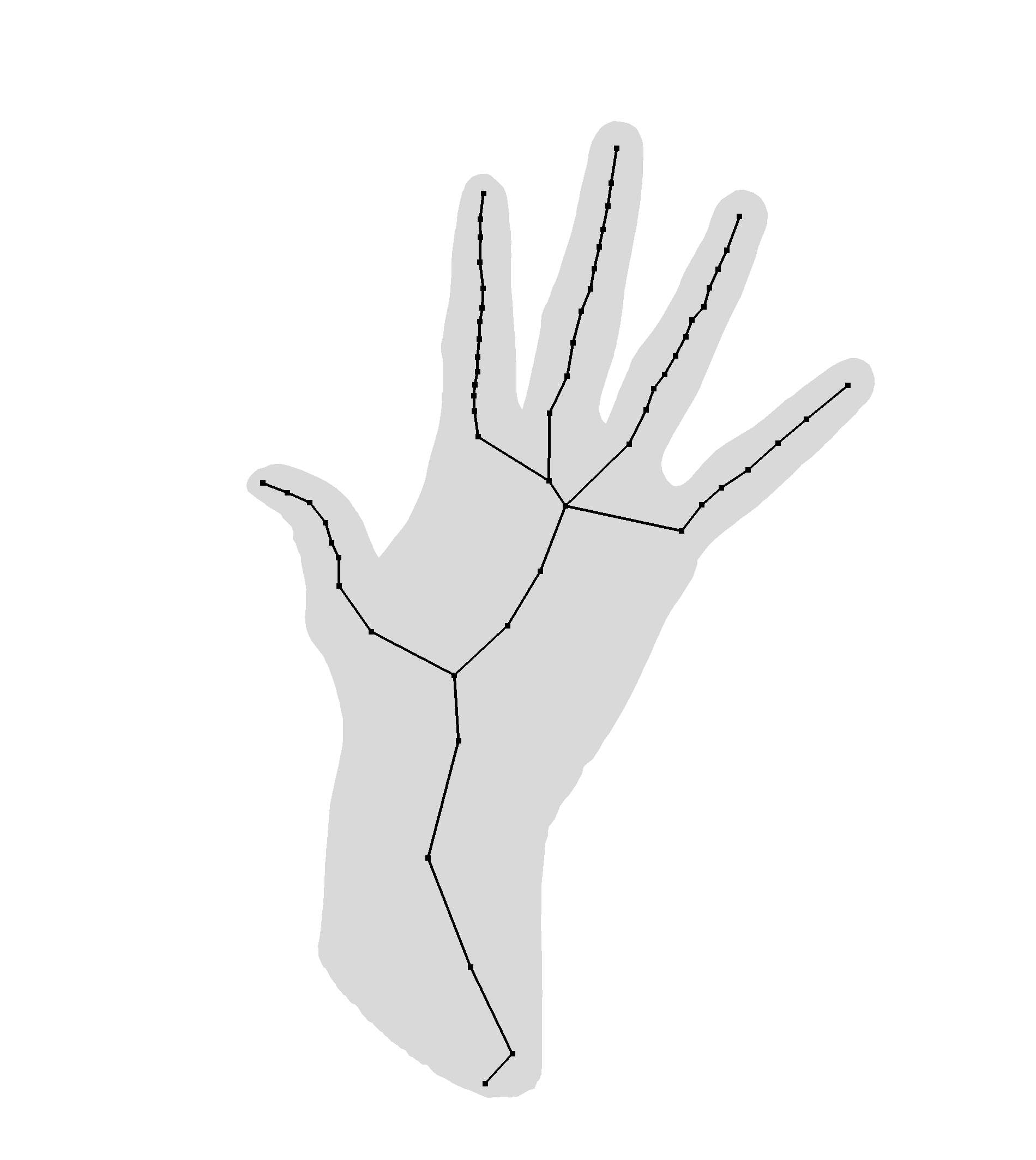}
    \includegraphics[width=0.19\textwidth]{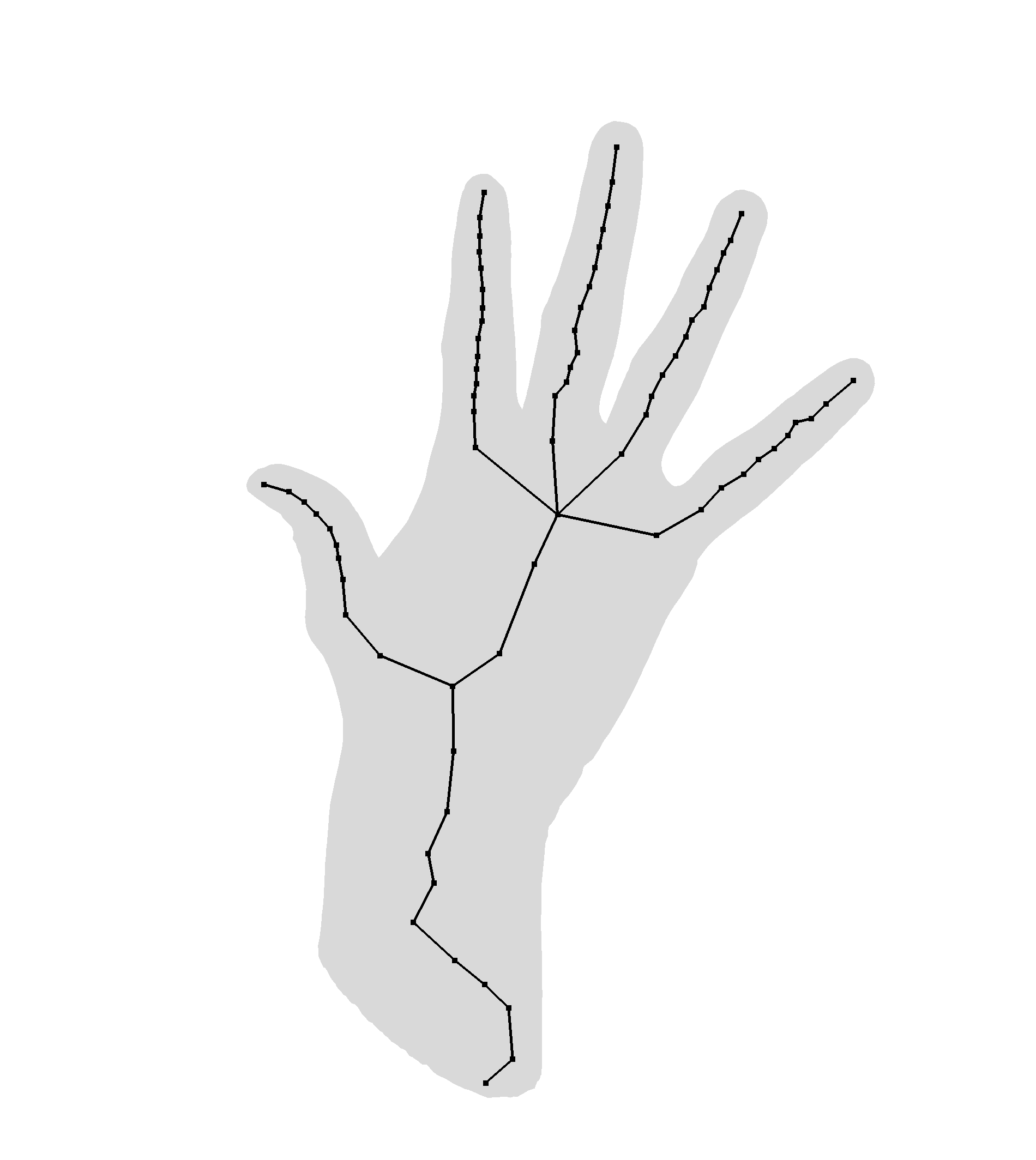}
    \includegraphics[width=0.19\textwidth]{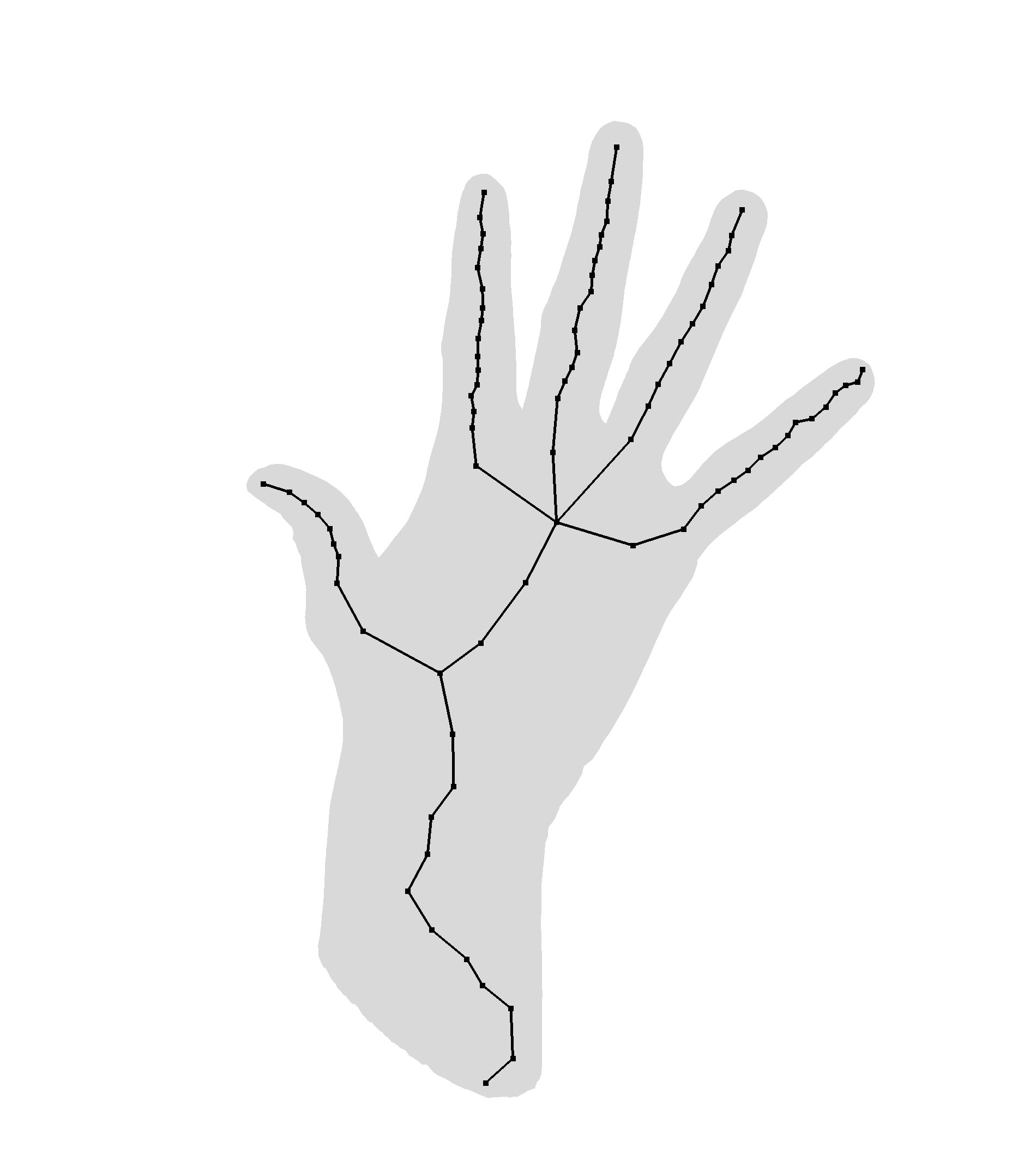}
    \caption{Skeletons found on \texttt{human\_hand.ply} for increasing patience thresholds. From left to right: $\alpha=8,16,32,64,128$.}
    \label{fig:alpha_hands}
\end{figure}

However, there may be a trade-off between skeleton quality and threshold value, which we explore qualitatively.

We visually examine the skeletons generated for our choices of $\alpha$ on \texttt{human\_hand.ply}, as seen in Figure~\ref{fig:alpha_hands}. For very low values of $\alpha$, the skeletons have few curves in areas that are relatively thick. With a low threshold, separators must be found on lower resolutions, which in turn means that few separators can be found. As we progress to higher values of $\alpha$, the level of detail of the skeleton rises, up to a certain point. Intuitively, if the threshold is high enough that the details can be captured on the higher levels of detail, then we gain nothing from the lower resolution levels.

\begin{figure}[p]
    \centering
    \includegraphics[width=0.31\textwidth]{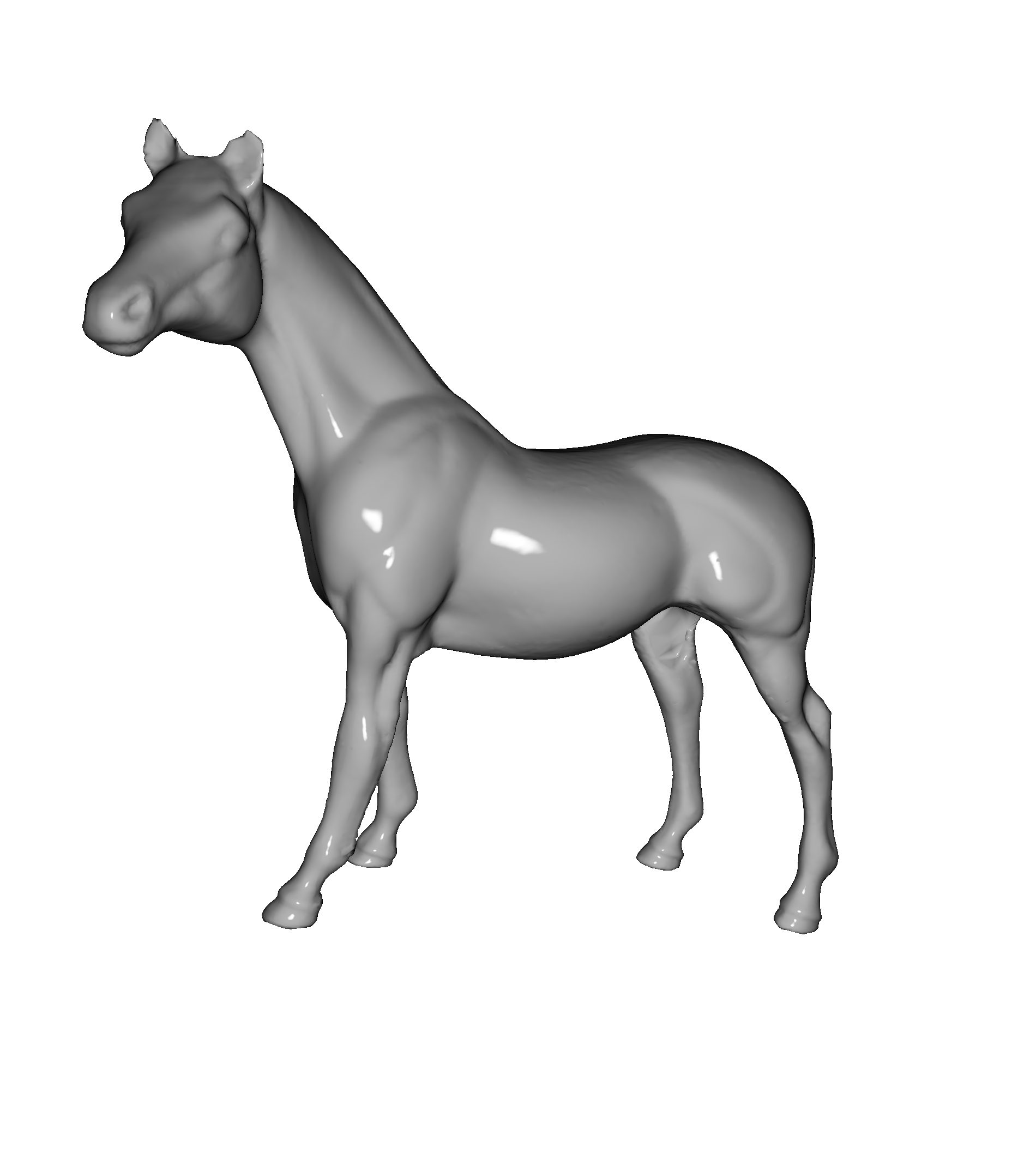}
    \includegraphics[width=0.31\textwidth]{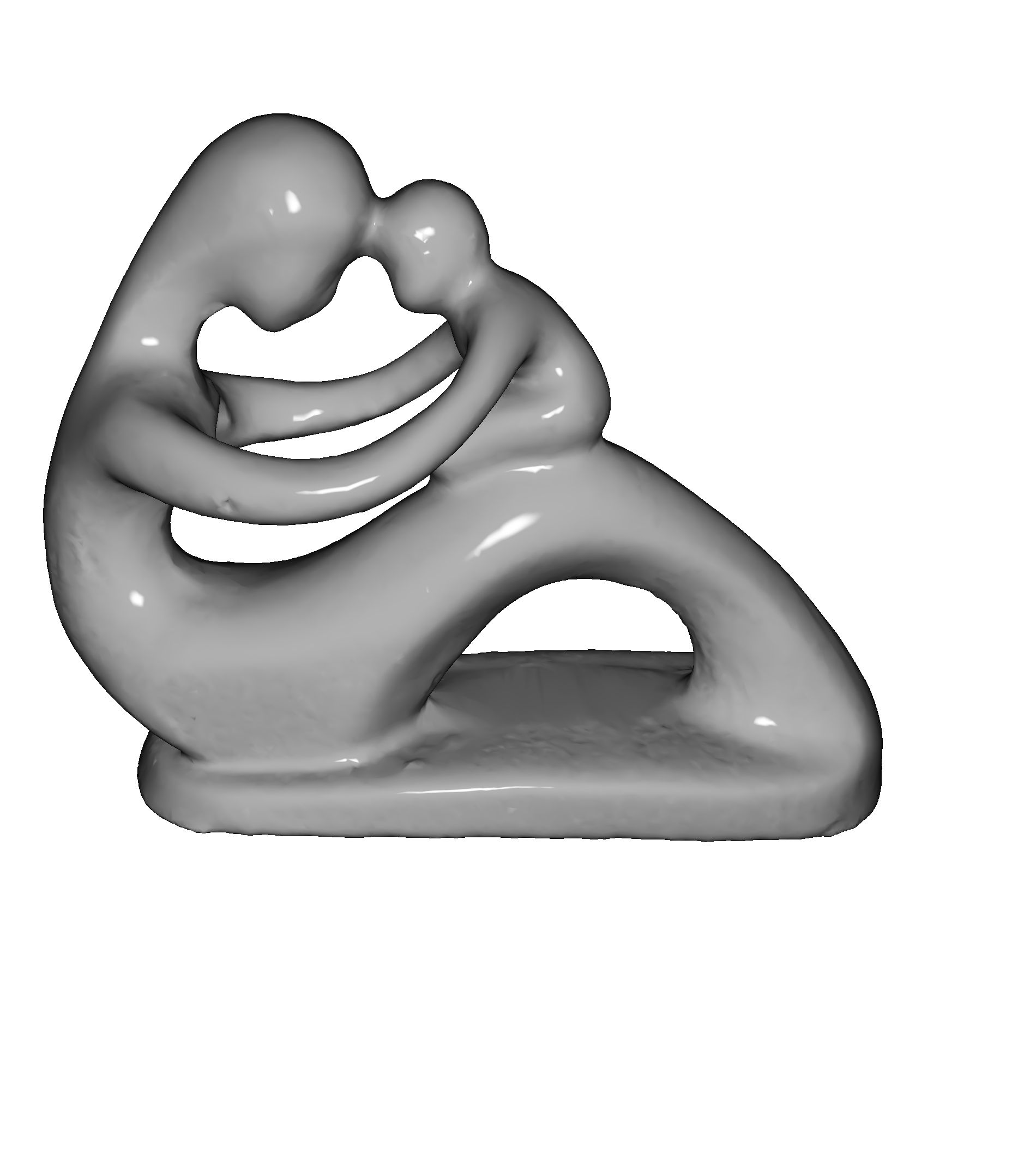}
    \includegraphics[width=0.31\textwidth]{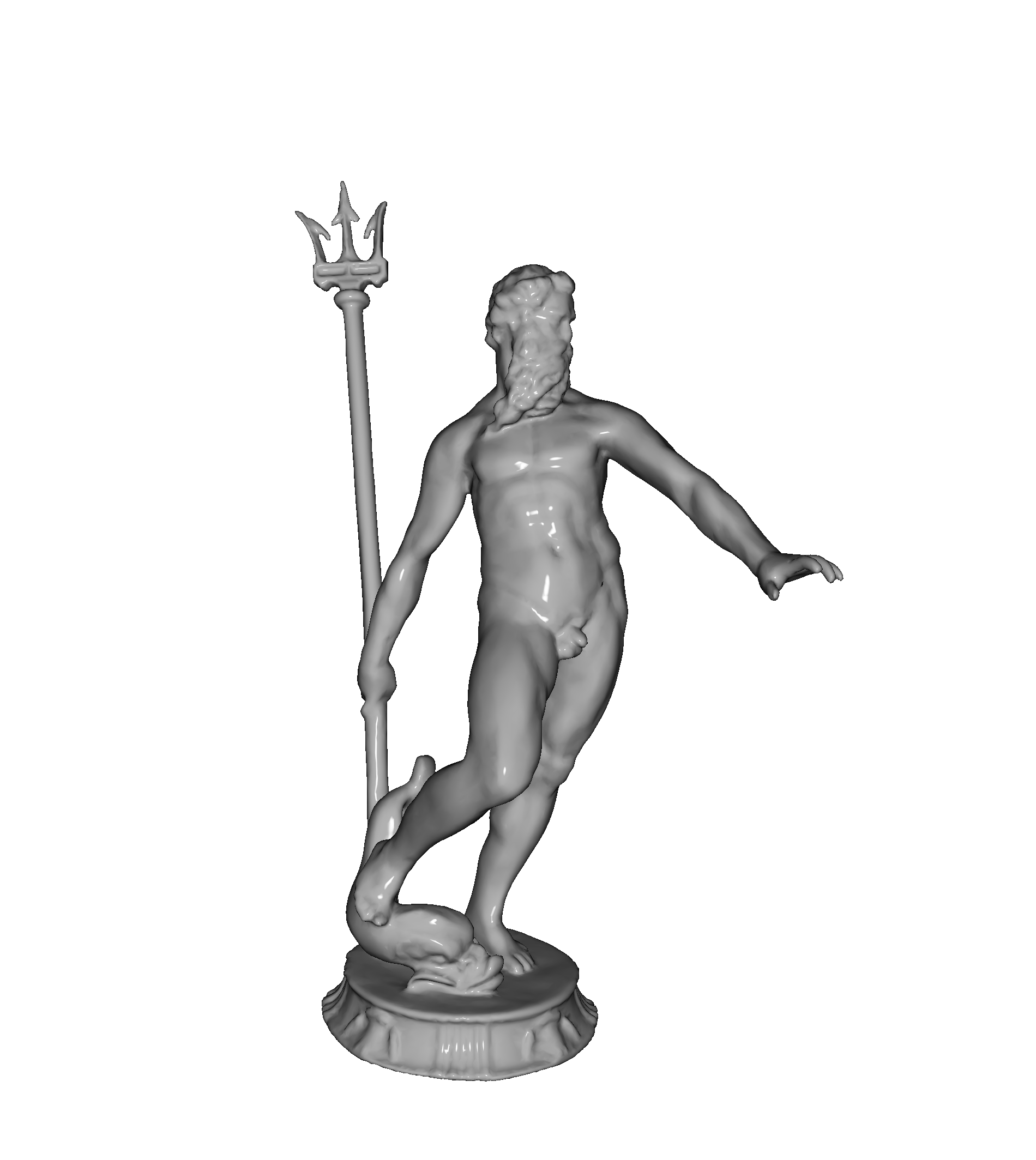}
    \includegraphics[width=0.31\textwidth]{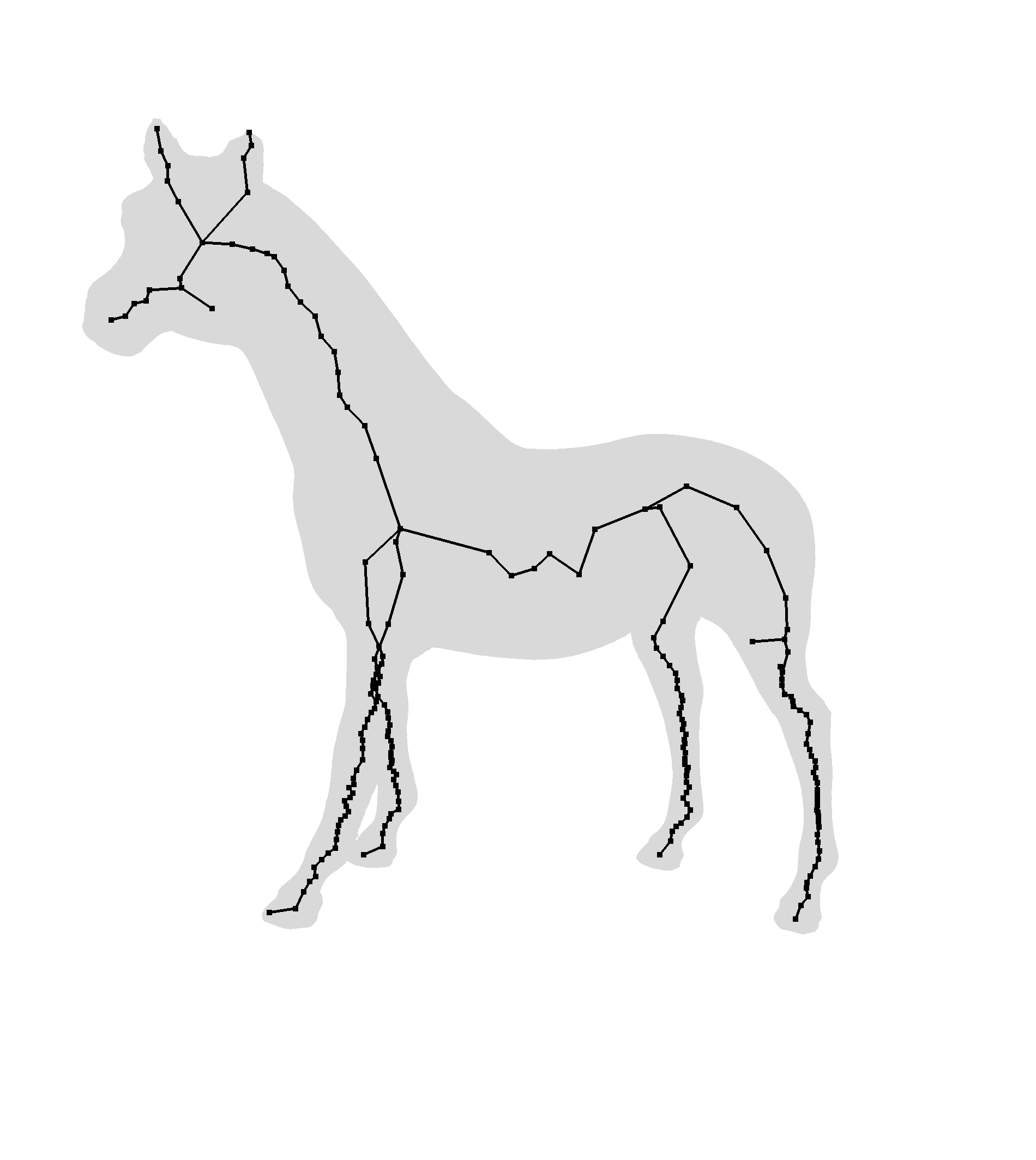}
    \includegraphics[width=0.31\textwidth]{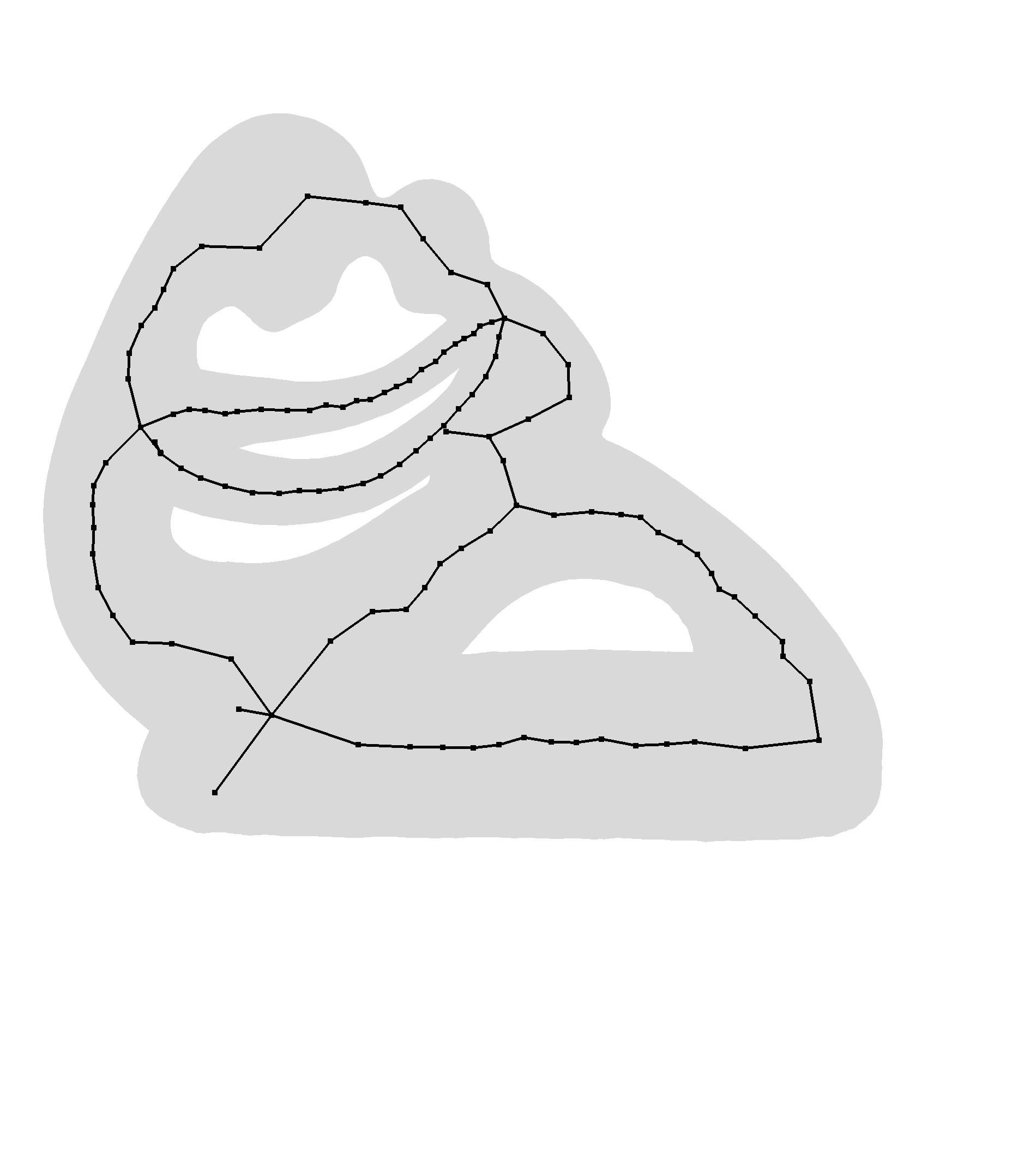}
    \includegraphics[width=0.31\textwidth]{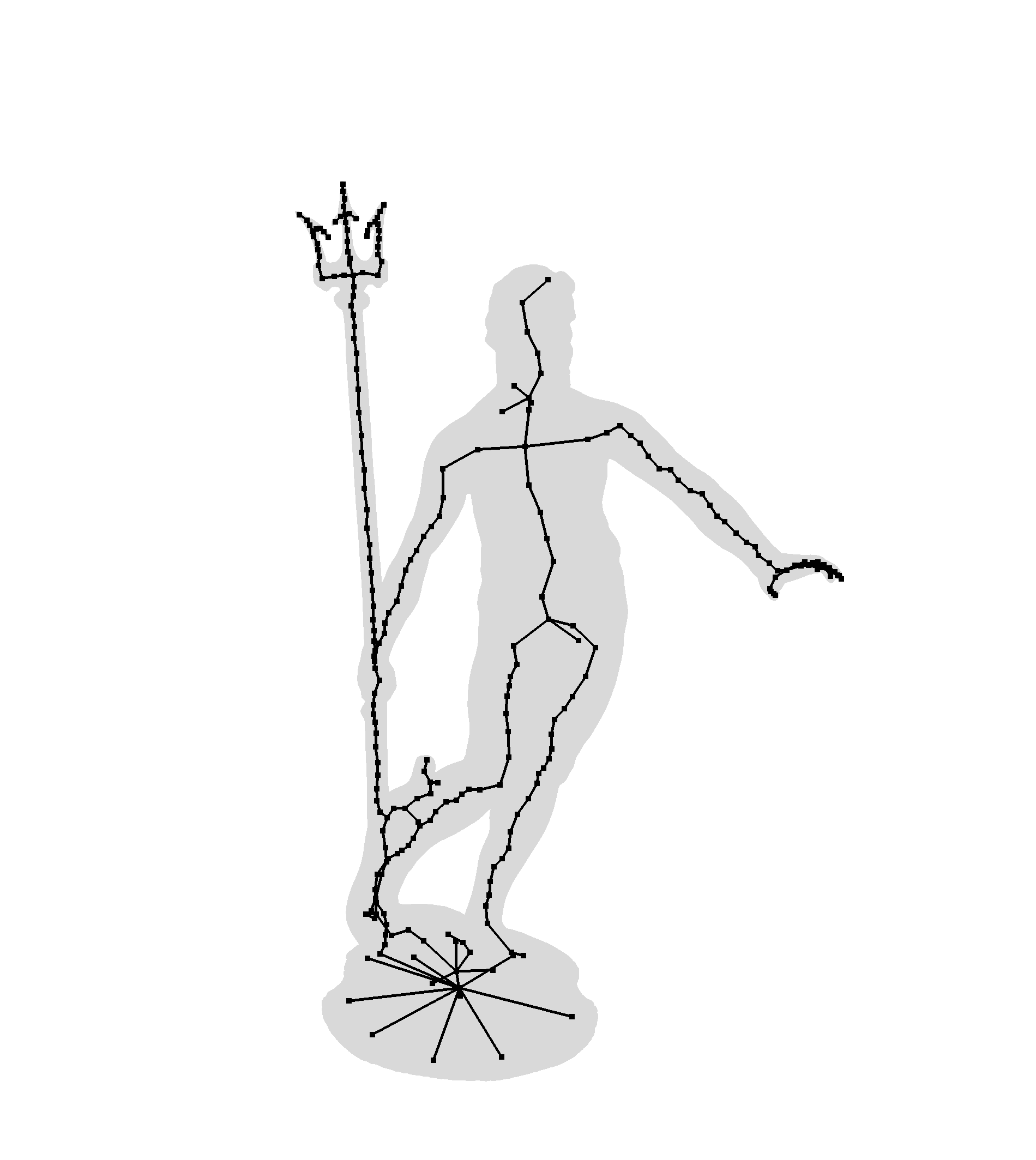}
    \includegraphics[width=0.31\textwidth]{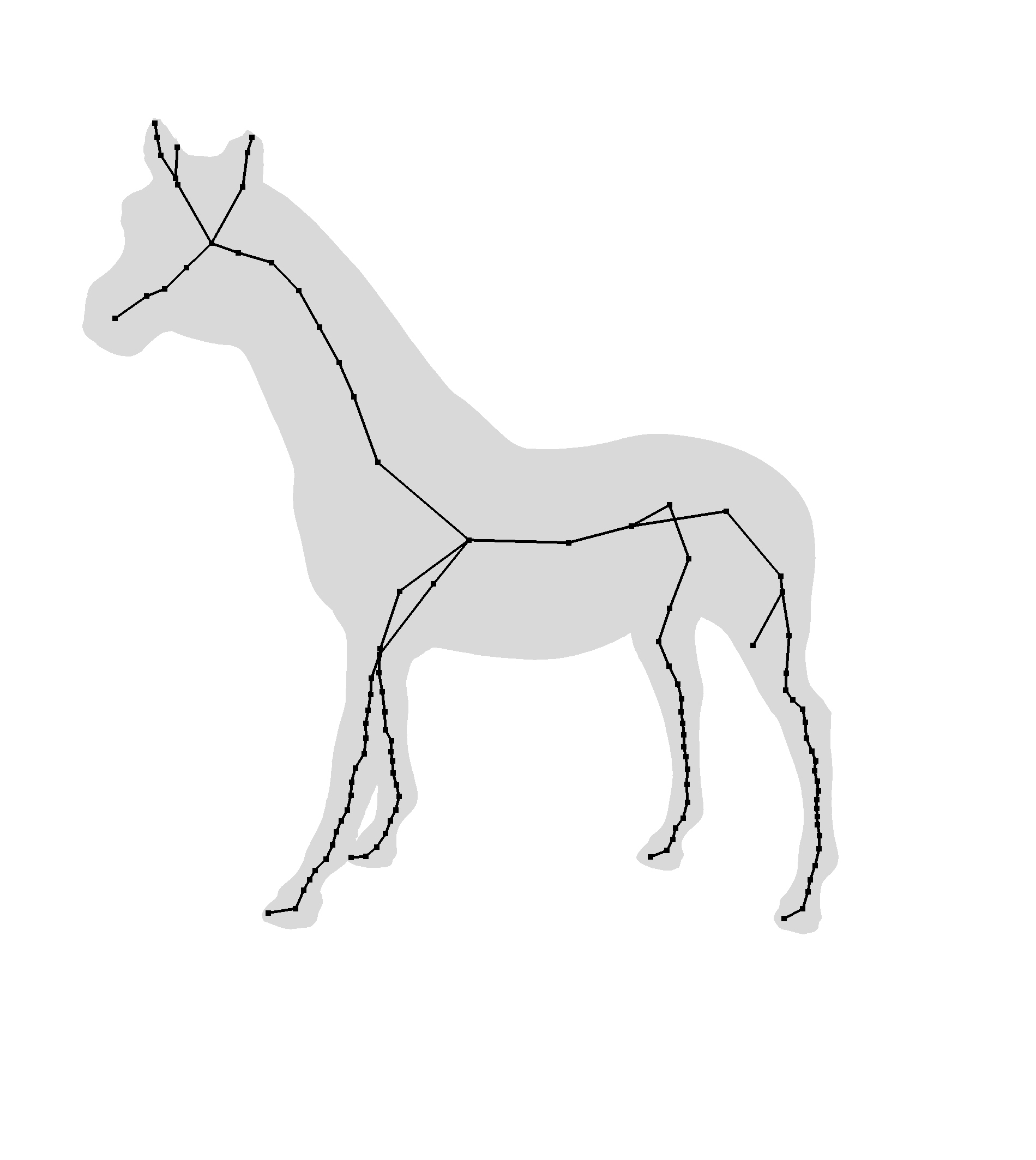}
    \includegraphics[width=0.31\textwidth]{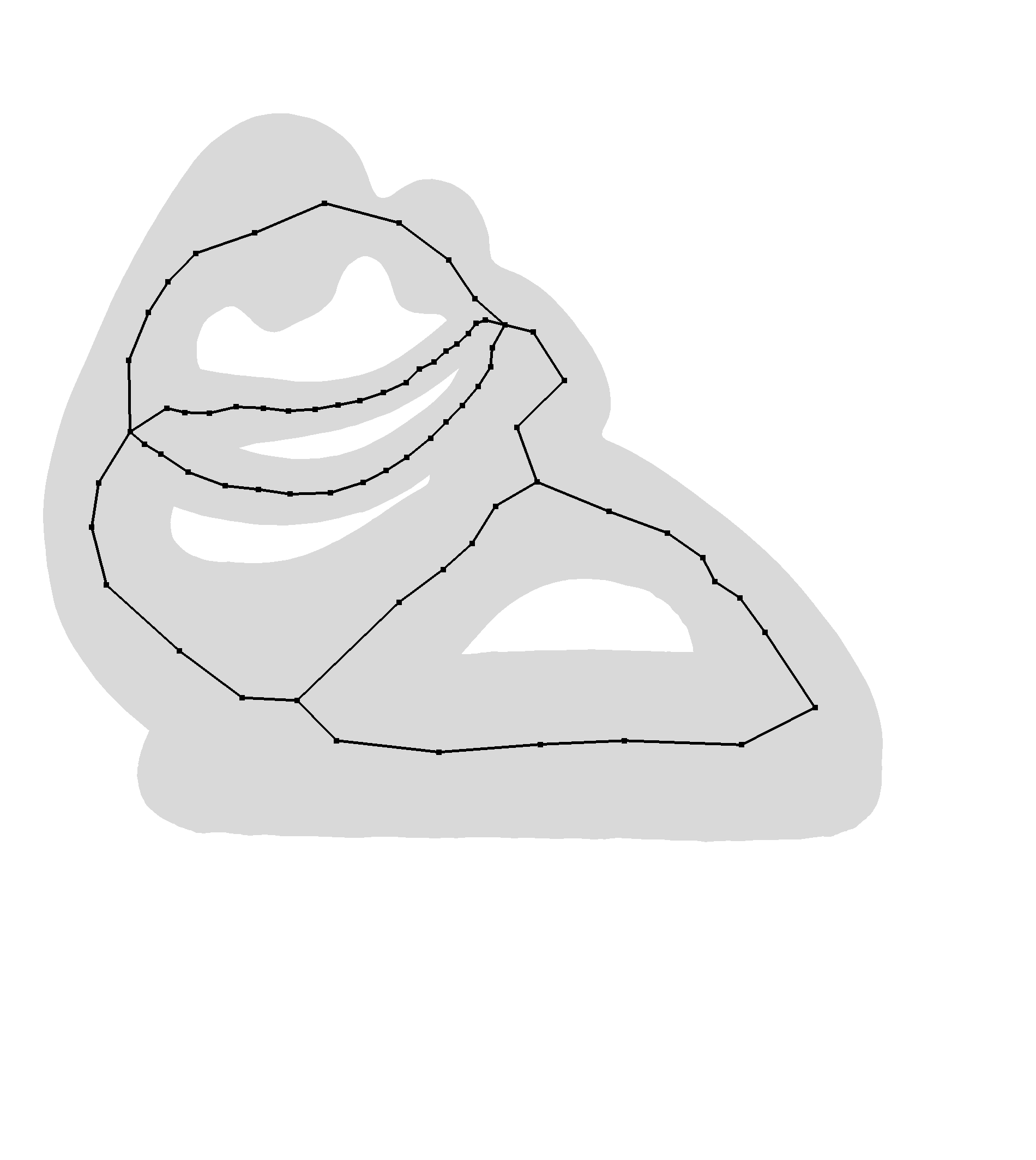}
    \includegraphics[width=0.31\textwidth]{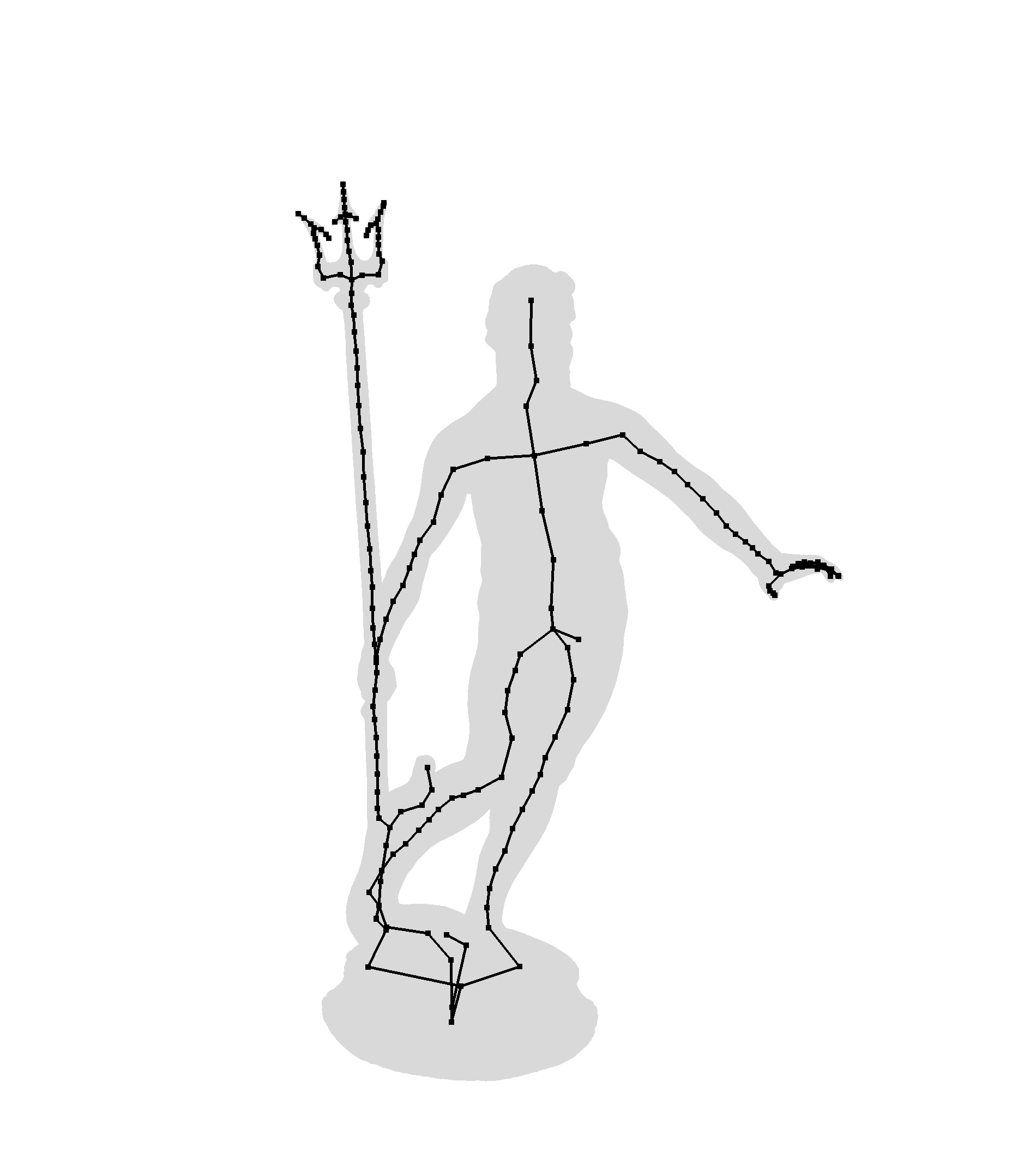}
    \includegraphics[width=0.31\textwidth]{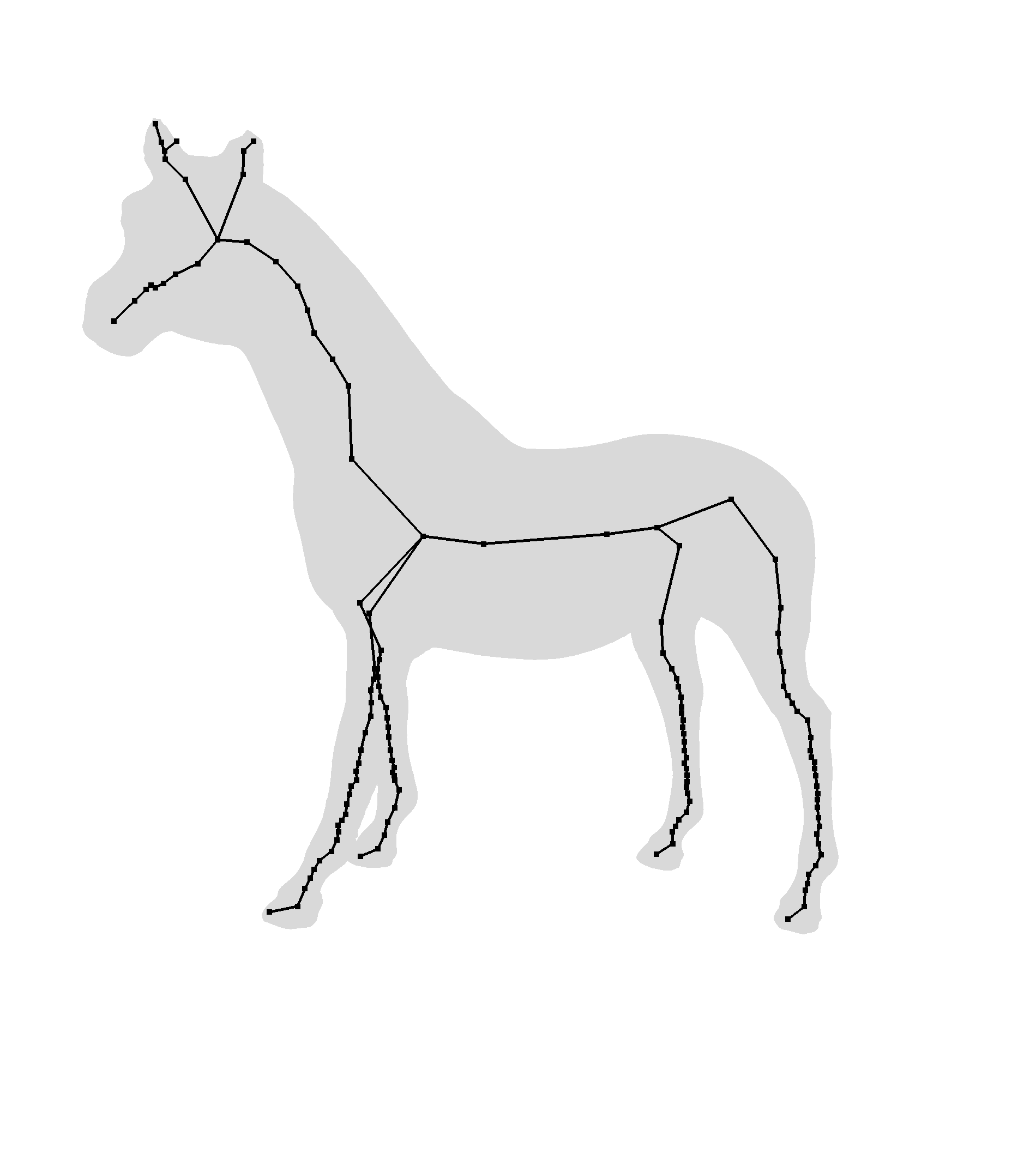}
    \includegraphics[width=0.31\textwidth]{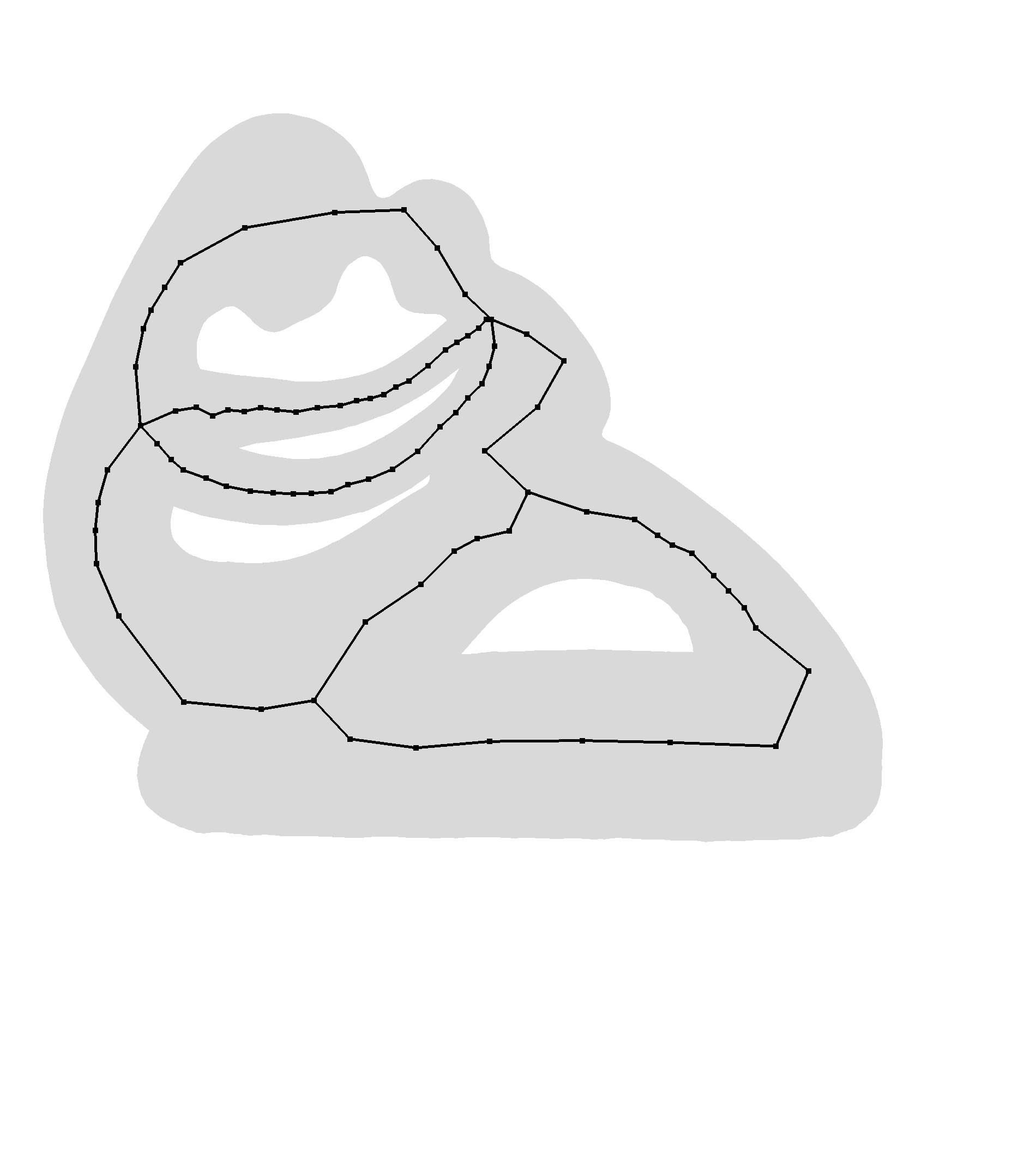}
    \includegraphics[width=0.31\textwidth]{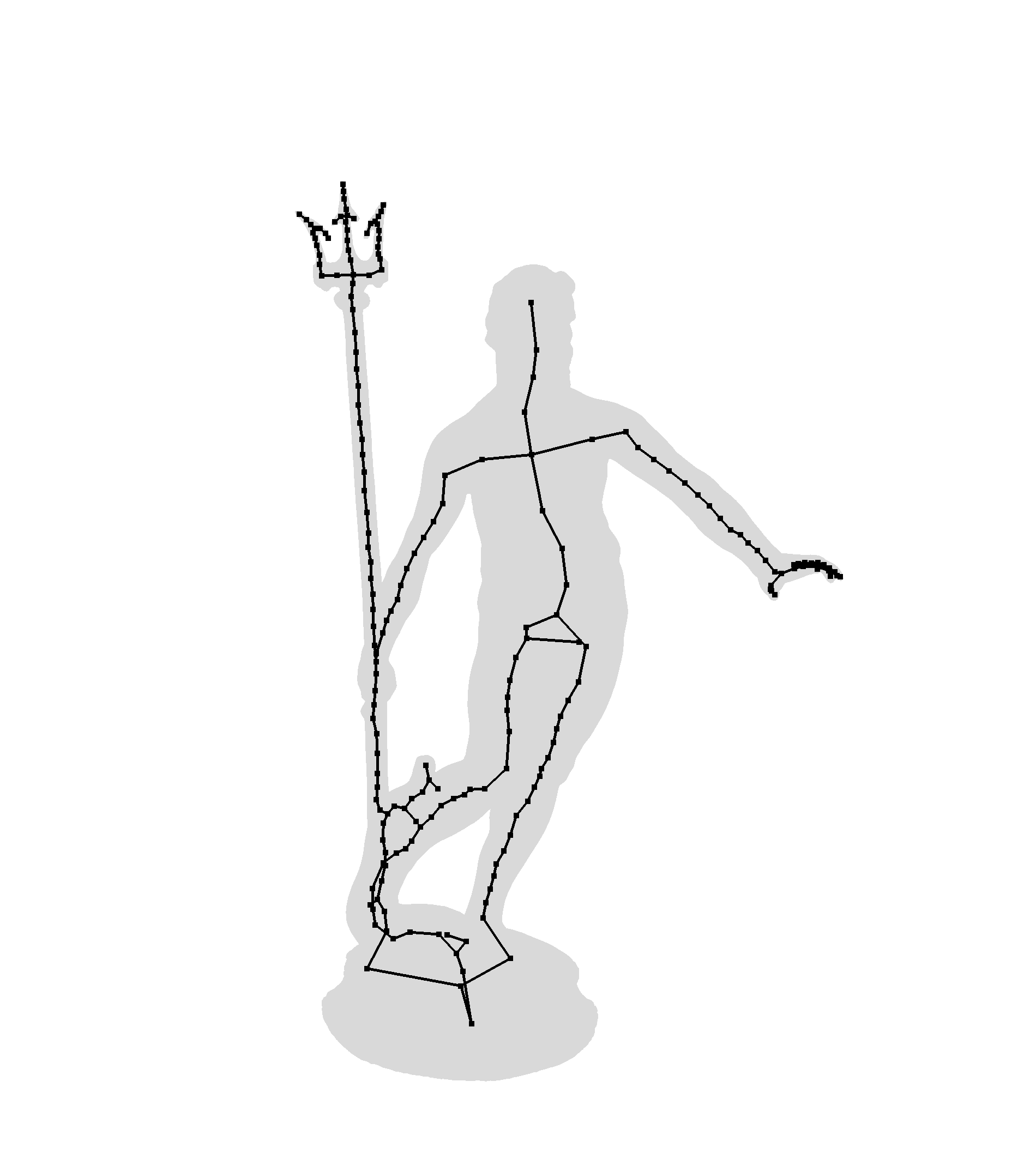}
    \caption{Each column indicates a different input, with each row showcasing a different method. From top to bottom: shaded renders of the input, skeletons obtained by LSS, skeletons obtained by LEM, skeletons obtained by LEMTS}
    \label{fig:fullskel}
\end{figure}

It could be argued that $\alpha=16$ or $\alpha=32$ generates the most visually appealing skeletons for this particular input, however we find that $\alpha=64$ offers the best trade-off for running time on other examined input such as that shown in Figure~\ref{fig:fullskel}.

Thus, we run the remainder of our tests using $\alpha=64$.

\paragraph*{Skeleton Quality}
In Figure~\ref{fig:fullskel} we show some of the skeletons obtained by LSS, LEM and LEMTS. Note that both LEM and LEMTS appear smoother, while also reaching the features that LSS finds in many cases. In addition, it seems that for these inputs our methods find less spurious features, giving a cleaner result. When comparing LEM and LEMTS the differences are subtle. On \texttt{horse.ply} it can be seen that the vertex where the front legs meet the body is positioned further to the left. This is due to LEMTS having a denser skeleton. On the other hand, LEMTS seems to not capture the structure of the groin area of \texttt{neptune.ply} as well as LEM. Although it appears as if LEMTS finds a cycle, this is not actually the case. The skeletal branch is, however, not positioned as one would expect.

\begin{table}[htb]
    \centering
    \begin{tabular}{c|c|c|c|c|c|c|c}
        input & algorithm & $\Delta$vertices &  $\Delta$leafs & $\Delta$branches & $\Delta$genus & $H(*,LSS)$ & $H(LSS,*)$ \\\hline
        \multirow{2}*{19465}	&	LEM & -563	&	-3	&	-35	&	2	&	0.0437028	&	0.0415616 \\
                                &  LEMTS& -416	&	11	&	-24	&	3	&	0.0355444	&	0.0382796 \\\hline
        \multirow{2}*{fertility}& LEM &	-46	&	-3	&	-1	&	0	&	0.273804	&	0.0874419 \\
                                & LEMTS&	-30	&	-3	&	-1	&	0	&	0.240033	&	0.165239 \\\hline
        \multirow{2}*{happy4}	&	LEM & -589	&	-409	&	-21	&	-100	&	0.199295	&	0.0887277 \\
                                & LEMTS & -560	&	-384	&	-22	&	-99	&	0.160472	&	0.100934 \\\hline
        \multirow{2}*{horse}	&	LEM & -95	&	-1	&	-1	&	0	&	0.112222	&	0.111066 \\
                                & LEMTS & -66	&	-2	&	-2	&	0	&	0.0883287	&	0.0851157 \\\hline
        \multirow{2}*{neptune}	&	LEM & -73	&	-16	&	-5	&	0	&	0.200924	&	0.0508065 \\
                                & LEMTS &-51	&	-15	&	-1	&	0	&	0.225756	&	0.0516766 \\
    \end{tabular}
    \caption{Excerpt of measurements on skeletons. The metrics denoted by $\Delta$ are relative to the skeletons of LSS, with negative values implying that LSS has more vertices, leafs, branches etc. Here $H(A,B)$ denotes the directed Hausdorff distance between $A$ and $B$, divided by the radius of a bounding sphere, and $*$ denotes skeletons generated by our multilevel algorithms.}
    \label{tab:skelqual}
\end{table}

In addition, we also showcase a small excerpt of measurements from Appendix~\ref{sec:skel}, which can be seen in Table~\ref{tab:skelqual}. Here it is clear that LEM and LEMTS produce slightly simpler skeletons with fewer vertices, leaves, and branches. However, from visual inspection of the models it is clear that (at least for the models in the table) the missing details in the skeleton correspond to features which are so subtle that the skeletal details might be considered spurious. For all inputs of the benchmark except \texttt{happy4.ply}, there is little deviation in the genus compared to LSS.

\begin{figure}[htb]
    \centering
    \includegraphics[width=0.24\textwidth]{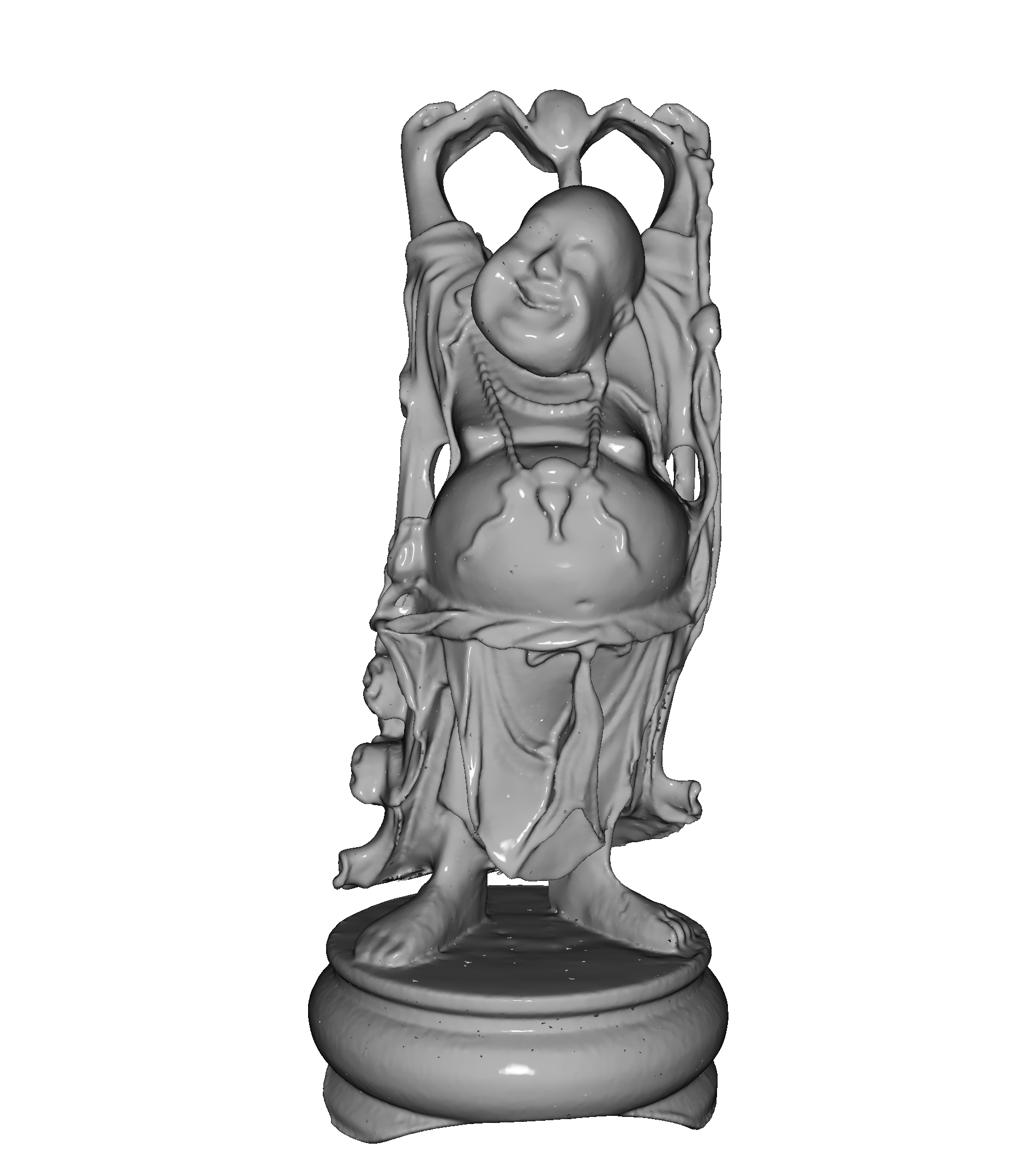}
    \includegraphics[width=0.24\textwidth]{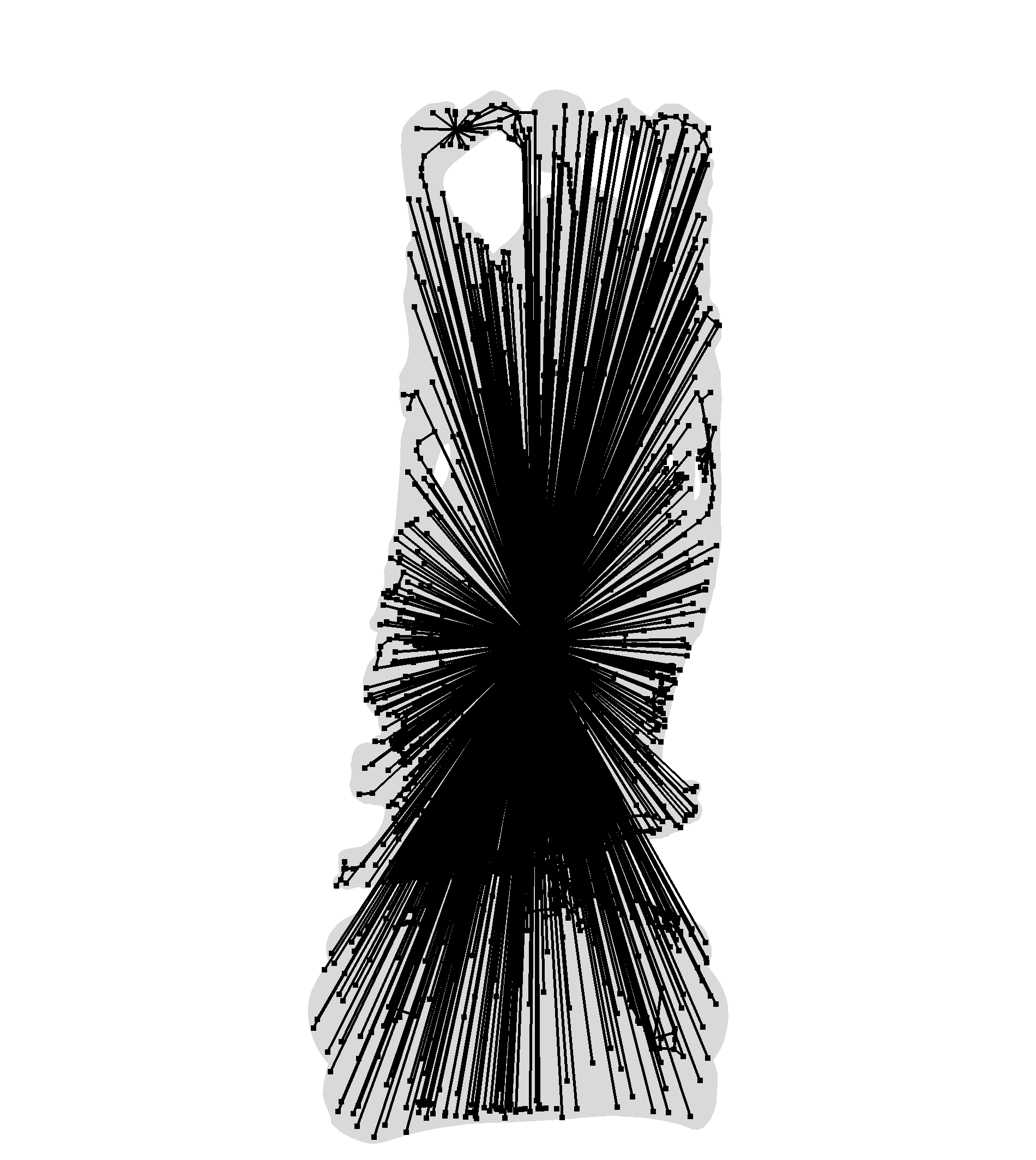}
    \includegraphics[width=0.24\textwidth]{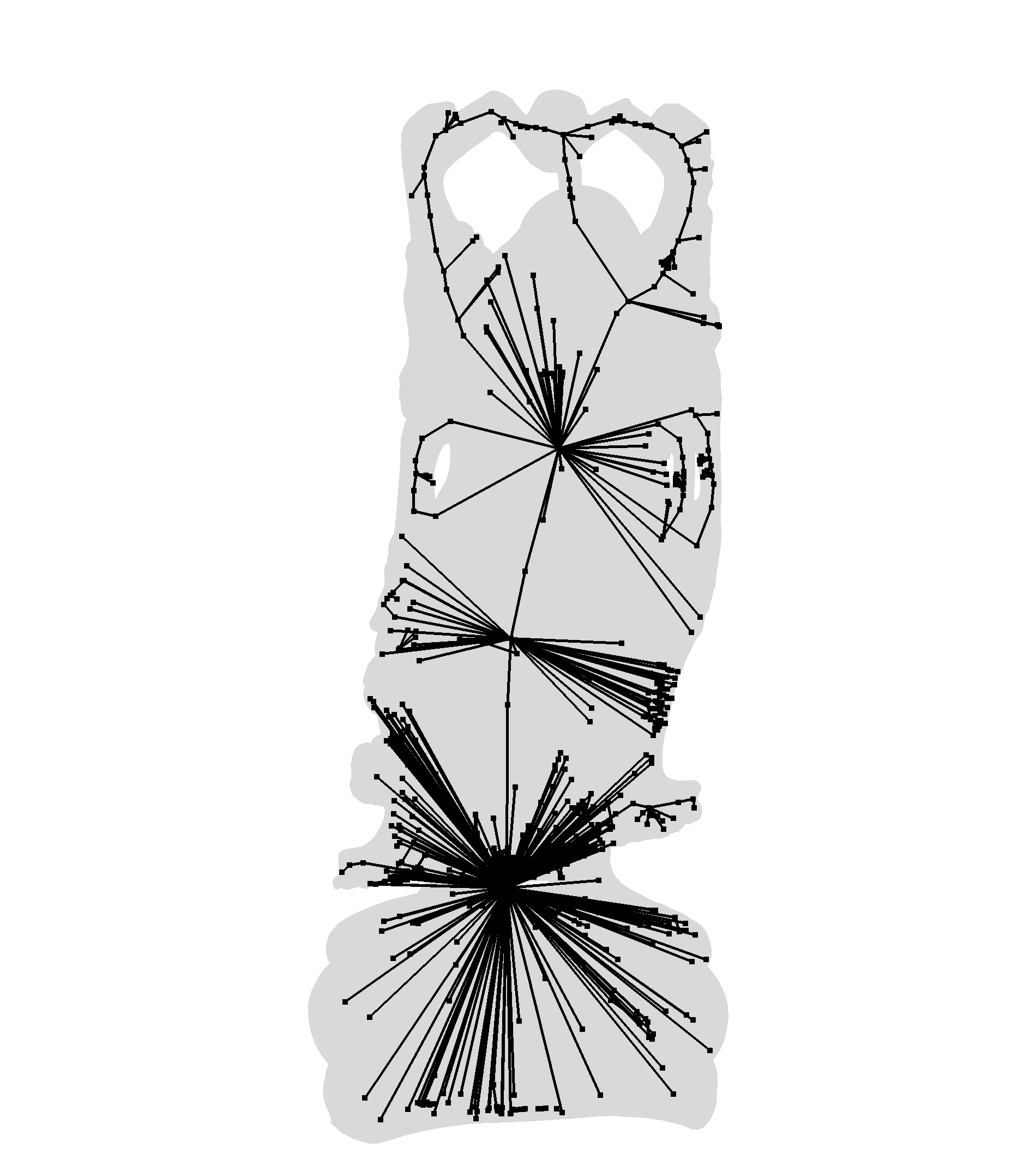}
    \includegraphics[width=0.24\textwidth]{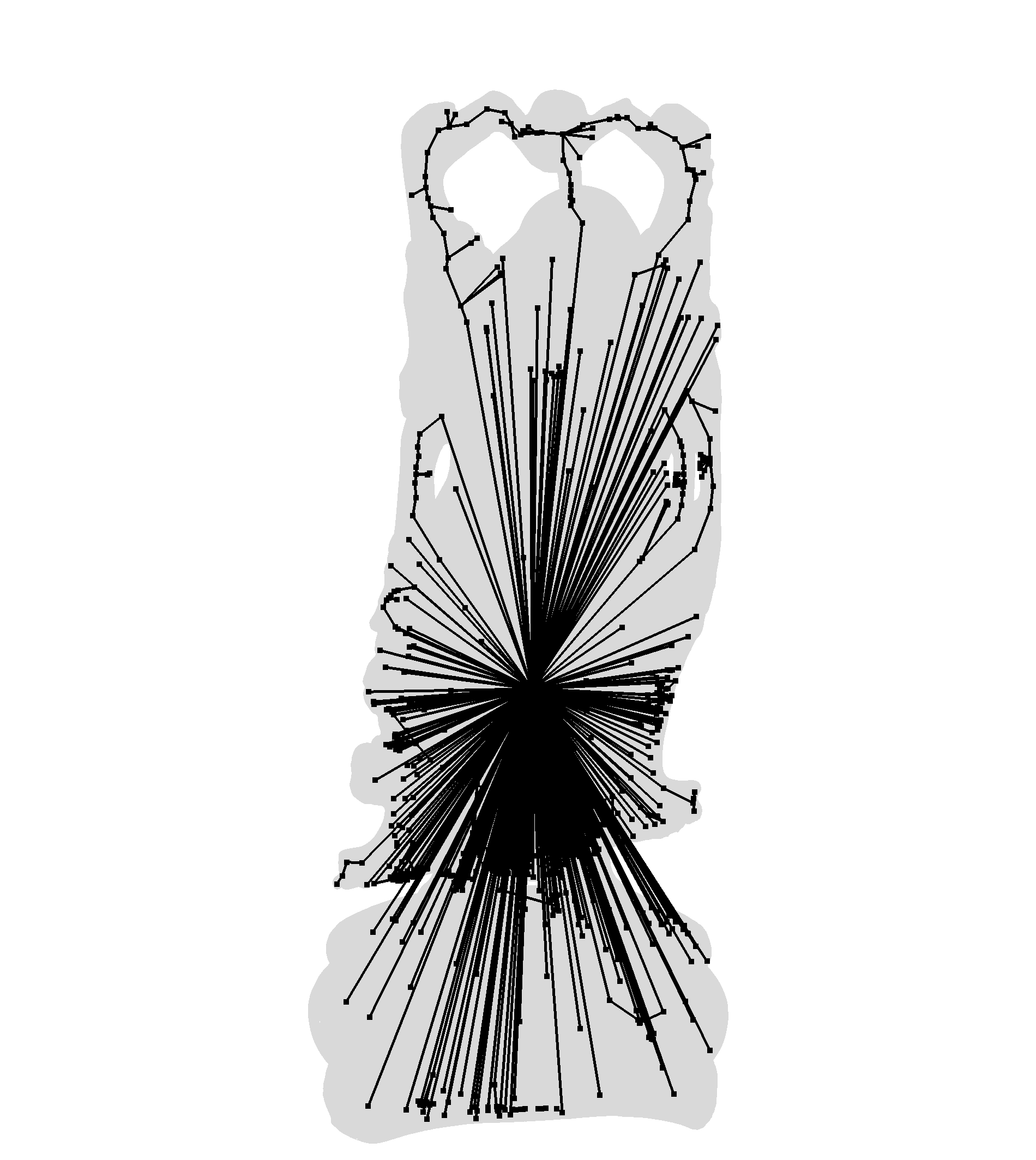}
    \caption{A triangle mesh with a large number of missing patches on the surface, \texttt{happy4.ply}, resulting in erroneous output for LSS, LEM, and LEMTS }
    \label{fig:happy}
\end{figure}

For context on the strange genus found on \texttt{happy4.ply}, we show the generated skeletons in Figure~\ref{fig:happy}. Of note is that the mesh has several missing patches, which seems to cause spurious small separators to be found on all of the local separator based methods. We consider this an error case for all of the methods examined.

\begin{figure}[p]
\begin{minipage}[t]{.48\textwidth}
    \centering
    \includegraphics[width=\textwidth]{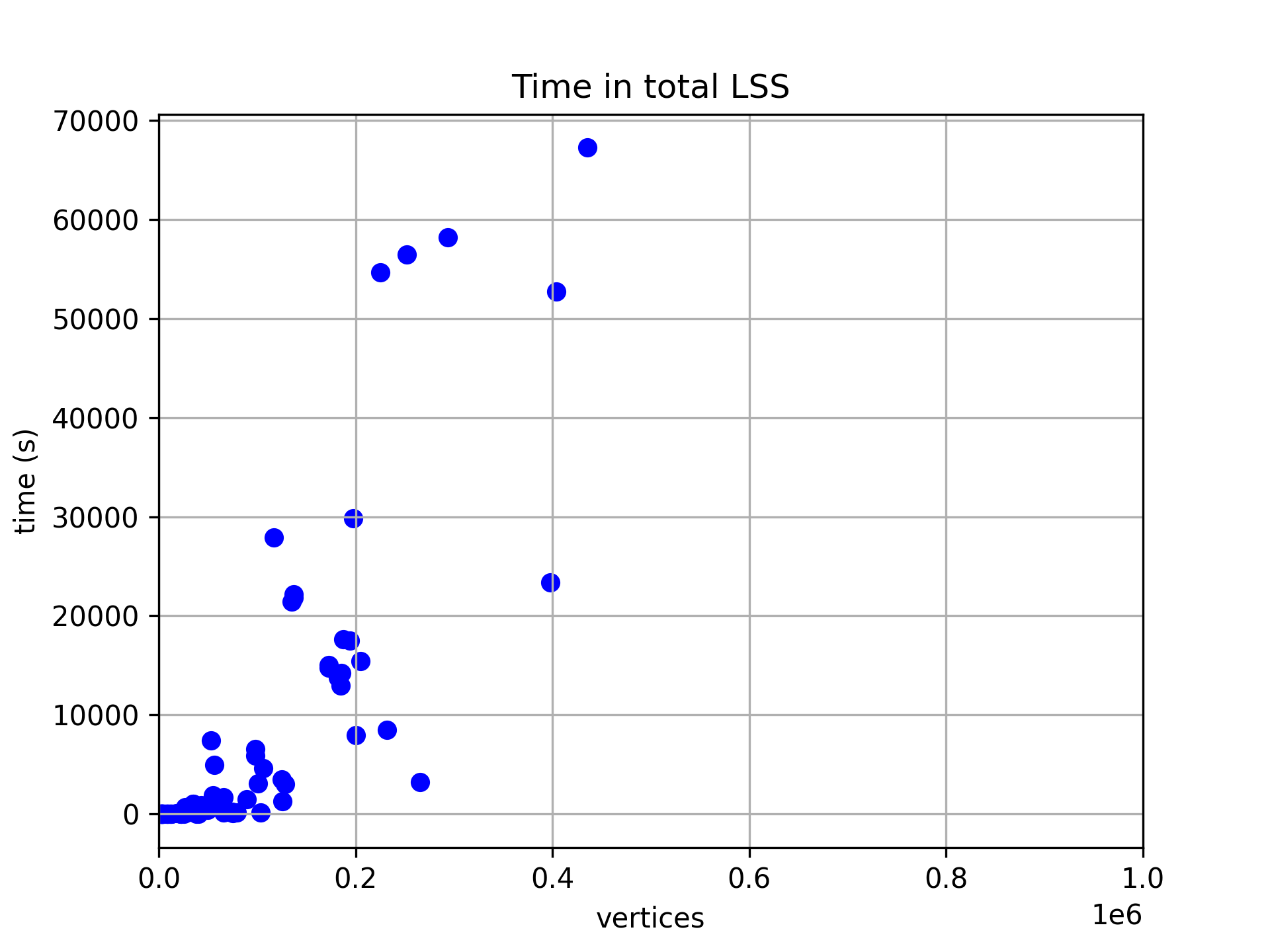}
    \includegraphics[width=\textwidth]{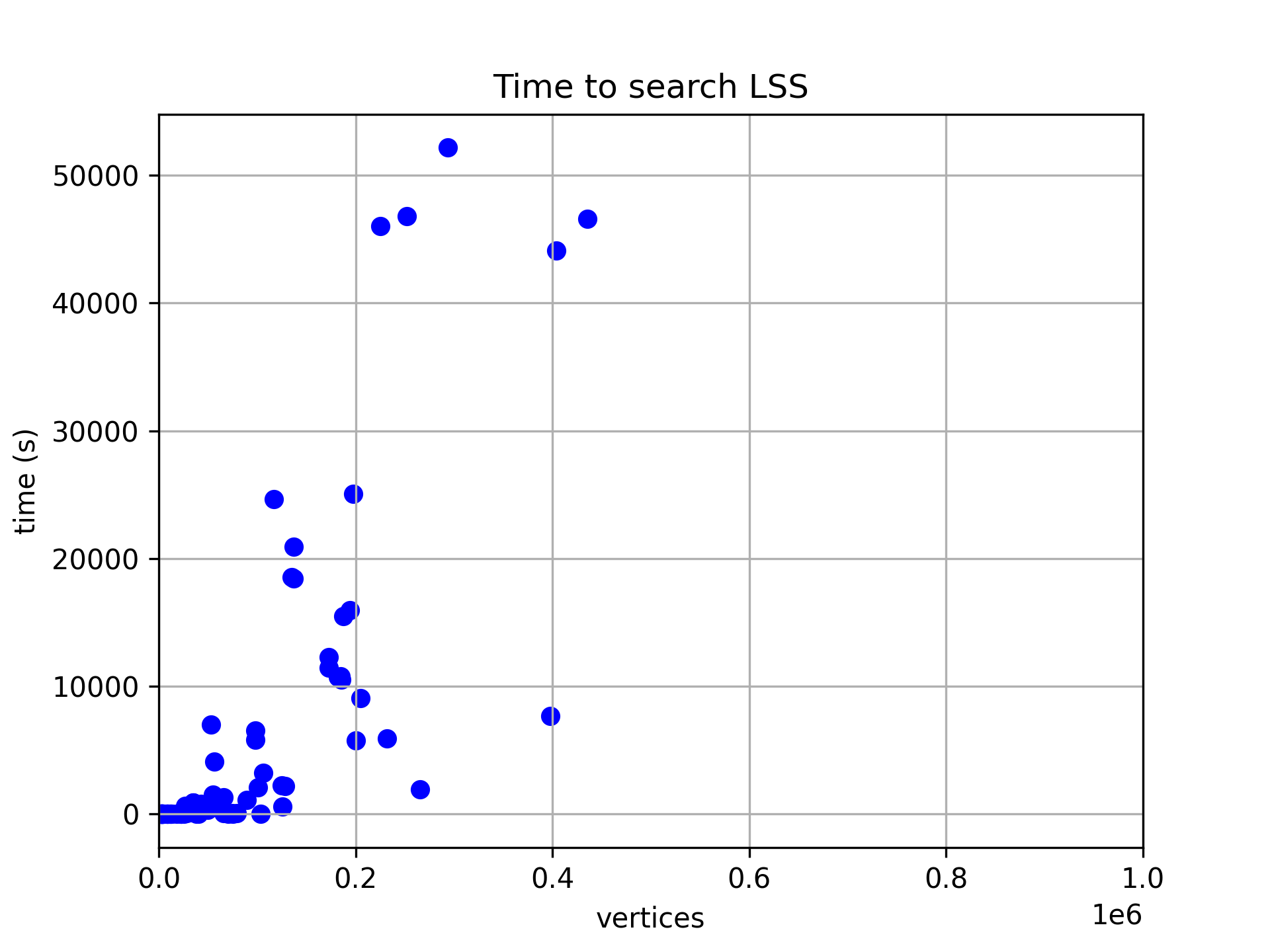}
    \subcaption{Running times of LSS.}
\end{minipage}
\begin{minipage}[t]{.48\textwidth}
    \centering
    \includegraphics[width=\textwidth]{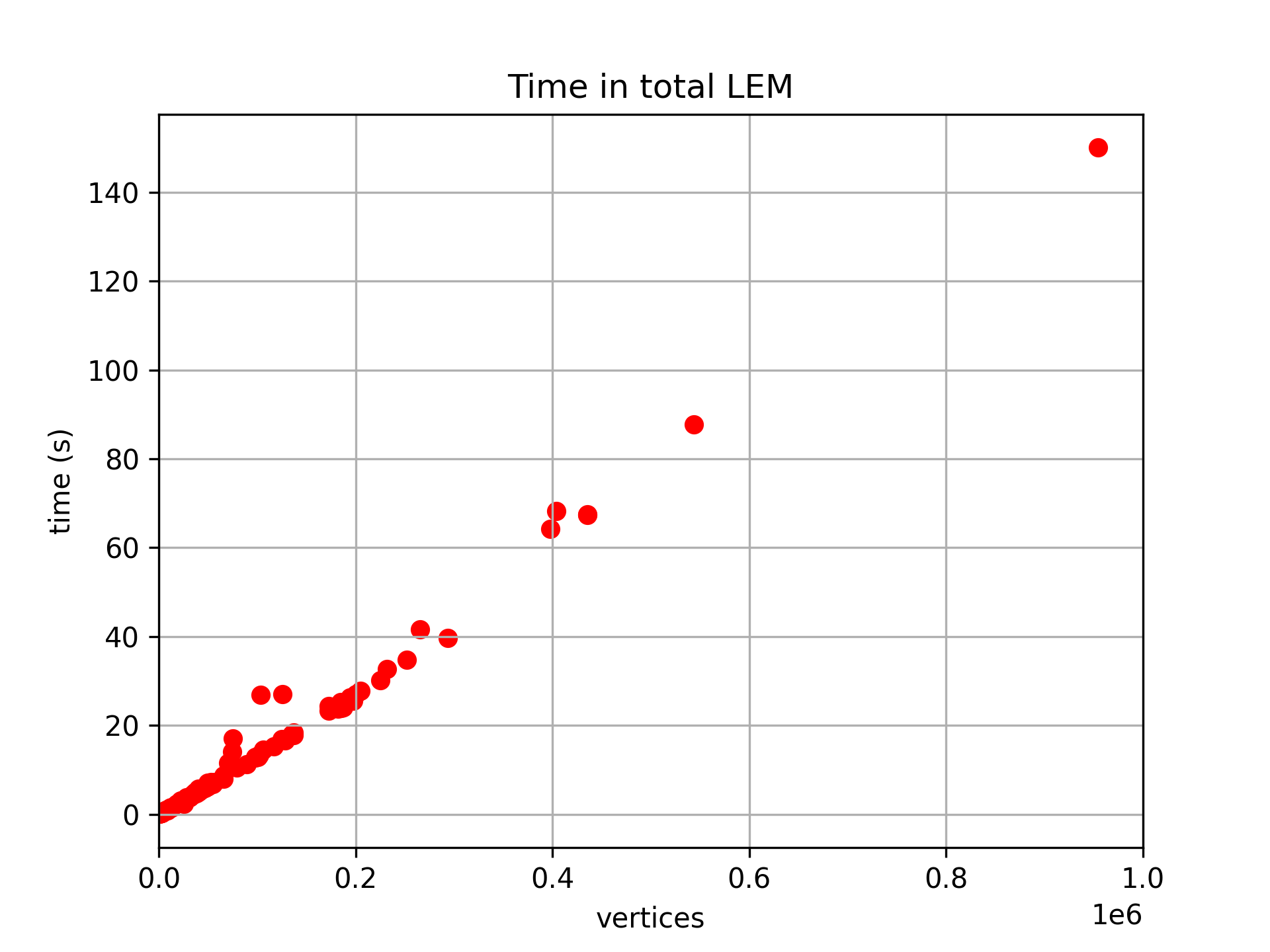}
    \includegraphics[width=\textwidth]{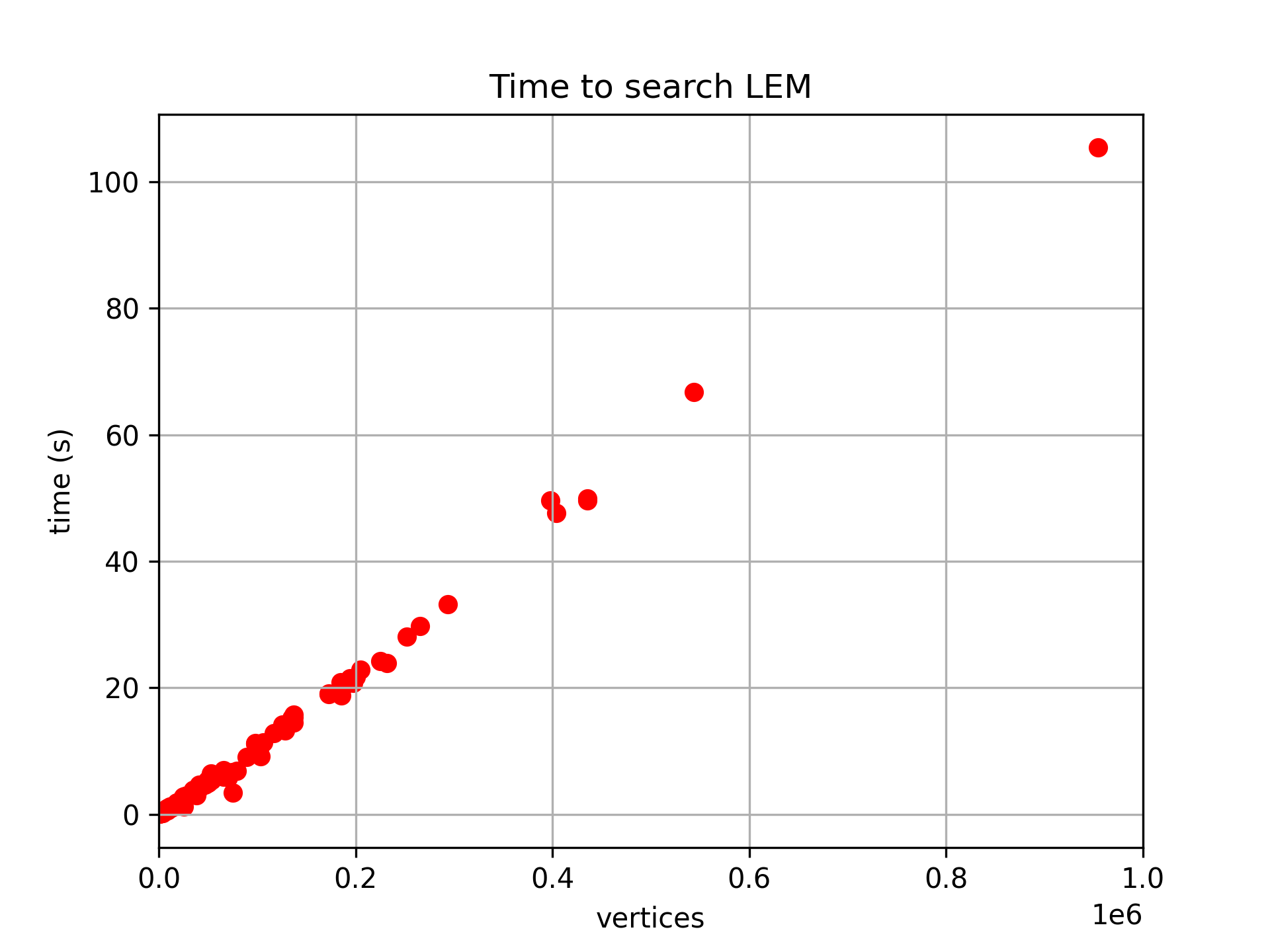}
    \subcaption{Running times of LEM.}
\end{minipage}
    \caption{Running times in total (top) and for searching (bottom) as function of the number of vertices on the Groningen Skeletonization Benchmark. Values over 20 hours omitted.}
    \label{fig:runtime_groningen}
\end{figure}

\paragraph*{Running Time}
Although the running time of local separator skeletonization methods depends very much on the search for separators, which in turn depends on the structure of the input, we give the running times of the examined methods in Figure~\ref{fig:runtime_groningen} as a function of the number of vertices in the input, over the entirety of the Groningen Skeletonization Benchmark~\cite{skelbench}.

Remarkably, we find that the multilevel algorithm not only outperforms LSS by several orders of magnitude, but also that it seems to be less dependant on the underlying structure of the triangle meshes, giving what appears to be a slightly superlinear curve. This effect is even more pronounced when considering only the time to search for separators. Under assumptions about the degree of the graphs, we showed that searching was $O(1)$ for a single separator and $O(|V|)$ in total. This experiment seems to confirm that this assumption is fitting for classical input, as is the case with the triangle meshes of the Groningen Skeletonization Benchmark.

\begin{table}[p]
    \centering
    \begin{tabular}{c|c|c|c|c|c|c}
    input   &   algorithm & coarsen (s) & search (s) & project (s) & pack (s) & total (s) \\\hline
    \multirow{3}*{19465}	&	LSS     & - & 25.8787	& - &	138.836	&	164.714 \\
                                &  LEM  &1.20002	&	9.18989	&	9.410914	&	12.4392	&	26.8306 \\
                                &  LEMTS&1.21294	&	9.41457	&	14.851733	&	9.8696	&	26.5518 \\\hline
    \multirow{3}*{dragon}   & LSS       & - &46615.4	& - & 	20637.9	&	67255.9 \\
                                & LEM   & 6.25851	&	49.6567	&	18.753684	&	3.42042	&	67.4554 \\
                                & LEMTS & 6.19801	&	50.2347	&	24.6420235	&	3.22507	&	69.4672 \\\hline
    \multirow{3}*{fertility}& LSS       & - &95.8574	& - &42.1409	&	136.664 \\
                                & LEM   & 0.234681	&	2.38572	&	0.7753682	&	0.0600592	&	3.03186 \\
                                & LEMTS & 0.244822	&	2.4305	&	1.8177949	&	0.0559833	&	3.44687 \\\hline
    \multirow{3}*{happy4}	&	LSS     & - & 7646.21	& - & 15731.9	&	23379.2 \\
                                & LEM & 6.89638	&	49.6684	&	12.7328477	&	2.22244	&	64.2245 \\
                                & LEMTS & 6.93433	&	50.1269	&	16.0581559	&	1.75872	&	65.4557 \\\hline
    \multirow{3}*{horse}	&	LSS     & - & 544.182	& - &	85.6915	&	628.3 \\
                                & LEM & 0.504898	&	5.10887	&	1.2762728	&	0.0669417	&	6.27845 \\
                                & LEMTS & 0.502791	&	5.17779	&	2.7604771	&	0.0542464	&	6.86374 \\\hline
    \multirow{3}*{neptune}	&	LSS     & - & 56.296	& - & 62.8555	&	119.232 \\
                                & LEM &0.312728	&	2.7382	&	1.0849813	&	0.230939	&	3.77785 \\
                                & LEMTS& 0.308124	&	2.76549	&	2.4149765	&	0.182575	&	4.22093 \\
    \end{tabular}
    \caption{Excerpt of running time measurements. In addition to measuring the total time, we also measure the time spent on each phase of the algorithm.}
    \label{tab:time}
\end{table}

In Table~\ref{tab:time} we show an excerpt of the running time measurements, including measurements of the phases of the algorithm. Here the vast gap in performance is clear, especially for \texttt{dragon.ply}, which is the largest input for which we have been able to run LSS, given a time frame of 20 hours. For this particular instance we achieve a running time that is almost a thousand times faster.

\begin{figure}[!btp]
    \centering
    \includegraphics[width=0.625\textwidth]{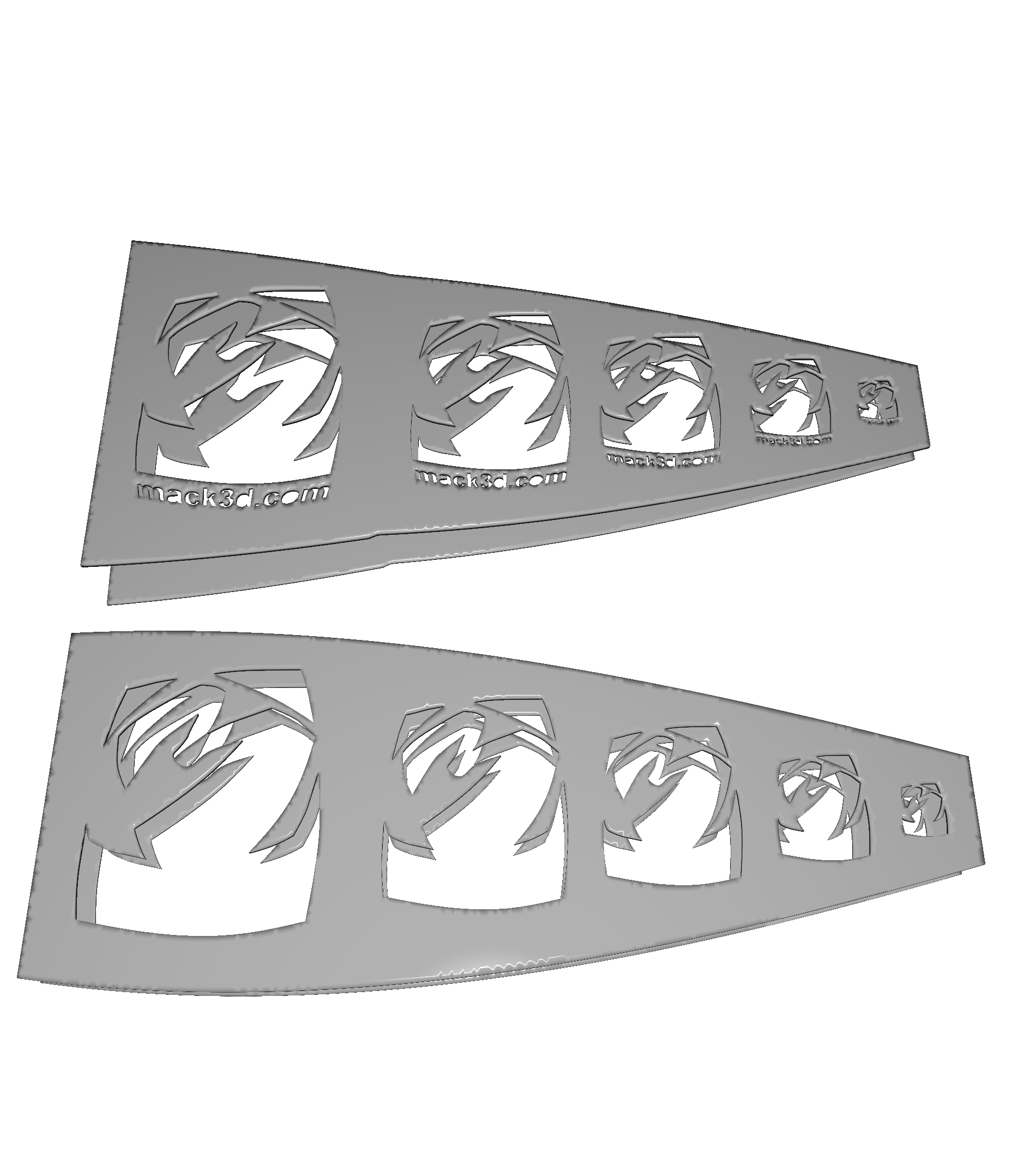} 
    \caption{The triangle mesh \texttt{19465.ply}, where packing makes up a large portion of the running time for (all) local separator based skeletonization algorithms.}
    \label{fig:19465}
\end{figure}

Of note is that LEMTS spends more time on projection, as expected, but less time on packing than LEM.
As stated previously, the search for separators is often the dominating phase, however there are types of input for which this is not the case, as evidenced by \texttt{19465.ply}. 
The mesh consists of flat sheets with small details engraved, as can be seen in Figure~\ref{fig:19465}. For both LSS and the multilevel algorithms, a large portion of the time is spent on packing and projection. This can occur if the separators are generally small and plentiful, so that many of them may quickly be found. For \texttt{19465.ply} these are particularly present around the imprinted text on the top sheets. It is worth noting that this would likely also be an example for which the structure of the input matters greatly for the running time of our multilevel algorithms.

\section{Conclusions and Future Work}
We have proposed a multilevel algorithm for computing local separator-based curve skeletons, and shown that the approach is very efficient. We obtain a practical running time that appears near linear in the number of vertices of the input (see Figure~\ref{fig:runtime_groningen}) with up to thousandfold improvement in running time while not deteriorating the quality of the output substantially, if at all.

This type of running time improvement makes separator-based skeletonization applicable as a tool in biomedical image analysis, including frame-by-frame skeletonization of videos~\cite{aasa}.

The application to video skeletonization motivates an unexplored line of related work, namely that of efficiently dynamically updating skeletons in a series of related shapes.

The multilevel approach offers great flexibility that has yet to be explored. It is easy to imagine coarsening schemes targeting specific structures of input, such as contracting clusters, rather than edges, on voxel input. 
These contraction schemes may provide new trade-offs between practical performance and skeleton quality.

When applying coarsening to scale-free graphs, as might be the case for data visualisation or areas of application that are not classical for skeletonization, we move into a domain known from the field of graph partitioning to cause trouble for matching contraction schemes~\cite{multipower}.
It is interesting to see if the improved practical performance, and the applicability to any spatially embedded graph, opens up for new areas of application of skeletons.



\bibliography{ref}

\appendix

\clearpage
\section{Time measurements}\label{sec:time}
\subsection{LSS}
\begin{longtable}{c|c|c|c|c|c}
    \centering
    input               &   vertices & edges    & search (s) & pack (s) & total (s) \\\hline
    17674	            &	124998	&	375000	&	2210.77	    &  	1267.03	&	3478.68 \\
    18020	            &	75122	&	225599	&	3.99153	    &	45.1806	&	49.2553 \\
    19465	            &	103351	&	310989	&	25.8787	    &	138.836	&	164.714 \\
    20304	            &	928	    &	2790	&	0.241744	&	0.0326365	&	0.276727 \\
    20306	            &	300	    &	894	    &	0.0331013	&	0.0025101	&	0.0371508 \\
    21362	            &	1828	&	5862	&	1.35847	    &	0.311618	&	1.67805 \\
    21464	            &	9041	&	26817	&	0.855021	&	0.163493	&	1.02239 \\
    21747	            &	2918	&	8766	&	0.0958573	&	0.0691565	&	0.167391 \\
    21788	            &	23062	&	69132	&	34.9016	    &	7.44311	&	42.3761 \\
    22081	            &	3970	&	11934	&	0.892292	&	0.485193	&	1.3761 \\
    22290	            &	944	    &	2832	&	1.28241	    &	0.0388288	&	1.32425 \\
    22601	            &	12554	&	37704	&	2.08853	    &	2.40571	&	4.50452 \\
    22669	            &	28738	&	86208	&	199.372	    &	25.7059	&	225.13 \\
    22701	            &	52784	&	158346	&	7008.77	    &	396.256	&	7405.48 \\
    22711	            &	2820	&	8460	&	1.23153	    &	0.252354	&	1.49249 \\
    22808	            &	9640	&	28920	&	4.3411	    &	0.936791	&	5.28941 \\
    22826	            &	56143	&	168453	&	4086.67	    &	867.461	&	4951.25 \\
    23091	            &	4120	&	12354	&	5.90686	    &	0.169803	&	6.0869 \\
    711\_dente	        &	9337	&	28005	&	34.1009	    &	4.06142	&	38.1319 \\
    armadillo	        &	172974	&	518916	&	11448.4	    &	3543.65	&	14997.0 \\
    armadillo\_vc	    &	106289	&	318861	&	3204.65	    &	1383.1	&	4592.3 \\
    asiandragon9a	    &	231606	&	694743	&	5918.86	    &	2587.44	&	8507.07 \\
    ballJoint	        &	137062	&	411180	&	20924.2	    &	1264.41	&	22189.5 \\
    bird2	            &	11718	&	35148	&	21.1993	    &	5.67294	&	26.8925 \\
    bird3	            &	46866	&	140592	&	502.613	    &	119.191	&	625.678 \\
    bird4	            &	187458	&	562368	&	15499.6	    &	2143.4	&	17643.8 \\
    bird	            &	11718	&	35148	&	21.455	    &	5.69072	&	27.1211 \\
    bmw2	            &	125424	&	375531	&	592.198	    &	723.553	&	1313.84 \\
    bmw	                &	70536	&	211048	&	30.796	    &	175.206	&	205.87 \\
    boeing	            &	3975	&	10811	&	0.0979876	&	0.0840983	&	0.185436 \\
    brain2	            &	294016	&	882048	&	52170.7	    &	6002.66	&	58175.3 \\
    bunny	            &	34834	&	104288	&	871.727	    &	128.254	&	1000.12 \\
    bun\_zipper	        &	34834	&	104288	&	875.806	    &	125.535	&	1004.16 \\
    bun\_zipper\_res2	&	8260	&	24480	&	25.3724	    &	4.08634	&	29.4296 \\
    bun\_zipper\_res3	&	2000	&	5775	&	1.02591	    &	0.143932	&	1.17524 \\
    bun\_zipper\_res4	&	518	    &	1425	&	0.0644569	&	0.00645594	&	0.0730992 \\
    chicken\_high	&	135142	&	405420	&	18539.4	&	2860.69	&	21424.2 \\
    cow10	&	181839	&	545481	&	10713.4	&	3048.3	&	13762.7 \\
    cow2	&	185730	&	557184	&	10498.4	&	3797.46	&	14222.4 \\
    cow3	&	252054	&	756090	&	46773.5	&	9664.54	&	56448.4 \\
    cube2	&	98306	&	294912	&	5827.69	&	35.6819	&	5863.63 \\
    dog	&	18114	&	54336	&	36.1876	&	13.7726	&	49.996 \\
    dragon	&	435680	&	1307182	&	46615.4	&	20637.9	&	67255.9 \\
    ele12a	&	50488	&	151470	&	430.976	&	107.081	&	538.198 \\
    ele\_fine	&	173026	&	519060	&	12251.0	&	2526.92	&	14778.7 \\
    elk2	&	35062	&	105186	&	132.307	&	98.1348	&	230.095 \\
    fertility	&	24994	&	75000	&	95.8574	&	42.1409	&	136.664 \\
    frog	&	37225	&	111669	&	325.885	&	102.875	&	431.331 \\
    gargoyle	&	25002	&	75000	&	221.986	&	68.5221	&	291.185 \\
    grandpiano	&	2904	&	8136	&	0.132531	&	0.0202437	&	0.154685 \\
    happy4	&	397569	&	1190310	&	7646.21	&	15731.9	&	23379.2 \\
    heart	&	23062	&	69132	&	34.0533	&	6.79733	&	40.8846 \\
    heptoroid	&	79056	&	237294	&	66.3978	&	109.493	&	176.037 \\
    hh2	&	197245	&	591230	&	25040.0	&	4864.86	&	29869.8 \\
    hh	&	49374	&	147871	&	649.029	&	180.944	&	830.178 \\
    hht	&	53754	&	161109	&	1115.03	&	214.659	&	1329.94 \\
    hippo	&	40367	&	121082	&	355.525	&	132.543	&	489.036 \\
    horse3	&	193934	&	581796	&	15966.7	&	1511.87	&	17478.4 \\
    horse9	&	89400	&	268187	&	1071.45	&	387.892	&	1458.51 \\
    horse	&	48485	&	145449	&	544.182	&	85.6915	&	628.3 \\
    hound2	&	200463	&	600913	&	5757.82	&	2253.16	&	7977.3 \\
    hound	&	12578	&	37592	&	7.98674	&	3.63663	&	11.6414 \\
    human\_hand	&	49374	&	147871	&	611.868	&	172.992	&	783.473 \\
    humerus	&	35002	&	105000	&	303.445	&	37.4675	&	340.978 \\
    ico	&	40962	&	122880	&	583.475	&	5.22691	&	588.763 \\
    kitten3	&	54838	&	164508	&	1491.24	&	389.112	&	1884.23 \\
    kitten	&	137098	&	411294	&	18429.4	&	3380.57	&	21810.9 \\
    lion2	&	65648	&	196696	&	1304.7	&	378.208	&	1683.55 \\
    lion	&	184738	&	550783	&	10751.5	&	2252.04	&	12985.5 \\
    lucy	&	100697	&	302085	&	2067.93	&	1035.66	&	3104.06 \\
    memento	&	26277	&	78825	&	100.99	&	36.0391	&	137.863 \\
    micscope2	&	265410	&	792121	&	1947.31	&	1260.84	&	3206.91 \\
    micscope	&	4146	&	12355	&	0.373911	&	0.0657359	&	0.443446 \\
    m	&	25631	&	71054	&	1.14553	&	1.7946	&	2.95392 \\
    mouse2	&	403652	&	1210944	&	44112.1	&	8690.16	&	52701.7 \\
    mouse	&	6311	&	18921	&	2.92597	&	0.691586	&	3.62578 \\
    neptune	&	28052	&	84168	&	56.296	&	62.8555	&	119.232 \\
    noisydino2	&	23370	&	69872	&	67.7212	&	30.5944	&	98.2667 \\
    noisydino8a	&	31497	&	94485	&	174.728	&	45.5758	&	219.857 \\
    noisydino	&	23370	&	69872	&	69.0721	&	30.8234	&	100.0 \\
    pig2	&	225282	&	675840	&	46042.1	&	8980.69	&	54639.5 \\
    pig	&	3522	&	10560	&	2.13615	&	0.539077	&	2.67718 \\
    pipe	&	2820	&	8460	&	1.20779	&	0.253799	&	1.46849 \\
    rcube2	&	98306	&	294912	&	6511.13	&	63.371	&	6574.24 \\
    rcube	&	24578	&	73728	&	281.882	&	3.76864	&	285.555 \\
    rhino2	&	128342	&	384790	&	2160.26	&	859.125	&	3025.39 \\
    rhino	&	8075	&	24058	&	3.43115	&	1.16439	&	4.60909 \\
    rockerArm2	&	43213	&	129639	&	792.765	&   118.601	&	911.377 \\
    rockerArm	&	40177	&	120531	&	629.687	&   62.2081	&	692.141 \\
    rotor	&	2328	&	7104	&	0.168337	&   0.0910922	&	0.261838 \\
    sacrum	&	204710	&	614226	&	9076.89	&	6283.98	&	15415.7 \\
    sandal2	&	40316	&	119568	&	27.8837	&	14.9089	&	42.8545 \\
    sandal3	&	22473	&	66039	&	35.9975	&	9.61242	&	45.7281 \\
    sandal	&	2636	&	7608	&	0.150706	&	0.0515755	&	0.205492 \\
    scapula	&	116930	&	350784	&	24658.3	&	3220.81	&	27880.1 \\
    screwdriver	&	27152	&	81450	&	635.749	&	48.1479	&	683.989 \\
    seabowl	&	45095	&	135303	&	569.478	&	119.554	&	688.461 \\
    ship1	&	2526	&	7220	&	0.0986438	&	0.0514934	&	0.152607 \\
    ship2	&	38122	&	113189	&	13.0851	&	20.6763	&	33.8212 \\
    ship3	&	74544	&	222199	&	56.9407	&	168.925	&	225.95 \\
    spider	&	4675	&	13929	&	0.524195	&	0.252	&	0.780605 \\
    tomgun2	&	65638	&	196496	&	82.568	&	78.9012	&	161.01 \\
    tomgun	&	4177	&	12305	&	0.388707	&	0.112197	&	0.504491 \\
    walleye2	&	49518	&	146582	&	309.593	&	81.8352	&	389.27 \\
    walleye	&	3343	&	9286	&	0.927982	&	0.128176	&	1.06049 \\
    \caption{Runtime measurements of LSS on the Groningen Skeletonization Benchmark}
    \label{tab:runtime_lss}
\end{longtable}

\setlength{\tabcolsep}{2pt}
\subsection{LEM}
\begin{longtable}{c|c|c|c|c|c|c|c}
    \centering
    input   &   vertices & edges    & coarsen (s) & search (s) & project (s) & pack (s) & total (s) \\\hline
    17674	&	124998	&	375000	&	1.42035	&	13.4979	&	3.4347296	&	0.349129	&	16.8551 \\
    18020	&	75122	&	225599	&	0.713029	&	3.41843	&	6.5007303	&	10.2568	&	17.022 \\
    19465	&	103351	&	310989	&	1.20002	&	9.18989	&	9.410914	&	12.4392	&	26.8306 \\
    20304	&	928	&	2790	&	0.00472449	&	0.101381	&	0.022000955	&	0.000595655	&	0.119704 \\
    20306	&	300	&	894	&	0.00202366	&	0.0381954	&	0.012915355	&	0.000244554	&	0.0477475 \\
    21362	&	1828	&	5862	&	0.00957011	&	0.211413	&	0.03324375	&	0.00390133	&	0.249427 \\
    21464	&	9041	&	26817	&	0.0507172	&	0.605148	&	0.3172494	&	0.0488438	&	0.850748 \\
    21747	&	2918	&	8766	&	0.013928	&	0.16035	&	0.12938649	&	0.0353876	&	0.272814 \\
    21788	&	23062	&	69132	&	0.187331	&	2.19023	&	0.57236465	&	0.0442584	&	2.69455 \\
    22081	&	3970	&	11934	&	0.0225922	&	0.3726	&	0.10956878	&	0.0134095	&	0.463392 \\
    22290	&	944	&	2832	&	0.00453149	&	0.138593	&	0.018698548	&	0.000293809	&	0.154601 \\
    22601	&	12554	&	37704	&	0.0887529	&	0.95333	&	0.4130805	&	0.120493	&	1.35994 \\
    22669	&	28738	&	86208	&	0.274313	&	2.98669	&	0.51025489	&	0.0317114	&	3.5369 \\
    22701	&	52784	&	158346	&	0.473056	&	6.44374	&	0.28137197	&	0.00285178	&	7.15982 \\
    22711	&	2820	&	8460	&	0.0140559	&	0.318642	&	0.12283238	&	0.00448683	&	0.395761 \\  
    22808	&	9640	&	28920	&	0.0650948	&	0.998223	&	0.20622033	&	0.0163558	&	1.17796 \\
    22826	&	56143	&	168453	&	0.512271	&	6.07731	&	1.0100605	&	0.0550804	&	7.14747 \\
    23091	&	4120	&	12354	&	0.0196667	&	0.455104	&	0.083162039	&	0.00300366	&	0.526664 \\  
    711\_dente	&	9337	&	28005	&	0.0641343	&	0.985219	&	0.22779954	&	0.0100632	&	    9 \\
    armadillo	&	172974	&	518916	&	2.51794	&	19.0322	&	5.0844824	&	0.48021	&	24.3899 \\
    armadillo\_vc	&	106289	&	318861	&	1.43063	&	11.3844	&	3.0456077	&	0.28918	&	14.4931 \\
    asiandragon10a	&	954227	&	2862226	&	14.0947	&	105.433	&	56.60956	&	7.17272	&	150.045 \\
    asiandragon9a	&	231606	&	694743	&	2.89486	&	23.9127	&	10.1410711	&	1.61975	&	32.5958 \\
    ballJoint	&	137062	&	411180	&	1.66799	&	15.7521	&	1.5901205	&	0.0301499	&	18.3389 \\
    bird2	&	11718	&	35148	&	0.092326	&	1.15743	&	0.35719044	&	0.0373765	&	1.44558 \\
    bird3	&	46866	&	140592	&	0.531434	&	4.70348	&	1.0330702	&	0.0722855	&	5.81154 \\
    bird4	&	187458	&	562368	&	2.23458	&	20.2136	&	2.6803021	&	0.134734	&	24.0237 \\
    bird	&	11718	&	35148	&	0.101527	&	1.13771	&	0.35191275	&	0.036676	&	1.42879 \\
    bmw2	&	125424	&	375531	&	1.53246	&	14.1645	&	7.991554	&	7.8399	&	26.958 \\
    bmw	&	70536	&	211048	&	0.699872	&	5.86059	&	3.613716	&	3.39631	&	11.5671 \\
    boeing	&	3975	&	10811	&	0.0174186	&	0.192201	&	0.140817	&	0.0562142	&	0.336693 \\
    brain2	&	294016	&	882048	&	3.57649	&	33.2128	&	5.1804536	&	0.266868	&	39.709 \\
    bunny	&	34834	&	104288	&	0.351915	&	3.78063	&	0.67709918	&	0.0480589	&	4.5399 \\
    bun\_zipper	&	34834	&	104288	&	0.359259	&	3.79813	&	0.66139868	&	0.046271	&	4.5309 \\
    bun\_zipper\_res2	&	8260	&	24480	&	0.06046	&	0.86436	&	0.18984358	&	0.0197775	&	1.04349 \\
    bun\_zipper\_res3	&	2000	&	5775	&	0.0103474	&	0.197153	&	0.058296746	&	0.00757159	&	0.245245 \\
    bun\_zipper\_res4	&	518	&	1425	&	0.00334376	&	0.0538328	&	0.019278557	&	0.000693185	&	0.0692462 \\
    chicken\_high	&	135142	&	405420	&	1.6475	&	15.2427	&	1.868914	&	0.041709	&	17.9348 \\
    cow10	&	181839	&	545481	&	2.13711	&	19.3315	&	4.4587347	&	0.226973	&	23.704 \\
    cow2	&	185730	&	557184	&	2.13695	&	18.7325	&	5.4557653	&	0.530463	&	23.8909 \\
    cow3	&	252054	&	756090	&	3.43701	&	28.0588	&	6.0824288	&	0.546523	&	34.7918 \\
    cow9	&	136988	&	410946	&	1.59892	&	14.5176	&	3.3032935	&	0.169401	&	17.7899 \\
    cube2	&	98306	&	294912	&	0.883138	&	11.2845	&	1.1164629	&	0.0178002	&	12.8058 \\
    dog	&	18114	&	54336	&	0.155731	&	1.79058	&	0.5426249	&	0.0481439	&	2.25427 \\
    dragon	&	435680	&	1307182	&	6.25851	&	49.6567	&	18.753684	&	3.42042	&	67.4554 \\
    dragon\_vrip	&	435680	&	1307182	&	6.06857	&	49.8861	&	18.780025	&	3.46152	&	67.4061 \\
    dragon\_vrip\_res2	&	102112	&	303088	&	1.26637	&	10.3709	&	2.9057934	&	0.333104	&	13.3148 \\
    dragon\_vrip\_res3	&	24417	&	71126	&	0.230862	&	2.22082	&	0.7889312	&	0.0935623	&	2.91563 \\
    dragon\_vrip\_res4	&	5743	&	16431	&	0.0387931	&	0.47057	&	0.21363361	&	0.0264042	&	0.63171 \\
    ele12a	&	50488	&	151470	&	0.53858	&	5.01851	&	1.2827198	&	0.0729709	&	6.25953 \\
    ele\_fine	&	173026	&	519060	&	2.29251	&	19.088	&	3.5553908	&	0.187306	&	23.2524 \\
    elk2	&	35062	&	105186	&	0.377056	&	3.53251	&	0.81446343	&	0.0563738	&	4.33083 \\
    fertility	&	24994	&	75000	&	0.234681	&	2.38572	&	0.7753682	&	0.0600592	&	3.03186 \\
    frog	&	37225	&	111669	&	0.362007	&	3.89609	&	1.0568452	&	0.153736	&	4.929 \\
    gargoyle	&	25002	&	75000	&	0.236856	&	2.72392	&	0.5215487	&	0.0393314	&	3.23942 \\
    grandpiano	&	2904	&	8136	&	0.0113828	&	0.175221	&	0.0559794	&	0.00671873	&	0.224557 \\
    happy4	&	397569	&	1190310	&	6.89638	&	49.6684	&	12.7328477	&	2.22244	&	64.2245 \\
    happy	&	543822	&	1631680	&	7.9146	&	66.7996	&	20.8506963	&	3.93399	&	87.752 \\
    heart	&	23062	&	69132	&	0.20292	&	2.16714	&	0.594186	&	0.0456437	&	2.69486 \\
    heptoroid	&	79056	&	237294	&	0.969235	&	6.84475	&	3.6109921	&	1.1477	&	10.5038 \\
    hh2	&	197245	&	591230	&	2.50199	&	20.7739	&	3.9906898	&	0.190638	&	25.4634 \\
    hh	&	49374	&	147871	&	0.517598	&	4.89417	&	1.366125	&	0.0889966	&	6.14942 \\
    hht	&	53754	&	161109	&	0.649063	&	5.36146	&	1.4218338	&	0.111029	&	6.83171 \\
    hippo	&	40367	&	121082	&	0.407343	&	4.28243	&	0.9159472	&	0.0617453	&	5.18609 \\
    horse3	&	193934	&	581796	&	2.46719	&	21.5099	&	4.4235634	&	0.160395	&	26.2848 \\
    horse9	&	89400	&	268187	&	1.02295	&	9.02118	&	2.4578347	&	0.153465	&	11.2601 \\
    horse	&	48485	&	145449	&	0.504898	&	5.10887	&	1.2762728	&	0.0669417	&	6.27845 \\
    hound2	&	200463	&	600913	&	2.21726	&	21.6746	&	5.4904756	&	0.53687	&	26.9504 \\
    hound	&	12578	&	37592	&	0.101115	&	1.15836	&	0.4332293	&	0.0557417	&	1.52113 \\
    human\_hand	&	49374	&	147871	&	0.563042	&	4.92293	&	1.2874627	&	0.084787	&	6.17145 \\
    humerus	&	35002	&	105000	&	0.368269	&	3.59696	&	0.9222292	&	0.0306101	&	4.42637 \\
    ico	&	40962	&	122880	&	0.385356	&	4.70133	&	0.2181534	&	0.000810135	&	5.27336 \\
    kitten3	&	54838	&	164508	&	0.564621	&	5.72572	&	0.92238507	&	0.0368314	&	6.8351 \\
    kitten	&	137098	&	411294	&	1.63015	&	15.2895	&	1.9815453	&	0.0444201	&	18.0658 \\
    lion2	&	65648	&	196696	&	0.761081	&	6.9304	&	1.6788871	&	0.16808	&	8.70898 \\
    lion	&	184738	&	550783	&	2.2956	&	20.9184	&	3.6100428	&	0.288113	&	25.2353 \\
    lucy	&	100697	&	302085	&	1.20264	&	10.2482	&	2.5055618	&	0.201589	&	12.8405 \\
    memento	&	26277	&	78825	&	0.251496	&	2.57331	&	0.9146878	&	0.0917045	&	3.3258 \\
    micscope2	&	265410	&	792121	&	3.0882	&	29.8056	&	13.284129	&	3.11965	&	41.5345 \\
    micscope	&	4146	&	12355	&	0.020966	&	0.407153	&	0.0991788	&	0.0213976	&	0.502653 \\
    m	&	25631	&	71054	&	0.172642	&	1.23407	&	0.7839142	&	0.511338	&	2.2808 \\
    mouse2	&	403652	&	1210944	&	4.65876	&	47.6706	&	20.374639	&	7.6661	&	68.2746 \\
    mouse	&	6311	&	18921	&	0.0383902	&	0.652853	&	0.23409506	&	0.0430953	&	0.847976 \\
    neptune	&	28052	&	84168	&	0.312728	&	2.7382	&	1.0849813	&	0.230939	&	3.77785 \\
    noisydino2	&	23370	&	69872	&	0.235437	&	2.25831	&	0.7979447	&	0.0873257	&	2.91716 \\
    noisydino8a	&	31497	&	94485	&	0.319535	&	3.00893	&	0.9531643	&	0.0592822	&	3.78642 \\
    noisydino	&	23370	&	69872	&	0.223275	&	2.25698	&	0.8086116	&	0.0826377	&	2.92994 \\
    pig2	&	225282	&	675840	&	2.76833	&	24.2396	&	5.5608734	&	0.502856	&	30.1216 \\
    pig	&	3522	&	10560	&	0.0212906	&	0.286789	&	0.12774142	&	0.0229754	&	0.395927 \\
    pipe	&	2820	&	8460	&	0.0142335	&	0.321047	&	0.1244618	&	0.00482331	&	0.396382 \\
    rcube2	&	98306	&	294912	&	0.956272	&	11.0931	&	1.2481161	&	0.0166792	&	12.7712 \\
    rcube	&	24578	&	73728	&	0.184148	&	2.7447	&	0.28438118	&	0.00394089	&	3.11112 \\
    rhino2	&	128342	&	384790	&	1.48467	&	13.2432	&	3.2735371	&	0.26516	&	16.5302 \\
    rhino	&	8075	&	24058	&	0.0515384	&	0.736705	&	0.26798898	&	0.0347169	&	0.949532 \\
    rockerArm2	&	43213	&	129639	&	0.458789	&	4.56635	&	0.8497291	&	0.0225036	&	5.48141 \\
    rockerArm	&	40177	&	120531	&	0.40493	&	4.16544	&	0.8525191	&	0.0211871	&	5.00707 \\
    rotor	&	2328	&	7104	&	0.0119959	&	0.178435	&	0.09086232	&	0.0205675	&	0.257278 \\
    sacrum	&	204710	&	614226	&	2.81857	&	22.8123	&	3.8453268	&	0.223871	&	27.7257 \\
    sandal2	&	40316	&	119568	&	0.344412	&	3.9165	&	1.8148215	&	0.581826	&	5.66137 \\
    sandal3	&	22473	&	66039	&	0.203555	&	2.18896	&	0.782279	&	0.250619	&	3.01813 \\
    sandal	&	2636	&	7608	&	0.0122295	&	0.209453	&	0.10131738	&	0.0163152	&	0.288116 \\
    scapula	&	116930	&	350784	&	1.50627	&	12.7815	&	1.6220242	&	0.0559348	&	15.2969 \\
    screwdriver	&	27152	&	81450	&	0.256815	&	2.90263	&	0.54800159	&	0.022825	&	3.46602 \\
    seabowl	&	45095	&	135303	&	0.499152	&	4.7644	&	1.0131984	&	0.0813877	&	5.8672 \\
    ship1	&	2526	&	7220	&	0.0123951	&	0.149451	&	0.09039014	&	0.0201223	&	0.226589 \\
    ship2	&	38122	&	113189	&	0.339199	&	2.99193	&	1.5399452	&	0.589327	&	4.64253 \\
    ship3	&	74544	&	222199	&	0.862242	&	6.60576	&	3.8058408	&	4.91441	&	14.06 \\
    spider	&	4675	&	13929	&	0.0256434	&	0.364411	&	0.16410401	&	0.0562838	&	0.529353 \\
    tomgun2	&	65638	&	196496	&	0.651883	&	5.95249	&	2.2170524	&	0.441005	&	8.03167 \\
    tomgun	&	4177	&	12305	&	0.0214907	&	0.294542	&	0.1132001	&	0.0142692	&	0.384086 \\
    walleye2	&	49518	&	146582	&	0.471817	&	5.38596	&	1.4151858	&	0.535818	&	7.05759 \\
    walleye	&	3343	&	9286	&	0.0165995	&	0.314689	&	0.08863707	&	0.0095865	&	0.390043 \\
    \caption{Runtime measurements of multilevel skeletonization using Light Edge Matchings on the Groningen Skeletonization Benchmark}
    \label{tab:runtime_lem}
\end{longtable}

\subsection{LEMTS}
\begin{longtable}{c|c|c|c|c|c|c|c}
    \centering
    input   &   vertices & edges    & coarsen (s) & search (s) & project (s) & pack (s) & total (s) \\\hline
    17674	&	124998	&	375000	&	1.37288	&	13.8259	&	6.1536268	&	0.252363	&	18.094 \\
    18020	&	75122	&	225599	&	0.70643	&	3.50639	&	12.0219301	&	9.39802	&	18.3356 \\
    19465	&	103351	&	310989	&	1.21294	&	9.41457	&	14.851733	&	9.8696	&	26.5518 \\
    20304	&	928	&	2790	&	0.00476734	&	0.105907	&	0.051260507	&	0.000753875	&	0.135355 \\
    20306	&	300	&	894	&	0.00249888	&	0.0371973	&	0.024086454	&	0.000285874	&	0.0506875 \\
    21362	&	1828	&	5862	&	0.00991051	&	0.210066	&	0.09045887	&	0.00351191	&	0.270029 \\
    21464	&	9041	&	26817	&	0.0559211	&	0.619575	&	0.69601524	&	0.0423914	&	0.996912 \\
    21747	&	2918	&	8766	&	0.014122	&	0.172405	&	0.29007067	&	0.0304931	&	0.338011 \\
    21788	&	23062	&	69132	&	0.193463	&	2.16468	&	1.28154913	&	0.0426115	&	2.93941 \\
    22081	&	3970	&	11934	&	0.0226519	&	0.380269	&	0.25070048	&	0.0123173	&	0.519388 \\
    22290	&	944	&	2832	&	0.00448527	&	0.139352	&	0.038550872	&	0.000262215	&	0.160744 \\
    22601	&	12554	&	37704	&	0.0975062	&	0.987717	&	0.851674	&	0.100677	&	1.54694 \\
    22669	&	28738	&	86208	&	0.286316	&	3.0546	&	1.04230053	&	0.0200472	&	3.79582 \\
    22701	&	52784	&	158346	&	0.461819	&	6.42563	&	0.44446955	&	0.00207974	&	7.17059 \\
    22711	&	2820	&	8460	&	0.014311	&	0.323639	&	0.29059017	&	0.00287403	&	0.458739 \\
    22808	&	9640	&	28920	&	0.0634594	&	1.03166	&	0.41110239	&	0.00978584	&	1.27532 \\
    22826	&	56143	&	168453	&	0.519806	&	5.99483	&	2.4316612	&	0.0505605	&	7.60915 \\
    23091	&	4120	&	12354	&	0.0194701	&	0.469513	&	0.189625955	&	0.00240466	&	0.575799 \\
    711\_dente	&	9337	&	28005	&	0.0686141	&	0.998655	&	0.53344135	&	0.00867415	&	1.30378 \\
    armadillo	&	172974	&	518916	&	2.46986	&	19.1877	&	9.4305278	&	0.327339	&	25.8982 \\
    armadillo\_vc	&	106289	&	318861	&	1.39404	&	11.5249	&	6.2142442	&	0.214773	&	15.7169 \\
    asiandragon10a	&	954227	&	2862226	&	14.0653	&	105.332	&	83.178314	&	4.52812	&	155.153 \\
    asiandragon9a	&	231606	&	694743	&	2.8553	&	24.0592	&	16.917632	&	1.13746	&	34.8331 \\
    ballJoint	&	137062	&	411180	&	1.67507	&	15.6877	&	3.3411102	&	0.020369	&	18.9884 \\
    bird2	&	11718	&	35148	&	0.0805309	&	1.14146	&	0.84050036	&	0.0315955	&	1.59471 \\
    bird3	&	46866	&	140592	&	0.50282	&	4.73525	&	2.300044	&	0.0632088	&	6.22924 \\
    bird4	&	187458	&	562368	&	2.40017	&	20.3605	&	5.5991965	&	0.101529	&	25.3824 \\
    bird	&	11718	&	35148	&	0.093867	&	1.13195	&	0.81265529	&	0.0290167	&	1.57061 \\
    bmw2	&	125424	&	375531	&	1.51884	&	14.236	&	12.5885782	&	6.99201	&	28.1195 \\
    bmw	&	70536	&	211048	&	0.691096	&	5.92067	&	6.792073	&	2.90765	&	12.4277 \\
    boeing	&	3975	&	10811	&	0.0177992	&	0.196806	&	0.26577117	&	0.0488982	&	0.379375 \\
    brain2	&	294016	&	882048	&	3.57092	&	33.3187	&	8.5313576	&	0.242088	&	40.9286 \\
    bunny	&	34834	&	104288	&	0.363232	&	3.79656	&	1.37544507	&	0.039346	&	4.79618 \\
    bun\_zipper	&	34834	&	104288	&	0.352807	&	3.84765	&	1.29270687	&	0.0410116	&	4.79398 \\
    bun\_zipper\_res2	&	8260	&	24480	&	0.0529122	&	0.878866	&	0.39589784	&	0.0167918	&	1.11783 \\
    bun\_zipper\_res3	&	2000	&	5775	&	0.010301	&	0.204269	&	0.12545401	&	0.00639514	&	0.279434 \\
    bun\_zipper\_res4	&	518	&	1425	&	0.00291506	&	0.0560733	&	0.04312842	&	0.000641622	&	0.0777027 \\
    chicken\_high	&	135142	&	405420	&	1.60237	&	15.1774	&	3.5583817	&	0.0341168	&	18.4927 \\
    cow10	&	181839	&	545481	&	2.17961	&	19.437	&	8.4240864	&	0.175242	&	25.198 \\
    cow2	&	185730	&	557184	&	2.18291	&	19.0676	&	10.1746708	&	0.399085	&	25.9429 \\
    cow3	&	252054	&	756090	&	3.47587	&	28.2994	&	9.7904799	&	0.352022	&	36.1018 \\
    cow9	&	136988	&	410946	&	1.61981	&	14.357	&	6.3390355	&	0.144272	&	18.8167 \\
    cube2	&	98306	&	294912	&	0.888989	&	11.229	&	2.1635131	&	0.0130268	&	13.1271 \\
    dog	&	18114	&	54336	&	0.157395	&	1.79984	&	1.2110295	&	0.0413447	&	2.50017 \\
    dragon	&	435680	&	1307182	&	6.19801	&	50.2347	&	24.6420235	&	3.22507	&	69.4672 \\
    dragon\_vrip	&	435680	&	1307182	&	6.15461	&	50.2762	&	25.094422	&	3.18479	&	69.3239 \\
    dragon\_vrip\_res2	&	102112	&	303088	&	1.19471	&	10.4939	&	4.989363	&	0.250379	&	14.0577 \\
    dragon\_vrip\_res3	&	24417	&	71126	&	0.230291	&	2.25178	&	1.39281064	&	0.0784439	&	3.1478 \\
    dragon\_vrip\_res4	&	5743	&	16431	&	0.037996	&	0.472024	&	0.42511038	&	0.0251362	&	0.713175 \\
    ele12a	&	50488	&	151470	&	0.53404	&	5.03626	&	2.7621634	&	0.0612321	&	6.73142 \\
    ele\_fine	&	173026	&	519060	&	2.37588	&	19.1998	&	6.5201373	&	0.151195	&	24.563 \\
    elk2	&	35062	&	105186	&	0.381521	&	3.61194	&	1.60635173	&	0.0517945	&	4.71299 \\
    fertility	&	24994	&	75000	&	0.244822	&	2.4305	&	1.8177949	&	0.0559833	&	3.44687 \\
    frog	&	37225	&	111669	&	0.389961	&	3.90161	&	2.3034716	&	0.128718	&	5.41558 \\
    gargoyle	&	25002	&	75000	&	0.256161	&	2.72948	&	1.14925343	&	0.0307409	&	3.48302 \\
    grandpiano	&	2904	&	8136	&	0.0110036	&	0.17332	&	0.12285781	&	0.00571585	&	0.257232 \\
    happy4	&	397569	&	1190310	&	6.93433	&	50.1269	&	16.0581559	&	1.75872	&	65.4557 \\
    happy	&	543822	&	1631680	&	8.00506	&	66.9676	&	25.5106596	&	3.69385	&	89.5235 \\
    heart	&	23062	&	69132	&	0.173748	&	2.15771	&	1.35531625	&	0.0421466	&	2.93709 \\
    heptoroid	&	79056	&	237294	&	0.916259	&	6.89268	&	7.5853171	&	0.811116	&	11.527 \\
    hh2	&	197245	&	591230	&	2.47622	&	20.8036	&	7.4632686	&	0.141546	&	26.6708 \\
    hh	&	49374	&	147871	&	0.549817	&	4.90281	&	2.5108662	&	0.0721409	&	6.57878 \\
    hht	&	53754	&	161109	&	0.622599	&	5.42222	&	2.9174229	&	0.0837678	&	7.35467 \\
    hippo	&	40367	&	121082	&	0.448026	&	4.35356	&	2.0381683	&	0.0544027	&	5.67898 \\
    horse3	&	193934	&	581796	&	2.42718	&	21.594	&	8.339175	&	0.119572	&	27.6041 \\
    horse9	&	89400	&	268187	&	0.997401	&	8.90845	&	5.3143602	&	0.116107	&	12.1142 \\
    horse	&	48485	&	145449	&	0.502791	&	5.17779	&	2.7604771	&	0.0542464	&	6.86374 \\
    hound2	&	200463	&	600913	&	2.27636	&	21.788	&	10.7396585	&	0.43615	&	28.8567 \\
    hound	&	12578	&	37592	&	0.104805	&	1.18453	&	0.9476307	&	0.05026	&	1.72187 \\
    human\_hand	&	49374	&	147871	&	0.540201	&	4.91299	&	2.5593691	&	0.0726399	&	6.53822 \\
    humerus	&	35002	&	105000	&	0.353858	&	3.68623	&	1.8732058	&	0.0239917	&	4.85062 \\
    ico	&	40962	&	122880	&	0.367218	&	4.71673	&	0.2244697	&	0.000601861	&	5.29382 \\
    kitten3	&	54838	&	164508	&	0.556148	&	5.78029	&	1.6071862	&	0.0306763	&	7.14854 \\
    kitten	&	137098	&	411294	&	1.64587	&	15.2583	&	3.3345443	&	0.0400726	&	18.4326 \\
    lion2	&	65648	&	196696	&	0.697095	&	6.87126	&	3.0194409	&	0.141892	&	8.98981 \\
    lion	&	184738	&	550783	&	2.26651	&	21.0421	&	6.0643192	&	0.200554	&	26.2529 \\
    lucy	&	100697	&	302085	&	1.15179	&	10.2905	&	4.5310875	&	0.166912	&	13.5347 \\
    memento	&	26277	&	78825	&	0.2703	&	2.57669	&	2.1355263	&	0.0789704	&	3.78361 \\
    micscope2	&	265410	&	792121	&	2.99505	&	29.5575	&	21.684033	&	2.22369	&	43.0669 \\
    micscope	&	4146	&	12355	&	0.0204097	&	0.403439	&	0.19514139	&	0.0205848	&	0.536987 \\
    m	&	25631	&	71054	&	0.185894	&	1.24326	&	1.4135061	&	0.40673	&	2.45396 \\
    mouse2	&	403652	&	1210944	&	4.61188	&	47.9421	&	31.821861	&	5.08602	&	69.811 \\
    mouse	&	6311	&	18921	&	0.0384411	&	0.662394	&	0.52321536	&	0.0358318	&	0.972776 \\
    neptune	&	28052	&	84168	&	0.308124	&	2.76549	&	2.4149765	&	0.182575	&	4.22093 \\
    noisydino2	&	23370	&	69872	&	0.236898	&	2.26549	&	1.7555544	&	0.0726686	&	3.26698 \\
    noisydino8a	&	31497	&	94485	&	0.30512	&	3.03009	&	2.0297282	&	0.0489356	&	4.20738 \\
    noisydino	&	23370	&	69872	&	0.239336	&	2.25737	&	1.7335359	&	0.0699845	&	3.25217 \\
    pig2	&	225282	&	675840	&	2.69757	&	24.441	&	10.4396925	&	0.39946	&	31.8039 \\
    pig	&	3522	&	10560	&	0.0200319	&	0.306373	&	0.29780277	&	0.0225411	&	0.477153 \\
    pipe	&	2820	&	8460	&	0.0143922	&	0.328995	&	0.28918437	&	0.00284439	&	0.457443 \\
    rcube2	&	98306	&	294912	&	0.975693	&	11.1107	&	2.1495174	&	0.0140855	&	13.1207 \\
    rcube	&	24578	&	73728	&	0.186935	&	2.79348	&	0.52618835	&	0.00214423	&	3.2175 \\
    rhino2	&	128342	&	384790	&	1.44105	&	13.5052	&	6.3808746	&	0.215573	&	17.6696 \\
    rhino	&	8075	&	24058	&	0.0539976	&	0.736505	&	0.57325506	&	0.0331398	&	1.06554 \\
    rockerArm2	&	43213	&	129639	&	0.463274	&	4.54615	&	1.99677959	&	0.01904	&	5.82642 \\
    rockerArm	&	40177	&	120531	&	0.392621	&	4.1855	&	1.82895265	&	0.0141119	&	5.34244 \\
    rotor	&	2328	&	7104	&	0.0118651	&	0.181803	&	0.17895838	&	0.0197181	&	0.2917 \\
    sacrum	&	204710	&	614226	&	2.73006	&	22.9159	&	6.8929541	&	0.177687	&	28.8622 \\
    sandal2	&	40316	&	119568	&	0.346656	&	4.06497	&	3.8798318	&	0.430996	&	6.39904 \\
    sandal3	&	22473	&	66039	&	0.191946	&	2.23852	&	1.6852481	&	0.210204	&	3.33725 \\
    sandal	&	2636	&	7608	&	0.0121956	&	0.212033	&	0.1907273	&	0.0166909	&	0.325527 \\
    scapula	&	116930	&	350784	&	1.51976	&	12.8966	&	3.0576659	&	0.0464446	&	15.87 \\
    screwdriver	&	27152	&	81450	&	0.246062	&	2.97392	&	1.08712076	&	0.0162187	&	3.69605 \\
    seabowl	&	45095	&	135303	&	0.470481	&	4.82754	&	2.1035185	&	0.0734754	&	6.33133 \\
    ship1	&	2526	&	7220	&	0.0122548	&	0.160322	&	0.17935572	&	0.0204451	&	0.27645 \\
    ship2	&	38122	&	113189	&	0.328891	&	3.04911	&	3.2117841	&	0.473475	&	5.20903 \\
    ship3	&	74544	&	222199	&	0.866528	&	6.60474	&	7.1166426	&	3.51893	&	13.8594 \\
    spider	&	4675	&	13929	&	0.0240519	&	0.385193	&	0.3277413	&	0.0539168	&	0.607832 \\
    tomgun2	&	65638	&	196496	&	0.65928	&	6.10129	&	4.5972224	&	0.347689	&	8.98068 \\
    tomgun	&	4177	&	12305	&	0.0210423	&	0.299244	&	0.23149471	&	0.0134622	&	0.434965 \\
    walleye2	&	49518	&	146582	&	0.450784	&	5.3944	&	2.9724058	&	0.528678	&	7.53503 \\
    walleye	&	3343	&	9286	&	0.015999	&	0.311624	&	0.21487931	&	0.00890859	&	0.43915 \\
    \caption{Runtime measurements of multilevel skeletonization using Light Edge Matchings and Thickened Separators on the Groningen Skeletonization Benchmark}
    \label{tab:runtime_lemts}
\end{longtable}

\setlength{\tabcolsep}{3pt}
\section{Skeleton Comparisons}\label{sec:skel}
\subsection{LSS and LEM}
\begin{longtable}{c|c|c|c|c|c|c}
    \centering
    input & $\Delta$vertices &  $\Delta$leafs & $\Delta$branches & $\Delta$genus & $H($LEM,LSS$)$ & $H($LSS,LEM$)$  \\\hline
    17674	&	-214	&	-6	&	-7	&	-2	&	0.242987	&	0.136002 \\
    18020	&	-132	&	0	&	2	&	0	&	0.0267864	&	0.0265365 \\
    19465	&	-563	&	-3	&	-35	&	2	&	0.0437028	&	0.0415616 \\
    20304	&	0	&	6	&	-1	&	-2	&	0.534996	&	0.365925 \\
    20306	&	1	&	1	&	0	&	0	&	0.13103	&	0.137122 \\
    21362	&	1	&	1	&	0	&	0	&	0.138338	&	0.0941336 \\
    21464	&	-94	&	-1	&	-1	&	0	&	0.0287383	&	0.0848934 \\
    21747	&	-4	&	0	&	0	&	0	&	0.0232842	&	0.0247181 \\
    21788	&	-39	&	-4	&	3	&	0	&	0.331518	&	0.253539 \\
    22081	&	-41	&	2	&	2	&	0	&	0.0191042	&	0.0364621 \\
    22290	&	-4	&	-8	&	2	&	-1	&	0.724273	&	0.584943 \\
    22601	&	-12	&	-4	&	6	&	0	&	0.0448619	&	0.0455526 \\
    22669	&	-106	&	1	&	1	&	0	&	0.144748	&	0.123809 \\
    22701	&	-4	&	2	&	-1	&	0	&	0.595045	&	0.353578 \\
    22711	&	-10	&	0	&	0	&	0	&	0.0529444	&	0.0371701 \\
    22808	&	-76	&	0	&	0	&	0	&	0.0371294	&	0.0360032 \\
    22826	&	-35	&	16	&	1	&	-4	&	0.263343	&	0.257184 \\
    23091	&	-29	&	0	&	-1	&	0	&	0.396678	&	0.28389 \\
    711\_dente	&	-10	&	-3	&	-2	&	0	&	0.474553	&	0.201984 \\
    armadillo	&	-149	&	-17	&	-6	&	0	&	0.220078	&	0.165342 \\
    armadillo\_vc	&	-130	&	-19	&	-10	&	0	&	0.249026	&	0.0830985 \\
    asiandragon10a	&	485	&	63	&	27	&	1	&	-	&	- \\
    asiandragon9a	&	-303	&	-27	&	-3	&	0	&	0.146472	&	0.0809961 \\
    ballJoint	&	-34	&	-9	&	-5	&	0	&	0.335091	&	0.128461 \\
    bird2	&	-40	&	-2	&	-2	&	0	&	0.0911887	&	0.0434792 \\
    bird3	&	-66	&	-6	&	-2	&	0	&	0.102782	&	0.0483936 \\
    bird4	&	-99	&	-6	&	-2	&	0	&	0.131071	&	0.0457664 \\
    bird	&	-35	&	-1	&	-1	&	0	&	0.064136	&	0.0591958 \\
    bmw2	&	-61	&	-5	&	5	&	0	&	0.181548	&	0.165533 \\
    bmw	&	-96	&	-10	&	0	&	0	&	0.140737	&	0.13919 \\
    boeing	&	-6	&	4	&	2	&	0	&	0.0700192	&	0.0958734 \\
    brain2	&	-91	&	-51	&	-11	&	2	&	0.380483	&	0.274007 \\
    bunny	&	-35	&	-13	&	-2	&	0	&	0.371084	&	0.218552 \\
    bun\_zipper\_res2	&	-23	&	-5	&	-5	&	-1	&	0.309635	&	0.228842 \\
    bun\_zipper\_res3	&	-10	&	-2	&	-1	&	-1	&	0.387004	&	0.224203 \\
    bun\_zipper\_res4	&	-1	&	0	&	0	&	0	&	0.237761	&	0.19944 \\
    bun\_zipper	&	-34	&	-12	&	-2	&	0	&	0.366736	&	0.184346 \\
    chicken\_high	&	-65	&	-25	&	-10	&	0	&	0.352596	&	0.128001 \\
    cow10	&	-204	&	-2	&	0	&	0	&	0.131613	&	0.0956342 \\
    cow2	&	-213	&	-1	&	0	&	0	&	0.0974546	&	0.0992171 \\
    cow3	&	-200	&	1	&	-2	&	0	&	0.161176	&	0.118553 \\
    cow9	&	145	&	15	&	3	&	1	&	-	&	- \\
    cube2	&	-36	&	-1	&	-1	&	0	&	0.545168	&	0.322115 \\
    dog	&	-58	&	-9	&	-4	&	0	&	0.0901815	&	0.058035 \\
    dragon	&	-277	&	-35	&	-31	&	-2	&	0.221926	&	0.124984 \\
    dragon\_vrip\_res2	&	390	&	174	&	51	&	1	&	-	&	- \\
    dragon\_vrip\_res3	&	307	&	152	&	39	&	1	&	-	&	- \\
    dragon\_vrip\_res4	&	152	&	65	&	24	&	1	&	-	&	- \\
    dragon\_vrip	&	777	&	222	&	71	&	23	&	-	&	- \\
    ele12a	&	-87	&	-3	&	-1	&	0	&	0.185378	&	0.123558 \\
    ele\_fine	&	-196	&	-4	&	-6	&	0	&	0.187097	&	0.157745 \\
    elk2	&	-53	&	-1	&	-2	&	0	&	0.212646	&	0.251487 \\
    fertility	&	-46	&	-3	&	-1	&	0	&	0.273804	&	0.0874419 \\
    frog	&	-86	&	-2	&	0	&	0	&	0.22128	&	0.141774 \\
    gargoyle	&	-56	&	-19	&	-10	&	0	&	0.318482	&	0.147751 \\
    grandpiano	&	-4	&	0	&	0	&	0	&	0.164264	&	0.190796 \\
    happy4	&	-589	&	-409	&	-21	&	-100	&	0.199295	&	0.0887277 \\
    happy	&	982	&	308	&	58	&	44	&	-	&	- \\
    heart	&	-41	&	-7	&	5	&	0	&	0.336127	&	0.243949 \\
    heptoroid	&	-245	&	0	&	0	&	0	&	0.0705061	&	0.068063 \\
    hh2	&	-100	&	-1	&	-2	&	0	&	0.193822	&	0.189663 \\
    hh	&	-61	&	0	&	-3	&	0	&	0.17783	&	0.141936 \\
    hht	&	-86	&	1	&	1	&	0	&	0.149574	&	0.174906 \\
    hippo	&	-80	&	-17	&	-12	&	0	&	0.20746	&	0.153264 \\
    horse3	&	-190	&	-2	&	-3	&	0	&	0.130538	&	0.123331 \\
    horse9	&	-148	&	-2	&	-1	&	0	&	0.110968	&	0.101046 \\
    horse	&	-95	&	-1	&	-1	&	0	&	0.112222	&	0.111066 \\
    hound2	&	-88	&	-20	&	1	&	0	&	0.17887	&	0.0557737 \\
    hound	&	-25	&	0	&	2	&	-1	&	0.0854626	&	0.0731542 \\
    human\_hand	&	-64	&	-3	&	-1	&	0	&	0.181229	&	0.0866464 \\
    humerus	&	-60	&	-4	&	-1	&	0	&	0.0959782	&	0.0774577 \\
    kitten3	&	-25	&	-5	&	0	&	0	&	0.234016	&	0.267631 \\
    kitten	&	-46	&	-8	&	-1	&	0	&	0.26524	&	0.197679 \\
    lion2	&	-179	&	-56	&	-17	&	0	&	0.171686	&	0.118093 \\
    lion	&	-246	&	-60	&	-18	&	1	&	0.206408	&	0.10091 \\
    lucy	&	-175	&	-93	&	-13	&	0	&	0.163102	&	0.133512 \\
    memento	&	-72	&	-1	&	-2	&	0	&	0.326454	&	0.226977 \\
    micscope2	&	-1447	&	1	&	-5	&	0	&	0.0999958	&	0.410087 \\
    micscope	&	-6	&	-6	&	-5	&	2	&	0.0163587	&	0.526192 \\
    mouse2	&	-202	&	-19	&	-5	&	0	&	0.133177	&	0.0925803 \\
    mouse	&	-2	&	-3	&	1	&	0	&	0.0942147	&	0.110964 \\
    m	&	44	&	15	&	6	&	1	&	0.0580312	&	0.181962 \\
    neptune	&	-73	&	-16	&	-5	&	0	&	0.200924	&	0.0508065 \\
    noisydino2	&	-59	&	-10	&	-7	&	0	&	0.0858338	&	0.0586712 \\
    noisydino8a	&	-63	&	-3	&	-3	&	0	&	0.0876304	&	0.0788599 \\
    noisydino	&	-51	&	-6	&	-5	&	0	&	0.0871886	&	0.0806523 \\
    pig2	&	-233	&	-3	&	-1	&	0	&	0.281781	&	0.13851 \\
    pig	&	-5	&	1	&	1	&	0	&	0.136812	&	0.185151 \\
    pipe	&	-9	&	0	&	0	&	0	&	0.0608525	&	0.0367033 \\
    rcube2	&	-26	&	-2	&	-2	&	0	&	0.82234	&	0.37449 \\
    rcube	&	-35	&	-1	&	-1	&	0	&	0.552556	&	0.481529 \\
    rhino2	&	-190	&	-52	&	-20	&	0	&	0.285996	&	0.107248 \\
    rhino	&	-28	&	-13	&	-6	&	0	&	0.207169	&	0.124026 \\
    rockerArm2	&	-46	&	-1	&	-1	&	0	&	0.254169	&	0.157712 \\
    rockerArm	&	-33	&	-1	&	-1	&	0	&	0.262787	&	0.147716 \\
    rotor	&	-29	&	3	&	1	&	0	&	0.0792849	&	0.0977375 \\
    sacrum	&	-147	&	-22	&	-12	&	-4	&	0.408094	&	0.169373 \\
    sandal2	&	-315	&	5	&	-6	&	-2	&	0.136338	&	0.0763513 \\
    sandal3	&	-95	&	16	&	-11	&	-6	&	0.148504	&	0.0690598 \\
    sandal	&	6	&	4	&	4	&	-2	&	0.102921	&	0.0923182 \\
    scapula	&	-81	&	-10	&	-9	&	0	&	0.325198	&	0.170676 \\
    screwdriver	&	-55	&	-2	&	-1	&	0	&	0.19814	&	0.0929329 \\
    seabowl	&	-88	&	-19	&	-5	&	-1	&	0.366275	&	0.173513 \\
    ship1	&	2	&	-9	&	6	&	4	&	0.248314	&	0.187957 \\
    ship2	&	-79	&	2	&	-1	&	-6	&	0.196807	&	0.20017 \\
    ship3	&	-167	&	13	&	-5	&	-5	&	0.166875	&	0.23599 \\
    spider	&	12	&	-2	&	-2	&	0	&	0.0914775	&	0.0515585 \\
    tomgun2	&	-89	&	51	&	-9	&	-2	&	0.0771206	&	0.0655649 \\
    tomgun	&	3	&	-5	&	-2	&	0	&	0.045953	&	0.0647683 \\
    walleye2	&	-49	&	0	&	4	&	-1	&	0.133166	&	0.041185 \\
    walleye	&	-1	&	-10	&	3	&	0	&	0.212548	&	0.0931524 \\
    \caption{Comparison between skeletons generated by LSS and LEM. The metrics denoted by $\Delta$ are relative to the skeletons of LSS, with negative values implying that LSS has more vertices, leafs, branches etc. Here $H($LEM,LSS$)$ denotes the directed Hausdorff distance between LEM and LSS, divided by the radius of a bounding sphere, and vice versa.}
    \label{tab:skelqual_lem}
\end{longtable}

\subsection{LSS and LEMTS}
\begin{longtable}{c|c|c|c|c|c|c}
    \centering
    input & $\Delta$vertices &  $\Delta$leafs & $\Delta$branches & $\Delta$genus & $H($LEMTS,LSS$)$ & $H($LSS,LEMTS$)$  \\\hline
    17674	&	-173	&	-9	&	-7	&	0	&	0.255787	&	0.140655 \\
    18020	&	-141	&	0	&	8	&	0	&	0.0369842	&	0.0388943 \\
    19465	&	-416	&	11	&	-24	&	3	&	0.0355444	&	0.0382796 \\
    20304	&	1	&	9	&	-2	&	-2	&	0.331099	&	0.371398 \\
    20306	&	2	&	1	&	0	&	0	&	0.141557	&	0.175343 \\
    21362	&	3	&	3	&	0	&	0	&	0.0831445	&	0.205519 \\
    21464	&	-37	&	-1	&	-1	&	0	&	0.0287383	&	0.0847032 \\
    21747	&	-3	&	0	&	0	&	0	&	0.0122935	&	0.0121544 \\
    21788	&	-21	&	-1	&	4	&	0	&	0.316065	&	0.241109 \\
    22081	&	-36	&	1	&	0	&	0	&	0.0184155	&	0.0292149 \\
    22290	&	-3	&	-9	&	1	&	-1	&	0.701622	&	0.593782 \\
    22601	&	-13	&	-8	&	3	&	0	&	0.0505034	&	0.041482 \\
    22669	&	-95	&	1	&	1	&	0	&	0.107136	&	0.18301 \\
    22701	&	-1	&	2	&	0	&	0	&	0.47311	&	0.324789 \\
    22711	&	-4	&	0	&	0	&	0	&	0.0496445	&	0.0350731 \\
    22808	&	-79	&	0	&	0	&	0	&	0.0509979	&	0.0508452 \\
    22826	&	-23	&	17	&	2	&	-2	&	0.142766	&	0.179552 \\
    23091	&	-28	&	0	&	0	&	0	&	0.194078	&	0.142976 \\
    711\_dente	&	-9	&	-4	&	-2	&	0	&	0.461223	&	0.304697 \\
    armadillo	&	-108	&	-17	&	-3	&	0	&	0.218784	&	0.145593 \\
    armadillo\_vc	&	-95	&	-17	&	-4	&	0	&	0.256114	&	0.149753 \\
    asiandragon10a	&	664	&	63	&	33	&	1	&	-	&	- \\
    asiandragon9a	&	-204	&	-26	&	0	&	0	&	0.149798	&	0.0878272 \\
    ballJoint	&	-30	&	-9	&	-5	&	0	&	0.345708	&	0.107102 \\
    bird2	&	-24	&	0	&	1	&	0	&	0.0646921	&	0.0660722 \\
    bird3	&	-45	&	-6	&	-2	&	0	&	0.106499	&	0.0485063 \\
    bird4	&	-79	&	-6	&	-1	&	0	&	0.154138	&	0.0779006 \\
    bird	&	-29	&	-1	&	-1	&	0	&	0.0540386	&	0.049263 \\
    bmw2	&	-49	&	-9	&	11	&	3	&	0.204429	&	0.154384 \\
    bmw	&	-25	&	17	&	22	&	1	&	0.143202	&	0.139063 \\
    boeing	&	1	&	21	&	2	&	-9	&	0.0911128	&	0.0844408 \\
    brain2	&	-71	&	-50	&	-5	&	-1	&	0.442813	&	0.243605 \\
    bunny	&	-29	&	-10	&	-1	&	-1	&	0.381509	&	0.185589 \\
    bun\_zipper\_res2	&	-21	&	-6	&	-4	&	0	&	0.332003	&	0.206632 \\
    bun\_zipper\_res3	&	-10	&	-4	&	-2	&	0	&	0.353706	&	0.228834 \\
    bun\_zipper\_res4	&	0	&	1	&	0	&	0	&	0.18867	&	0.390443 \\
    bun\_zipper	&	-35	&	-10	&	-3	&	-1	&	0.373315	&	0.186499 \\
    chicken\_high	&	-56	&	-25	&	-10	&	0	&	0.345959	&	0.104203 \\
    cow10	&	-129	&	-2	&	0	&	0	&	0.132759	&	0.0965579 \\
    cow2	&	-132	&	-1	&	1	&	0	&	0.145401	&	0.187487 \\
    cow3	&	-126	&	0	&	0	&	0	&	0.188258	&	0.10453 \\
    cow9	&	217	&	15	&	8	&	1	&	-	&	- \\
    cube2	&	-35	&	-1	&	-1	&	0	&	0.502338	&	0.334772 \\
    dog	&	-44	&	-9	&	-4	&	0	&	0.0854458	&	0.072164 \\
    dragon	&	-222	&	-36	&	-26	&	-3	&	0.255397	&	0.11414 \\
    dragon\_vrip\_res2	&	412	&	170	&	52	&	1	&	-	&	- \\
    dragon\_vrip\_res3	&	323	&	155	&	44	&	1	&	-	&	- \\
    dragon\_vrip\_res4	&	149	&	64	&	22	&	1	&	-	&	- \\
    dragon\_vrip	&	824	&	227	&	74	&	19	&	-	&	- \\
    ele12a	&	-54	&	-3	&	0	&	0	&	0.174368	&	0.182298 \\
    ele\_fine	&	-158	&	-4	&	-3	&	0	&	0.202761	&	0.163576 \\
    elk2	&	-35	&	-1	&	-1	&	0	&	0.189953	&	0.170198 \\
    fertility	&	-30	&	-3	&	-1	&	0	&	0.240033	&	0.165239 \\
    frog	&	-65	&	-2	&	2	&	0	&	0.213764	&	0.176573 \\
    gargoyle	&	-54	&	-21	&	-11	&	0	&	0.379466	&	0.147014 \\
    grandpiano	&	0	&	2	&	0	&	0	&	0.129011	&	0.162077 \\
    happy4	&	-560	&	-384	&	-22	&	-99	&	0.160472	&	0.100934 \\
    happy	&	1020	&	334	&	57	&	41	&	-	&	- \\
    heart	&	-21	&	-2	&	5	&	0	&	0.326899	&	0.232986 \\
    heptoroid	&	-132	&	0	&	2	&	0	&	0.0975426	&	0.0763043 \\
    hh2	&	-70	&	2	&	-1	&	0	&	0.19555	&	0.170324 \\
    hh	&	-44	&	0	&	0	&	0	&	0.169264	&	0.116556 \\
    hht	&	-60	&	1	&	-1	&	0	&	0.177192	&	0.16365 \\
    hippo	&	-64	&	-13	&	-6	&	0	&	0.19232	&	0.132381 \\
    horse3	&	-125	&	0	&	-1	&	0	&	0.119945	&	0.110306 \\
    horse9	&	-96	&	-2	&	-1	&	0	&	0.113884	&	0.103661 \\
    horse	&	-66	&	-2	&	-2	&	0	&	0.0883287	&	0.0851157 \\
    hound2	&	-54	&	-16	&	2	&	0	&	0.190078	&	0.0818325 \\
    hound	&	-22	&	-3	&	1	&	0	&	0.0846057	&	0.0696693 \\
    human\_hand	&	-46	&	-2	&	-1	&	0	&	0.147885	&	0.0809871 \\
    humerus	&	-37	&	-6	&	-3	&	0	&	0.112056	&	0.0725165 \\
    kitten3	&	-18	&	-5	&	0	&	0	&	0.221592	&	0.306183 \\
    kitten	&	-29	&	-5	&	1	&	0	&	0.271288	&	0.201393 \\
    lion2	&	-147	&	-55	&	-10	&	0	&	0.170177	&	0.121737 \\
    lion	&	-203	&	-59	&	-19	&	1	&	0.209613	&	0.102484 \\
    lucy	&	-173	&	-95	&	-17	&	0	&	0.181825	&	0.147681 \\
    memento	&	-47	&	-1	&	-2	&	0	&	0.335567	&	0.234729 \\
    micscope2	&	-1188	&	7	&	5	&	0	&	0.107463	&	0.514925 \\
    micscope	&	-8	&	0	&	-6	&	-1	&	0.0139525	&	0.73694 \\
    mouse2	&	-100	&	-15	&	2	&	0	&	0.122646	&	0.079835 \\
    mouse	&	-3	&	-4	&	1	&	0	&	0.105162	&	0.14296 \\
    m	&	82	&	44	&	13	&	-1	&	0.0563222	&	0.19182 \\
    neptune	&	-51	&	-15	&	-1	&	0	&	0.225756	&	0.0516766 \\
    noisydino2	&	-59	&	-10	&	-6	&	0	&	0.0938331	&	0.0921727 \\
    noisydino8a	&	-49	&	-3	&	-2	&	0	&	0.0855693	&	0.0707761 \\
    noisydino	&	-49	&	-4	&	-5	&	0	&	0.0789956	&	0.073725 \\
    pig2	&	-141	&	-3	&	0	&	0	&	0.275567	&	0.124252 \\
    pig	&	-8	&	1	&	0	&	0	&	0.124846	&	0.152941 \\
    pipe	&	-4	&	0	&	0	&	0	&	0.0496445	&	0.0350761 \\
    rcube2	&	-24	&	-1	&	-1	&	0	&	0.6971	&	0.583039 \\
    rcube	&	-37	&	-2	&	-2	&	0	&	0.487541	&	0.225971 \\
    rhino2	&	-144	&	-46	&	-18	&	0	&	0.240027	&	0.211737 \\
    rhino	&	-14	&	-7	&	-2	&	0	&	0.398101	&	0.126916 \\
    rockerArm2	&	-36	&	0	&	0	&	0	&	0.240848	&	0.250368 \\
    rockerArm	&	-25	&	-1	&	-1	&	0	&	0.212998	&	0.0977234 \\
    rotor	&	-23	&	7	&	5	&	0	&	0.0882883	&	0.0897212 \\
    sacrum	&	-126	&	-28	&	-8	&	-1	&	0.427385	&	0.157824 \\
    sandal2	&	-241	&	10	&	-4	&	-2	&	0.0668588	&	0.0665051 \\
    sandal3	&	-73	&	17	&	-10	&	-6	&	0.0947857	&	0.0566952 \\
    sandal	&	7	&	4	&	3	&	-2	&	0.123407	&	0.0806454 \\
    scapula	&	-65	&	-9	&	-8	&	0	&	0.332365	&	0.244491 \\
    screwdriver	&	-36	&	-1	&	0	&	0	&	0.203488	&	0.104393 \\
    seabowl	&	-77	&	-20	&	-4	&	-1	&	0.32748	&	0.206797 \\
    ship1	&	2	&	-6	&	1	&	2	&	0.215595	&	0.182269 \\
    ship2	&	-74	&	1	&	1	&	1	&	0.170628	&	0.168638 \\
    ship3	&	-164	&	-1	&	-4	&	-2	&	0.151358	&	0.160943 \\
    spider	&	31	&	-2	&	-2	&	0	&	0.0956197	&	0.0793788 \\
    tomgun2	&	-33	&	75	&	7	&	-2	&	0.0602913	&	0.032726 \\
    tomgun	&	10	&	-5	&	1	&	1	&	0.0462826	&	0.0491785 \\
    walleye2	&	-16	&	13	&	18	&	-3	&	0.150528	&	0.0501536 \\
    walleye	&	-2	&	-11	&	2	&	0	&	0.212825	&	0.0933191 \\
    \caption{Comparison between skeletons generated by LSS and LEM. The metrics denoted by $\Delta$ are relative to the skeletons of LSS, with negative values implying that LSS has more vertices, leafs, branches etc. Here $H($LEMTS,LSS$)$ denotes the directed Hausdorff distance between LEMTS and LSS, divided by the radius of a bounding sphere, and vice versa}
    \label{tab:skelqual_lemts}
\end{longtable}
\section{On Computing Local Separators}
\subsection{Dynamic Connectivity}\label{sec:dyncon}
To speed up the check for disconnection of the front when computing local separators, Bærentzen and Rotenberg suggests~\cite{lsalg} using a dynamic connectivity data structure such as the one by Holm, de Lichtenberg and Thorup~\cite{dyncon}. This data structure uses a hierarchy of Euler Tour Forests to give $O(\log^2|V|)$ amortized update and query time. In practice, it can be beneficial to stop using the hierarchy, once forests reach small size. 
\begin{figure}[htb]
    \centering
    \includegraphics[width=\textwidth]{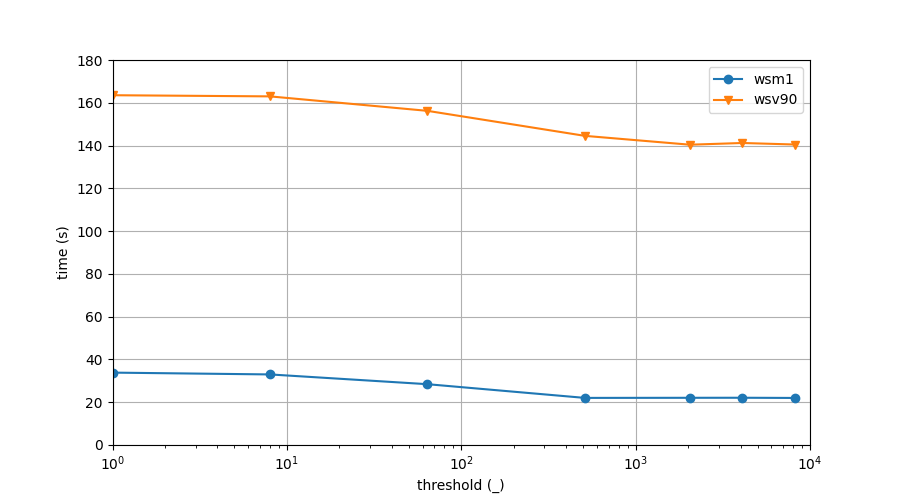}
    \caption{Running time of local separator skeletonization using increasing thresholds in the dynamic connectivity data structure. The tested input wsm1 is a mesh while wsv is a voxel grid representing the same shape.}
    \label{fig:thresholdtest}
\end{figure}

We have implemented the structure, integrated it into the local separator skeletonization implementation of GEL and tested the performance for various thresholds for when to stop using the hierarchy. The measurements are presented in Figure~\ref{fig:thresholdtest}. What we found is that the performance increases with the size of the threshold, up to a point where the entirety of the hierarchy exists on a single level. In this manner, the data structure simply reduces to maintaining an augmented Euler Tour Forest. This is likely due to the specific access pattern when computing local separators, where edges are inserted and removed only once in a manner where separators that were inserted early are often removed before those inserted recently.

\begin{figure}[htb]
    \centering
    \includegraphics[width=0.48\textwidth]{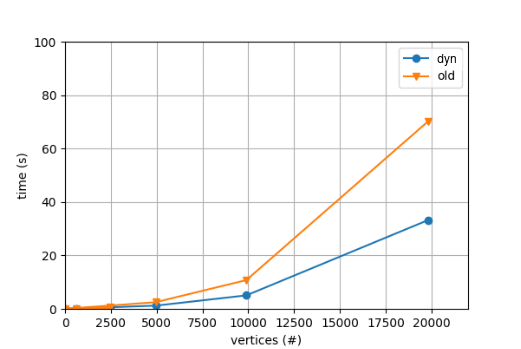}
    \includegraphics[width=0.48\textwidth]{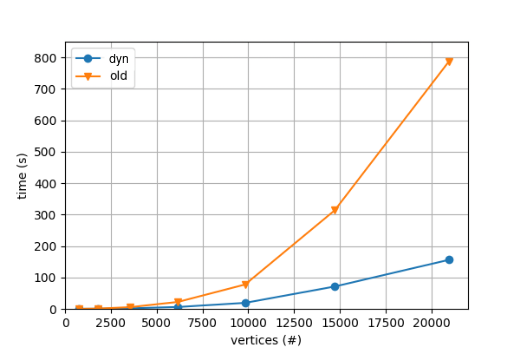}
    \caption{Running times of local separator skeletonization without (old) and with (dyn) the use of dynamic connectivity. Left shows performance on meshes while right shows performance on voxels.}
    \label{fig:dynconrun}
\end{figure}

We also examine and compare running times when using dynamic connectivity. In Figure~\ref{fig:dynconrun} we compare the old method to the one using dynamic connectivity (dyn), and find that for both meshes and voxels there is a large improvement. It is worth noting that this improvement seems even greater for voxels.

\subsection{Analysis of Minimising Separators}\label{sec:analysisofshrinking}
Given a local separator $\Sigma$ on a graph $G$ with $n$ vertices and $m$ edges we describe the time it takes to minimise $\Sigma$.

To smooth the attribute of a vertex, we have to consider the positions of that vertex' neighbours, giving a time complexity on the order of the degree of the vertex. Since we do this for every vertex in the separator, the time is proportional to the sum of degrees of vertices in the separator which is then bounded by two times the number of edges in the neighbourhood of the separator, by the handshaking lemma. Let $m'$ be the number of edges in the closed neighbourhood of $\Sigma$, the time to smooth is then $O(m')$. Computing the heuristic after smoothing takes $O(1)$ time per vertex, totalling in $O(|\Sigma|)$ time. We then sort the vertices according to the smoothed attributes in $O(|\Sigma|\log|\Sigma|)$ time.
    
After sorting we iteratively move vertices from the separator to front components. In the worst case we move only a constant number of vertices in each iteration, resulting in $O(|\Sigma|)$ total iterations over a collection of $O(|\Sigma|)$ vertices. Assuming we can move a vertex from one set to another in constant time, this step takes $O(|\Sigma|^2)$ time.
    
The total time it takes is then $O(|\Sigma|^2+m')$. Once again we can pessimistically bound both sets on the size of the graph s.t. $|\Sigma|=O(n)$ and $m'=O(m)$ but also $m=O(n^2)$. The time it takes to shrink a single separator is then bounded by $O(n^2)$ in the worst case.

\end{document}